\documentclass[a4paper,10pt]{report}
\usepackage[utf8x]{inputenc}
\usepackage[english]{babel}

\usepackage[dvips]{graphicx}
\usepackage{wrapfig,epsfig,epsf}
\usepackage{graphics}

\usepackage{amsmath}
\usepackage{multicol}

\begin{document}

\title{ Analytical models of relativistic accretion disks }

\author{ 
V.V. Zhuravlev$^{1}$} 

\date{ \it \small
Sternberg Astronomical Institute \\
Moscow MV Lomonosov State University
\footnote {Corresponding e-mail: zhuravlev@sai.msu.ru}
}

\maketitle

\label{firstpage}

\begin{abstract}

We present not a literature review
but a description, as detailed and consistent as possible, of two 
analytic models of disk accretion onto a rotating black hole: a standard 
relativistic disk and a twisted relativistic disk. Although one of these models
is much older than the other, both are of topical current interest for black hole
studies. The way the exposition is presented, the reader with only a limited
knowledge of general relativity and relativistic hydrodynamics can --- with little or no  use of additional sources -- 
gain good insight into many technical details lacking in the original papers. 

\vspace{0.5cm}

{\bf Keywords:} accretion, accretion discs, black hole physics,
hydrodynamics

\end{abstract}

\tableofcontents

\chapter{Relativistic standard accretion disk}

In Chapter 1, the model of a standard accretion 
disk around a rotating black hole is presented with general relativity effects rigorously taken into account. This model was first described in
paper \cite{NT} and since then has been used in many studies to obtain a convincing evidence 
of the existence of black holes in both stellar binary systems and active galactic nuclei. It remains topical because a full account for general relativistic properties of matter motion in the disk and generation of its emission allows
the position of the inner disk radius and hence the 
black hole spin to be inferred from observations. Observational appearances of relativistic disks were modeled for the first time in
\cite{Cunn_1975} (see \cite{Gierl_2001}, \cite{Li_2005} for comprehensive reviews, references therein and their citations list).
In addition, the standard accretion disk underlies more complicated theories of warped (twisted) accretion disks which are formed when the accreting matter moves outside 
the equatorial plane of a rotating black hole. 
Such a theory is presented in \cite{ZhI} and is discussed in more detail in Chapter 2.

Everywhere below, the natural units $G=c=1$ are used. 
If the mass is measured in units of the black hole mass, $M$, 
the unit of length is half of the Schwarzschild gravitational radius, $R_g/2$, such that $R_g/2 = GM/c^2=1$
and the unit of time is the light crossing time of the unit of length. 

In addition, Latin indices
$i,j,k...$ taking values from 0 to 3 are used to denote components of vectors, with  the zero component standing for the time coordinate. Also, wherever needed the Einstein summation convention is used. 
  
\section{Space-time near rotating black holes}
\subsection{Kerr metric}
\label{sec_metric}

Properties of spacetime near a rotating black hole are described by axially symmetric and stationary metric of the form (see, e.g., \cite{NFr_1986},
paragraph 4.2):

\begin{equation}
\label{Kerr}
ds^2 = \frac{\varrho^2\Delta}{\varSigma^2} dt^2 - \frac{\varSigma^2 \sin^2\theta}{\varrho^2} (d\phi - \omega dt)^2 - 
\frac{\varrho^2}{\Delta} dR^2 - \varrho^2 d\theta^2, 
\end{equation}
where the signature $(1,-1,-1,-1)$ is chosen and the coefficients are
$$
\varSigma^2 = (R^2+a^2)^2 - a^2 \Delta \sin^2 \theta, 
$$
$$
\varrho^2 = R^2 + a^2 \cos^2 \theta \quad \mbox{and} \quad \Delta = R^2 - 2R + a^2
$$
The coordinates  $\{ t, \phi, R, \theta\}$ are called the Boyer-Lindquist coordinates. Far away from 
the gravitating body, the spatial part of these coordinates in the limit of the zero
black hole spin parameter, $a$, transits into the usual spherical coordinates, where 
$\phi$ is the azimuthal angle. For non-zero $a$, it transits into generalized spherical coordinates in which the surfaces of constant radial distance, 
$R=const$, represent spheroids with the aspect ratio $R/(R^2+a^2)^{1/2} $.
 
The space-time described by (\ref{Kerr}), is axially symmetric with respect to the line 
$\theta=0$, called the black hole rotation axis; the plane corresponding to $\theta=\pi/2$, is called the black hole equatorial axis. 

In (\ref{Kerr}) an important quantity appears:  
\begin{equation}
\label{LT}
\omega = \frac{2aR}{\Sigma^2},
\end{equation}
which has the dimension of frequency. This is the angular velocity that every freely moving observer, without exception, acquires in the direction of the black hole rotation. 

As is described in the literature on the structure of rotating black holes (see, e.g., \cite{Chandr2},
paragraph 58), metric (\ref{Kerr}) has several special hypersurfaces, including the event horizon and the ergosphere. 
However, as we show below, for the astrophysical problem under consideration, of most importance is the dynamics of free circular motion of particles in the equatorial plane of the gravitating body. 
This motion has additional features in comparison to the corresponding Newtonian problem. Note also, that in Chapter 2 weakly elliptical orbits slightly inclined to the equatorial plane are considered. 

\vspace{0.2cm}

We consider a standard, and hence geometrically thin, accretion disk. Such a disk is basically flat. 
By definition, this is a stationary flow of matter with mirror symmetry with respect to its middle plane and axial symmetry with respect to the line perpendicular to this plane. 
Clearly, such a model flow can be described by dynamical equations in axially symmetric metric only if the disk symmetry plane coincides with the equatorial plane of the black hole. 
To tackle the problem, only the form of metric near the plane $\theta=\pi/2$ is sufficient. Passing to cylindrical coordinates using the standard transformation
$$
r = R \sin \theta, \quad z = R \cos \theta,
$$
all metric coefficients $g_{ik}$ in (\ref{Kerr}) 
can be expanded in a power series in small ratio  $z/r \ll 1$. 
For geometrically thin disks, the corrections to $g_{ik}$
due to non-equatorial motion up to $(z/r)^2$ are sufficient. Indeed, one of the basic equations describing the disk, namely, 
the projection of the relativistic analog of the Euler equation onto the direction normal to the disk plane, must be odd with respect to the coordinate reflection $z\to -z$ due to the mirror symmetry of the disk. 
This means that in its series expansion in $(z/r)$ only odd powers of $(z/r)$ must be present. By the main assumption on the smallness of  $(z/r)$ only the first term in this expansion should be kept. 
This, in turn, corresponds to series expansion of $g_{ik}$ 
up to quadratic terms, since only first derivatives of $g_{ik}$, characterizing the 'strength' of the gravitational field, enter dynamical equations. 

Note, however, that hydrodynamic equations also contain a second covariant derivative of the velocity field (see below), and hence the final expressions can involve second derivatives of $g_{ik}$ with respect to $z$, which may seem to require that we keep terms of the order of  $(z/r)^3$ in $g_{ik}$.
But this is not required, because, as follows from the explicit form of the stress-energy tensor, such terms can appear only 
when multiplied by some of the viscous coefficients, which in turn cannot be greater than of the order of
$(z/r)$ being proportional to the characteristic mixing length in the fluid. The latter is initially assumed to be less than the disk thickness. 

As regards other equations, namely (see below): two projections of the relativistic analog of the Euler equation on to the disk plane, 
the energy balance equation and the rest energy conservation law -- the same symmetry considerations imply that they are even under the coordinate reflection
$z\to -z$; therefore, the leading term is of the zeroth order in $(z/r)$ in the metric expansion. 

Using these expansions and expressions for the coordinate differentials,
$$
dR = \left ( 1-\frac{1}{2}\frac{z^2}{r^2} \right ) dr + \frac{z}{r} dz,
$$
$$
d\theta = \frac{z}{r} \frac{dr}{r} - \left ( 1 - \frac{z^2}{r^2} \right ) \frac{dz}{r}, 
$$
we find the metric in the following form (see also \cite{RH_1995}):

\begin{multline}
\label{Kerr_rz}
ds^2 = \left [  1 - \frac{2}{r} + \frac{z^2}{r^3}\left ( 1 + \frac{2a^2}{r^2} \right ) \right ] dt^2 - 
\left [ r^2 + a^2 + \frac{2a^2}{r} - \frac{a^2 z^2}{r^2}\left ( 1 + \frac{5}{r} + \frac{2a^2}{r^3} \right ) \right ] d\phi^2 + \\
\frac{2a}{r} \left [ 2 - \frac{z^2}{r^2} \left ( 3 + \frac{2a^2}{r^2}\right ) \right ] d t d\phi - \\
\left \{ 1 - \frac{z^2}{r^2 D} \left [ \frac{3}{r} - \frac{4}{r^2} - \frac{a^2}{r^2} 
\left ( 3 - \frac{6}{r} + \frac{2a^2}{r^2} \right ) \right ] \right \} \frac{dr^2}{D} - 
\frac{2 z}{r D} \left ( \frac{2}{r} - \frac{a^2}{r^2} \right ) dr dz - \\
\left [ 1 + \frac{z^2}{r^2 D} \left ( \frac{2}{r} - \frac{2a^2}{r^3} + \frac{a^4}{r^4} \right ) \right ] dz^2,
\end{multline}
where the notation 

$$
D = 1 - \frac{2}{r} + \frac{a^2}{r^2}
$$
is introduced. Below, we also use (with a few exceptions) the notations introduced in the original paper 
by Novikov and Thorne \cite{NT} for the relativistic correction coefficients. 
 
Finally, the inverse of the matrix $g^{ik}$ corresponding to double-contravariant tensor has the form:
\begin{equation}
\label{g^ik}
g^{ik} = \left | 
\begin{array}{cc}
(g_{tt} g_{\phi\phi} - g_{t\phi}^2)^{-1} \times 
 
\left | 
\begin{array}{ll}
g_{\phi\phi}  & -g_{t\phi} \\
-g_{t\phi} & -g_{tt}
\end{array}
 \right |  
& 0 \\ 
0 & (g_{rr} g_{zz} - g_{rz}^2)^{-1} \times 
\left | 
\begin{array}{ll}
g_{zz}  & -g_{rz} \\
-g_{rz} & -g_{rr}
\end{array}
\right |
\end{array}
 \right |
\end{equation}

\subsection{Circular equatorial geodesics}
\label{sec_geod}
The expression for circular equatorial geodesics can be conveniently found from the extremum condition for the distance along them. Here we follow the exposition from  \cite{HEL_2006}, 
(see paragraphs 13.10 and 13.13).
Indeed, for time-like trajectories the functional

$$
S = \int L ds = \int g_{ik}\, \frac{d x^i}{d s} \frac{d x^k}{d s} d s,
$$
should be minimal, which is equivalent to the Euler-Lagrange equations for $L$:
\begin{equation}
\label{lagr_geod}
\frac{d}{d s} \left ( \frac{\partial L}{\partial \dot x^i} \right ) - \frac{\partial L}{\partial x^i} = 0,
\end{equation}
where $U^i_g \equiv d x^i/ds \equiv \dot x^i$ is the four-velocity in the Boyer-Lindquist coordinates. As $L$ does not explicitly depend on $t$ and $\phi$, 
the following quantities are conserved:
$$
g_{t i} U^i_g = k,
$$
$$
g_{\phi i} U^i_g = -h,
$$
where $k$ and $h$ have the meaning of the time and azimuthal covariant velocity components, respectively. 

In explicit form, using the components $g_{ik}$ from (\ref{Kerr_rz}) at $z=0$, we find:
\begin{equation}
\label{geod_t}
\left ( 1-\frac{2}{r} \right ) \dot t + \frac{2a}{r} \dot\phi = k,
\end{equation}
\begin{equation}
\label{geod_phi}
\frac{2a}{r} \dot t - \left ( r^2 + a^2 + \frac{2a^2}{r} \right ) \dot \phi = -h.
\end{equation}

We temporarily assume that the motion is not necessarily circular and $U_g^r\neq 0$. 
Instead of the $r$-component of the Euler-Lagrange equations, it is more convenient to use the condition of the normalization of the four-velocity of particles with non-zero mass:
\begin{equation}
\label{geod_norm0}
g^{tt} k^2 - 2g^{t\phi} k h + g^{\phi\phi} h^2 + g^{rr} (U_r)^2 = 1.
\end{equation}
This yields the following equation for $k$ and $h$:
\begin{equation}
\label{geod_norm}
\frac{\dot r^2}{2} + V_{eff}(r) = \frac{k^2-1}{2},
\end{equation}
where we introduce the effective potential 
\begin{equation}
\label{V_eff}
V_{eff} = -\frac{1}{r} + \frac{h^2-a^2(k^2-1)}{2r^2} - \frac{(h-ak)^2}{r^3}.
\end{equation}

The conditions for circular motion include, first, $\dot r=0$ and, second, $\ddot r=0$ (for the particle to stay in a circular orbit). 
The latter condition is equivalent to the vanishing of the derivative of $V_{eff}$ with respect to $r$:
\begin{equation}
\label{geod_norm1}
1 + \frac{a^2(k^2-1)-h^2}{r} + \frac{3(h-ak)^2}{r^2} = 0.
\end{equation}

Equation (\ref{geod_norm}) with $\dot r=0$ and equation (\ref{geod_norm1}) allow us to determine $k$ and $h$ as functions of $r$ and then, 
using (\ref{geod_t}) and (\ref{geod_phi}), to find $U^t_g$ and $U^\phi_g$.

To solve the first problem, let us introduce the new variable $\mu\equiv h-ak$ and, to facilitate manipulations, make the change $u\equiv 1/r$. 
Then equation (\ref{geod_norm}) taken with $\dot r=0$ and equation (\ref{geod_norm1}) yield the following equation for $\mu$:
\begin{equation}
\label{mu}
u^2[(3u-1)^2-4a^2u^3] \mu^4 - 2u [ (3u-1)(a^2 u-1) -2u a^2 (u-1) ] \mu^2 + (au-1)^2 = 0.
\end{equation}

The solution of (\ref{mu}) for a stable circular prograde orbit has the form:
\begin{equation}
\label{mu_sol}
\mu= - \frac{a\sqrt{u} - 1}{[u(1-3u+2au^{3/2})]^{1/2}}.
\end{equation}

Using (\ref{mu_sol}) and (\ref{geod_norm}) taken at $\dot r=0$, we find the constants  $h$ and $k$, as well as the components $U^i_g$:
\begin{equation}
\label{geod_vel}
{U_g}^t = C^{-1/2} \, B, \quad {U_g}^\phi = (r^3 C)^{-1/2}, \quad {U_g}^r = {U_g}^z = 0,  
\end{equation}
where
\begin{equation}
\label{B_C}
B = 1 + \frac{a}{r^{3/2}}, \quad C = 1 - \frac{3}{r} + \frac{2a}{r^{3/2}}
\end{equation}

It is easy to check that modulus of this vector is equal to unity:
$$
g_{ik}\, {U_g}^i {U_g}^k = 1
$$

The angular velocity measured by the clock of an infinite observer (who measures the coordinate time $t$), 
corresponding to such a motion, is 
\begin{equation}
\label{kepl_freq}
\Omega = \frac{d\phi}{dt} = r^{-3/2} B^{-1}
\end{equation}
It follows that in the Schwarzschild case this value exactly coincides with the Keplerian angular velocity.

\subsection{Radius of the innermost (marginally) stable orbit}
\label{sec_rms}

This is determined by the condition 
that the stable circular motion is no longer possible when the minimum of the function $V_{eff}(r,h(r_c),k(r_c))$ disappears at $r=r_c$,
where $r_c$ is the radius of a circular orbit. This is equivalent to the condition
$$
\left . \frac{d^2 V_{eff}}{dr^2} \right |_{r=r_c} = 0,
$$
which leads to the quartic equation
\begin{equation}
\label{eq_r_ms}
z^4 - 6z^2 + 8az - 3a^2 = 0,
\end{equation}
where $z\equiv r^{1/2}$.
 
Using the Ferrari method (see, e.g., \cite{KK_1973}), we
write the corresponding auxiliary cubic equation:
\begin{equation}
\label{eq_r_ms_cub}
y^3 - 12y^2 + 12(3+a^2) y- 64a^2 = 0. 
\end{equation}
The real root of equation (\ref{eq_r_ms_cub}) is related to the Cardano solution 
to the corresponding incomplete cubic equation and is given by
\begin{equation}
\label{root_kardano}
y_1 = -2 (1-a^2)^{1/3} [ (1+a)^{1/3} + (1-a)^{1/3}] + 4.
\end{equation}

Next, having $y_1$, it is possible to use the Ferrari solution to write the quadratic equation that gives two real roots of (\ref{eq_r_ms}):
\begin{equation}
\label{quadr_eq}
p^2 + \sqrt{y_1} p + \frac{1}{2} \left ( -6 + y_1 - \frac{8a}{\sqrt{y_1}} \right ) = 0
\end{equation}
The larger root of (\ref{quadr_eq}), $p_1$, determines the boundary of the stable circular motion of test particles in the equatorial plane, which we denote as $r=r_{ms}$. 
Thus, 
\begin{equation}
\label{r_ms}
r_{ms} = p_1^2 = 3 + \frac{4a}{\sqrt{y_1}} - (-y_1^2/4 + 4a\sqrt{y_1} + 3y_1)^{1/2} 
\end{equation}
It is easy to check that the result (\ref{r_ms}) coincides with the expression 
presented in \cite{PT_1974},
(see formula (15k) therein), taking into account that 
the auxiliary values $Z_{1,2}$ in \cite{PT_1974} take the form
$Z_1 \equiv 3-y_1/2$ and $Z_2 \equiv 4a/\sqrt{y_1}$ for $a\ge 0$ in our notation.

In the case of the Schwarzschild metric, $a=0$, we recover the well-known result that 
the circular motion becomes unstable for $r<6$, i.e. at distances smaller than three gravitational radii from the black hole. For slow rotation, 
$1 \gg a > 0$, we have $r_{ms}\approx 6 - 4\sqrt{6}a/3$, and hence
the zone of stable motion shifts closer towards the event horizon. In the limit case $a=1$ we find $r_{ms}=1$, i.e. the marginally stable circular orbit coincides with the gravitational radius of the extreme-spin black hole. 

During accretion, gas elements in the disk slowly approach $r_{ms}$ by loosing their angular momentum due to the action of viscous forces. 
Once the gas elements fall into the region with $r<r_{ms}$, due to instability of the circular motion they need not lose the angular momentum any more to approach the black hole. 
This means that the matter falls freely inside $r_{ms}$, and the standard accretion disk model assumes that $r_{ms}$ is the inner disk radius.

\section{Choice of the reference frame}

\subsection{Bases in general relativity}
\label{sec_bases}
Mechanical laws, formulated in the form of vector equations, can be written in the symbolic form irrespective to any observer or any reference frame. 
But to represent some physical quantity describing a natural phenomenon in the form of a set of numerical values, the measurement procedure should be specified. 
In Newtonian mechanics, this means that the observer introduces a coordinate system, and then at each point of space he/she arbitrarily constructs three basis vectors. 
The coordinate system and the basis vectors can evolve in time, which is the same at all points. 
A tool measuring time, together with the coordinate grid and a vector basis, form a reference frame, in which any physical value (scalar, vector or tensor) can be measured, i.e. it can be represented by a number or a set of numbers. 

The situation in relativistic mechanics is different: because it is not possible to consider the time independently, it becomes the fourth component of the space-time continuum. 
Therefore, the choice of the reference frame reduces to the construction of a coordinate system and four basis vectors determined at each point of space-time. 
In general, this set of basis vectors is usually referred to as a tetrad. 
There is no a universal observer any more; instead, a set of observers moving along a certain family of world lines is considered. 
If one of the tetrad orts, conventionally corresponding to the time direction, is tangent at each point to these world lines, the tetrad is said to be 'transported' by the observers.  
The last statement can be easily understood, because in such a basis, the four-velocity of each observer at any time has a non-zero projection only on the 'time' ort, in other words, 
the observers are at rest in this basis, i.e. transport it with them.

\subsubsection{Coordinate representation}

Thus, the choice of the coordinate system and the choice of the tetrad are independent procedures. Nevertheless, if there is a coordinate system,
$x^i$, the tetrad is frequently chosen in such a way that each basis vector, ${\bf e}_{i}$,
is tangent to the corresponding coordinate line. Here the moduli of orts of these so-called coordinate bases are chosen such that their pairwise scalar products are equal to the corresponding metric coefficients:
\begin{equation}
\label{scalar_g_ik}
({\bf e}_i \cdot {\bf e}_k) = g_{ik}
\end{equation}

We recall that in differential geometry (see paragraphs 3.1-3.4 \cite{HEL_2006}),
such coordinate orts are introduced as objects isomorphic to the partial derivatives of an arbitrary scalar function on the manifold with respect to the coordinates, 
\begin{equation}
\label{tan_vect}
{\bf e}_i \equiv \frac{\partial}{\partial x^i}.
\end{equation}

Any (tangent) vector is a linear combination of the coordinate orts, and the components of this linear combination are called contravariant components of the vector.

In addition to ${\bf e_i}$ the so-called dual basis, 
${\bf e^i}$, is introduced with the orts defined as
\begin{equation}
\label{duality}
({\bf e_i}\cdot{\bf e^j}) = {\delta_i}^j, 
\end{equation}
where ${\delta_i}^j$ is the Kronecker symbol. The condition (\ref{duality}) implies that each ort of the dual basis has unit projection on the corresponding ort 
of the coordinate basis and is orthogonal to all other orts of the coordinate basis. 

The dual coordinate orts, in turn, are introduced as objects isomorphic to the coordinate differentials, 
\begin{equation}
\label{tan_vect_dual}
{\bf e}^j \equiv d x^i.
\end{equation}

Next, if we use the fact that any tangent vector ${\bf A}$ can be alternatively presented as a linear combination of dual coordinate orts, whose coefficients are referred to 
as covariant vector components, we obtain the well-known rule of lowering vector indices:

\begin{equation}
\label{indices}
A_k = A_i ({\bf e}^i \cdot {\bf e}_k) = (A_i {\bf e}^i \cdot {\bf e}_k) = 
(A^i {\bf e}_i \cdot {\bf e}_k) = A^i ({\bf e}_i \cdot {\bf e}_k) = A^i g_{ik},
\end{equation}
In a similar way it is easy to show that if we introduce the notation $g^{ik} \equiv ({\bf e}^i \cdot {\bf e}^k)$,  then due to the duality of bases, the matrix $g^{ik}$ 
is inverse to the matrix $g_{ik}$, and the rule of raising of vector indices holds. Similar representation in coordinate bases can be extended 
to a more general case of tensors.  

\subsubsection{Tetrad representation}

In this and subsequent sections we mostly follow the exposition given in paragraph 7 of \cite{Chandr1}.
Assume that now we want to project the same vectors and tensors on an arbitrary tetrad defined by relations
\begin{equation}
\label{tetrade_gen}
{\bf e}_{(a)} = {e_{(a)}}^i \, \frac{\partial}{\partial x^i},
\end{equation}
where ${e_{(a)}}^i$ are some functions of coordinates, and the indices labeling the tetrad orts are in parentheses. 

From the duality condition (\ref{duality}) we can introduce the dual tetrad:
\begin{equation}
\label{dual_tetrade_gen}
{\bf e}^{(a)} = {e^{(a)}}_i \,d x^i,
\end{equation}
where ${e^{(a)}}_i$ is the matrix inverse to ${e_{(a)}}^i$.  

In these matrices, there are two kinds of indices: coordinate and tetrad ones. 
The coordinate indices can be lowered or raised using the metric (\ref{Kerr}).
We can impose an additional constraint on the tetrad:
\begin{equation}
\label{tetr_orth}
{e_{(a)}}^i {e_{(b)}}_i = \eta_{(a)(b)}, \quad {e^{(a)}}^i {e^{(b)}}_i = \eta^{(a)(b)}, 
\end{equation}
where
\begin{equation}
\label{eta_inv}
\eta_{(a)(c)} \, \eta^{(c)(b)} = {\delta_{(a)}}^{(b)}
\end{equation}
are mutually inverse matrices and $\eta_{(c)(b)} = diag(1,-1,-1,-1)$ is the Minkowski matrix. 
In other words, we require that the original and dual tetrads be orthonormal in four-dimensional pseudo-Euclidean space. 

Using above relations, it is straightforward to show that
\begin{equation}
\label{g_tetr}
{e^{(a)}}_i \, {e_{(a)}}_j = g_{ij}, 
\end{equation}
and therefore the following alternative expression for the interval squared holds:
\begin{equation}
\label{kerr_tetr}
ds^2 = \eta_{(a)(b)} ({e^{(a)}}_i \, dx^i) \, ({e^{(b)}}_k \, dx^k) =\eta_{(a)(b)} \, {\bf e}^{(a)} \, {\bf e}^{(b)}, 
\end{equation}
which is useful below. 

Note that the values in parentheses in the right-hand side of (\ref{kerr_tetr}) 
can be considered as infinitesimal shifts along the corresponding orts of the tetrad; 
therefore, in the introduced tetrad representation with an orthonormal tetrad, the square of interval takes exactly the form as in the Minkowski space-time of special relativity.  
Similarly, expressions (\ref{tetrade_gen}) can be thought of as directional derivatives along the tetrad’s orts, and these have exactly the form that the usual partial derivatives with respect to coordinates 
in the coordinate basis take when changing from the coordinate basis to the tetrad one.    

Using the definitions and relations given above, it is easy to see how the tetrad components of vectors are expressed through the coordinate components. Tetrad components of a vector are written as
\begin{equation}
\label{tetrade_vec}
A_{(a)} = e^i_{(a)} A_i, \quad A^{(a)} = e^{(a)}_i A^i = \eta^{ab}A_{(b)}.
\end{equation}
Conversely, 
$$
A_i = e^{(a)}_i A_{(a)}, \quad A^i = e^i_{(a)} A^{(a)}.
$$

Similar expressions can be written for tensor of any valence. For example, for a two-covariant tensor, we have
$$
T_{(a)(b)} = e^i_{(a)} e^j_{(b)} T_{ij} = 
e^i_{(a)} T_{i(b)},
$$
and conversely,
\begin{equation}
 \label{tetrade_ten}
T_{ij} = e^{(a)}_i e^{(b)}_j T_{(a)(b)} = 
e^{(a)}_i T_{(a)j}.
\end{equation}

Note in conclusion that relations (\ref{tetrade_vec}) and isomorphism (\ref{tan_vect_dual})
can be used to find contravariant components of the four-velocity in the tetrad representation:

\begin{equation}
\label{tetr_vel}
U^{(a)} = \frac{{\bf e}^{(a)}}{ds}.
\end{equation}
That is, it is again a unit tangent vector along the world line, but now its components are given by small shifts along the corresponding orts of the dual basis. 
Using (\ref{tetrade_vec}) it is easy to find the relation between the conventional coordinate components of four-velocity, $U^i = dx^i/ds$,
and its tetrad components. Covariant tetrad components are derived from contravariant ones using the standard rule in special relativity: 
lowering a spatial index is equivalent to changing the sign of the corresponding component. 

\subsubsection{Covariant derivative in tetrad representation}

Let us calculate the directional derivative along a tetrad ort from a contravariant component  of a vector:
\begin{equation}
\label{der_vec}
A_{(a),(b)} = e^i_{(b)} \frac{\partial}{\partial x^i}
A_{(a)} = e^i_{(b)} \frac{\partial}{\partial x^i}
e^j_{(a)} A_j = e^i_{(b)} [ e^j_{(a)} A_{j;i} 
+ A_k e^{k}_{(a);i} ],
\end{equation}
where the semicolon denotes the usual covariant derivative in the coordinate basis. 

Expression
(\ref{der_vec}) can be recast to the form:
\begin{equation}
\label{der_vec_1}
A_{(a),(b)} = e^j_{(a)} A_{j;i} e^i_{(b)} + 
e_{(a)k;i} e^i_{(b)} e^k_{(c)} A^{(c)}, 
\end{equation}
whence 
\begin{equation}
\label{der_vec_2}
e^j_{(a)} A_{j;i} e^i_{(b)} = A_{(a),(b)} -
\gamma_{(c)(a)(b)} A^{(c)},
\end{equation}
where $\gamma_{(c)(a)(b)}$ are the so-called Ricci rotation coefficients,
\begin{equation}
\label{ricci}
\gamma_{(a)(b)(c)} = e_{(b)k;i} e^i_{(c)} e^k_{(a)}
\end{equation}
An important point is that for orthonormal bases satisfying
(\ref{tetr_orth}), the coefficients $\gamma_{(a)(b)(c)}$ are antisymmetric in first two indices. 
Indeed, 
$$
0 = (\eta_{(b)(a)})_{,i} = (e_{(b)k} e^k_{(a)})_{;i} = e_{(b)k;i} e^k_{(a)} + e_{(b)k} e^k_{(a);i} = 
e_{(b)k;i} e^k_{(a)} + e_{(b)}^k e_{(a)k;i}.
$$
Comparing this relation with (\ref{ricci}) proves the stated property of the Ricci coefficients. 

Finally, let us discuss one more useful property of coefficients (\ref{ricci}): to calculate these coefficients 
only partial derivatives of the components of the tetrad basis orts are needed, 
and therefore the Christoffel symbols are not required. 
Indeed, we consider auxiliary combinations
\begin{equation}
\label{lambdas1}
\lambda_{(a)(b)(c)} = e_{(b)i,j} [e^i_{(a)} e^j_{(c)} - 
e^j_{(a)} e^i_{(c)} ],
\end{equation}
and rewrite them in the form
\begin{equation}
\label{lambdas2}
\lambda_{(a)(b)(c)} = e^i_{(a)} e^j_{(c)} [e_{(b)i,j} - e_{(b)j,i} ],
\end{equation}
In the last expression, the ordinary partial derivatives can be substituted by covariant ones, because
the additional terms with Christoffel symbols are symmetric in $i,j$. 
Then expression (\ref{lambdas2}) is equal to the difference
$\gamma_{(a)(b)(c)} - \gamma_{(c)(b)(a)}$. 

But in such a case, 
\begin{equation}
\label{gammas}
\gamma_{(a)(b)(c)} = 1/2[\lambda_{(a)(b)(c)} + 
\lambda_{(c)(a)(b)} -\lambda_{(b)(c)(a)} ]
\end{equation}
and, using (\ref{lambdas1}), it is possible to calculate the Ricci rotation coefficients by taking partial derivatives of components of the tetrad basis orts. 

We now consider formula (\ref{der_vec_2}). The left-hand side represents simply the projection on tetrad basis of a rank-2 covariant tensor obtained by taking the derivative of some vector field. 
Therefore, this combination has meaning of the covariant derivative of a vector taken in a non-coordinate basis.

Next, the right-hand side of (\ref{der_vec_2}) has exactly the same form as the covariant derivative in a coordinate basis,  
with the only difference that it involves tetrad indices (which can be raised or lowered, including for $\gamma_{(a)(b)(c)}$,
using the Minkowski metric). It can be shown that the same holds for contravariant components of a vector field and for tensor fields in general. 

Thus, as the components of a covariant derivative in a tetrad basis has the same form as in a coordinate basis, it is convenient to use the same notations and terms that are used in the coordinate basis. 
In particular, the Ricci rotation coefficients are simply referred to as the connection symbols in a given basis. 
We emphasize once again that they should not be confused with the Christoffel symbols, 
which represent another limit case of connection coefficients in a coordinate basis and have different index symmetry.

\subsection{Tetrad transported by rotating observers} 
\label{sec_tetrad}
We construct a tetrad basis related at each point of space-time to observers moving around 
a black hole in equatorial circular orbits with angular velocity $\Omega$.

Strongly at $z=0$ this is the free motion along geodesics found in Section \ref{sec_geod}. However, for a small deviation from the equatorial plane, such a motion, corresponding to the constant $z$, 
is possible only if there is some external supporting force; in the case of a gas disk, for example, this force is due to the pressure gradient. 

To start the construction, we direct the time ort of the tetrad along the world line under discussion. Using the four-vector of the geodesic found in Section
\ref{sec_geod},  write it in the form
$$
{\bf e}_{(t)} = ( {U_g}^t  + Z_0) \,\, \frac{\partial}{\partial t} + {U_g}^\phi \,\, \frac{\partial}{\partial \phi},
$$
where we add the correction factor $Z_0(z/r)$ to the time coordinate component, since the modulus of the vector ${\bf e}_{(t)}$ 
should be equal to unity away from the equatorial plane as well, whereas the vector ${\bf U}_g$ itself is unitary only at $z=0$.
Clearly, with account for this correction, ${\bf e}_{(t)}$ would correspond to the four-velocity of the real motion. 
Calculation of the modulus of the vector ${\bf e}_{(t)}$ in metric (\ref{Kerr_rz}) shows that it is equal to unity under the following condition:
$$
Z_0 = - \left ( \frac{z}{r}\right )^2 \frac{H}{2rGC^{1/2}},
$$
where we introduce the relativistic correction coefficients
\begin{equation}
G = 1 - \frac{2}{r} + \frac{a}{r^{3/2}}
\end{equation}

\begin{equation}
H = 1 - \frac{4a}{r^{3/2}} + \frac{3a^2}{r^2}
\end{equation}

Thus, the ort ${\bf e}_{(t)}$ 
is transported by the observer rotating around the black hole with a frequency equal to the $\phi$-component of ${\bf e}_{(t)}$, which is independent of $z$. 
This frequency corresponds to the free circular motion in the equatorial plane of the black hole, and rotation occurs in planes of constant $z$.

We now calculate the time ort of the dual basis. According to the convention rule for raising and lowering coordinate indices, we have
$$
{\bf e}^{(t)} = ( {U_g}^t g_{tt} + {U_g}^\phi g_{t\phi} ) \,\, d t + 
                ( {U_g}^t g_{\phi t} + {U_g}^\phi g_{\phi \phi} ) \,\, d\phi.
$$

Next, we consider the part of metric (\ref{Kerr_rz}) containing differentials $dr$ and $dz$.
It can be rewritten in the form (see the result (\ref{kerr_tetr}) in the previous section):
$$
ds_{rz}^2 = - \left [ {\bf e}^{(r)}\right ]^2 - \left [ {\bf e}^{(z)} \right ]^2, 
$$
where
$$
{\bf e}^{(r)} = |g_{rr}|^{1/2} \,\, dr - \frac{g_{rz}}{|g_{rr}|^{1/2}} \,\, dz ,
$$
$$
{\bf e}^{(z)} = \left (  |g_{zz}| - \frac{{g_{rz}}^2}{|g_{rr}|} \right )^{1/2} \,\, dz \quad \mbox{---}
$$
are the radial and vertical orts of the dual basis, respectively.
The coordinate components of vectors ${\bf e}^{(t)}, {\bf e}^{(r)}$ and ${\bf e}^{(z)}$ 
satisfy the orthonormality condition (\ref{tetr_orth}), as can be easily checked by direct substitution.

The orthonormality condition for a tetrad can now be used to determine the fourth ort corresponding to the azimuthal direction. 

From three orthonormality conditions for three already known vectors, we obtain that for these conditions to be consistent, the following relation should hold:

$$
{e^{(\phi)}}_r = {e^{(\phi)}}_z = 0, 
$$
and the time and azimuthal components should be related as
$$
{e^{(\phi)}}_\phi = - \, {e^{(\phi)}}_t \,\, \frac{{e_{(t)}}^t}{{e_{(t)}}^\phi}. 
$$

Finally, the normalization condition for ${\bf e}^{(\phi)}$ yields a quadratic equation for
${e^{(\phi)}}_t$, and the sign of the solution is dictated by the additional requirement of the choice of a right-hand triple of space orts of the tetrad. 

We thus obtain the dual tetrad basis with the leading corrections in $(z/r)$ due to out-of-equatorial-plane motion in the form

\begin{multline} 
\label{e_dual_1}
{\bf e}^{(t)} \, = \, C^{-1/2} \left \{ G + \left (\frac{z}{r}\right )^2 \frac{1}{2rG} 
\left ( D + \frac{2a}{r^{3/2}}\left ( F - \frac{a}{r^{3/2}} + \frac{a^2}{r^2} \right ) \right ) \right \}\,\, dt 
\,- \\  \, 
C^{-1/2} \left \{ r^{1/2} F + \left ( \frac{z}{r} \right )^2 \frac{a}{rG} Z_1 \right \} \,\, d\phi,
\end{multline}

\begin{multline} 
\label{e_dual_2}
{\bf e}^{(\phi)} \, = \, - \left \{ \left( \frac{D}{rC} \right )^{1/2} + \frac{1}{2} \left ( \frac{z}{r} \right )^2 
\frac{1-a/r}{r^{3/2}} (DC)^{-1/2} \right \} \,\, dt \,+  \\
\, \left \{ 
r B \left( \frac{D}{C} \right )^{1/2} + 
\frac{1}{2}\left ( \frac{z}{r} \right )^2 \left [ ( 1-a/r ) \frac{B}{(DC)^{1/2}} - 
\frac{H}{G} \left ( \frac{D}{C} \right )^{1/2} \right ]  \right \} \,\, d\phi,
\end{multline}

\begin{multline} 
\label{e_dual_3}
{\bf e}^{(r)} \, = \, D^{-1/2} \left \{ 1 - \frac{1}{2D}\left ( \frac{z}{r} \right )^2 Z_2\right \}
\,\, dr \, + \frac{z}{r} D^{-1/2} \left ( \frac{2}{r} - \frac{a^2}{r^2} \right ) \,\, dz,
\end{multline}

\begin{equation} 
\label{e_dual_4}
{\bf e}^{(z)} \, = \, \left ( 1 + \frac{z^2}{r^3} \right ) \,\, dz. 
\end{equation}

To obtain the original basis, which we use to write equations of motion, it suffices to calculate the inverse to the matrix ${e^{(a)}}_i$, which yields

\begin{equation}
\label{e_1}
{\bf e}_{(t)} \, = \, C^{-1/2} \left [ B - \left ( \frac{z}{r} \right )^2 \frac{H}{2rG} \right ] \,\, 
\frac{\partial}{\partial t}  + (r^3 C)^{-1/2} \,\, \frac{\partial} {\partial \phi},
\end{equation}

\begin{equation}
\label{e_2}
{\bf e}_{(\phi)}\, = \, \left \{ \frac{F}{(rCD)^{1/2}}\, + \, 
O \left (\frac{z^2}{r^2} \right ) 
\right \} 
\,\, \frac{\partial}{\partial t} \, + 
\left \{ \frac{G}{r(DC)^{1/2}} \, + \, 
O\left (\frac{z^2}{r^2} \right )
\right \} \,\, \frac{\partial}{\partial \phi}, 
\end{equation}

\begin{equation}
\label{e_3}
{\bf e}_{(r)} \, = \, \left \{ D^{1/2} + \frac{1}{2} \left ( \frac{z}{r} \right )^2 \frac{Z_2}{D^{1/2}} \right \} \,\,\frac{\partial }{\partial r},
\end{equation}

\begin{equation}
\label{e_4}
{\bf e}_{(z)} \, = \, -\,  \frac{z}{r^2}  (2-a^2/r)\,\, \frac{\partial}{\partial r} \,\, +
\,\, \left ( 1 - \frac{z^2}{r^3} \right ) \frac{\partial}{\partial z}.
\end{equation}

\vspace{1cm}

The following notations for relativistic correction coefficients are introduced in the expressions for the original and dual bases:
\begin{equation}
F = 1 - \frac{2a}{r^{3/2}} + \frac{a^2}{r^2}
\end{equation}

\begin{equation}
Z_1 = 3 - \frac{5}{r} - \frac{a}{r^{1/2}} + \frac{3a}{r^{3/2}} - \frac{3a^2}{r^3} + \frac{a^2}{r^2} + \frac{2a^3}{r^{7/2}}
\end{equation}

\begin{equation}
Z_2 = \frac{3}{r} - \frac{4}{r^2} - \frac{a^2}{r^2} \left ( 3 - \frac{6}{r} + \frac{2a^2}{r^2} \right )
\end{equation}

Here, we omit terms $\sim O(z^2/r^2)$ in the expression for the azimuthal ort of the original basis due to their complexity; 
in addition, as we will see below, these terms are not required in the standard accretion disk model.

For the reader's convenience, we here preserve the notations introduced in paper \cite{NT} for the coefficients $B, C, D, F, G$, 
but use the standard style of Latin letters, which is more familiar to the reader. In addition, the coefficient $H$ 
is equivalent to the coefficient $C$ introduced in \cite{RH_1995}. 
We also draw attention to the fact that two other coefficients, $A$ and $B$, introduced in the same paper, are equivalent to our coefficients $D$ and $C$, respectively.
It can be checked that the original and dual bases presented in \cite{NT} coincide with bases derived here at $z=0$.

Using formulas (\ref{e_dual_1}-\ref{e_dual_4}) and (\ref{tetr_vel}), it is easy to deduce that solution (\ref{geod_vel}) 
indeed yields $U^{(a)}=(1,0,0,0)$ in the equatorial plane.

\subsubsection{The connection coefficients}

Using (\ref{lambdas1}) and then (\ref{gammas}) and knowing the matrices of the original and dual bases given above, we can calculate the connection coefficients $\gamma_{(a)(b)(c)}$.

Of the 64 coefficients, 16 are equal to zero due to anti-symmetry of $\gamma_{(a)(b)(c)}$ 
in the first two indices. For the same reason, of the other coefficients, only half (i.e. 24) have to be found. 
Because we are interested in the region near the equatorial plane of the black hole, 
it makes sense to separate these coefficients into two groups: those that are
$\sim (z/r)^0$ in the leading order, and those that are proportional to the first power of $(z/r)$. As mentioned in Section \ref{sec_metric}, 
the latter coefficients must appear in the vertical projection of the relativistic Euler equation, while the former emerge in other equations. 

It can be shown that 

\noindent
1)  if there is no index $(z)$ among the indices of $\gamma_{(a)(b)(c)}$, then
$\gamma_{(a)(b)(c)}\sim (z/r)^0$, 

\noindent
2) if only one such index is present, then
$\gamma_{(a)(b)(c)}\sim (z/r)$, and, finally,

\noindent
3) if two indices $(z)$ appear in $\gamma_{(a)(b)(c)}$, then this coefficient is of the second order in $(z/r)$.

Indeed, we examine formula (\ref{lambdas1}). Here the brackets contain the original basis components, which are summed with the coordinate derivatives of the dual basis components 
(the raising of a tetrad index can only change the sign of the component).

\noindent
In case (1) $(a),(b),(c)\neq(z)$.
As the $(t)-$, $(\phi)-$ and $(r)-$orts of the original basis have no $z$-component,
only terms which do not contain derivatives with respect to $z$ of the dual basis components and have no 
$z$-component of the dual $(r)-$ort make a non-zero contribution to
$\gamma_{(a)(b)(c)}$.
Only in these two cases can the contribution
$\sim (z/r)$ appear, and hence we prove the statement (1).

Now, in (\ref{lambdas1}) let $(b)=(z)$ and $(a),(c)\neq (z)$.
Then the non-zero contribution can only be due to terms containing the $z$-component of the $(t)-$, $(\phi)-$ and $(r)-$orts
of the original basis which are absent. Therefore, to check case (2) we should consider only 
the variant when in (\ref{lambdas1})  $(a)=(z)$ or $(c)=(z)$. Here, the terms containing separately either  
$r-$ or $z-$components of the $(z)-$ort of the original basis contribute.
In the first variant, the proportionality to $\sim (z/r)$ is due to exactly the component $e^r_{(z)}$, while 
in the second, it is due to the derivative with respect to $z$ of one of the dual basis components that is always even function of $z$, as can be easily verified. 

We leave it to the reader to prove the statement (3).

The counting shows that there must be 9 connection coefficients without the index $(z)$, and hence an even function of $z$,
and 12 coefficients with the index $(z)$ and hence odd function of $z$. 
The calculation indicates that only 4 coefficients of the first type are non-zero, namely:

\begin{equation}
\label{conn1}
\gamma_{(t)(\phi)(r)} = -\frac{1}{2}\frac{H}{r^{3/2}C}
 \quad 
\gamma_{(t)(r)(\phi)} = -r^{-3/2}
\end{equation}
\begin{equation}
\label{conn2}
\gamma_{(\phi)(r)(t)} = -r^{-3/2} \quad
\gamma_{(\phi)(r)(\phi)}= -\frac{1}{r}\frac{d}{dr}\left ( r D^{1/2} \right )
\end{equation}

To compute coefficients (\ref{conn1},\ref{conn2}) it suffices to use bases taken without corrections in $z$.
When constructing the standard disk model, the following facts are also important. 
First, the direct calculation shows that another 5 connection coefficients of this type are zero through corrections of the order of
$\sim (z/r)^2$ inclusive. 
This is a rigorous result, since the coefficients $\gamma_{(a)(b)(c)}$ under discussion have no derivatives of the basis components with respect to $z$, and therefore the possible unaccounted for corrections due to terms
$\sim(z/r)^3$ in $e^{(r)}_z$ and $e_{(z)}^r$ cannot contribute. Second, direct calculation similarly shows that
$\gamma_{(t)(z)(z)}=0$ through the order $\sim(z/r)^2$.

The calculation of all non-zero coefficients of the second type is much more cumbersome task. But as we will see below, the only coefficient of this type that is needed has the form
\begin{equation} \nonumber
\gamma_{(z)(t)(t)} = \frac{z}{r^3} \frac{H}{C}.
\end{equation}

We note that all connection coefficients of the type $\gamma_{(a)(t)(t)}$
vanish in the equatorial plane $z=0$. This is consistent with the requirement that the four-velocity  
$U^{(a)}=(1,0,0,0)$ must satisfy the geodesic equation at $z=0$:

\begin{equation}
\label{geodez_tetr}
\frac{D U^{a}}{Ds} = U^{b}\, {\bf e}_{(b)} \left ( U^{(a)} \right ) + \eta^{(a)(c)}\gamma_{(c)(b)(d)} U^{(b)} U^{(d)} = 
\gamma_{(a)(t)(t)} = 0.
\end{equation}

\subsection{Relativistic hydrodynamic equations}
\label{sec_rel_eq}
Everywhere below, we only use the tetrad components of vectors, tensors and covariant derivatives. 
Therefore, starting from this Section, we will substitute the tetrad notation by the standard one, which is familiar when using the coordinate basis. 
This means that from now on we do not put tetrad indices in parentheses and 
denote them by Latin letters $i,j,k$
\footnote{If one of the symbols $t,\phi,r,z$, appears among the indices, it means that the corresponding index takes this value.}; also, we denote the connection coefficients by $\Gamma$. 

The stress-energy tensor of a viscous fluid with energy flux has the form (see, e.g., paragraph 4.3 in \cite{Mihalas} or paragraph 22.3 in \cite{MTW})
\begin{equation}
\label{Tik} 
T^{ik} = (\rho + \epsilon + p) U^i U^k - p \eta^{ik} + 2\eta \sigma^{ik} + \zeta \Theta P^{ik}
 - U^i q^k - U^k q^i, 
\end{equation}
where $\rho$, $\epsilon$, $p$, $\eta$, $\zeta$ are the rest-energy density, internal energy density, pressure and two viscosity coefficients, respectively, as measured  in the local comoving fluid volume;
${\bf q}$ is the energy flux inside the fluid as measured by the local comoving observer. 

The shear tensor is 
\begin{equation}
\label{shear} 
\sigma^{ik} = \frac{1}{2}\left ( {U^i}_{;j} P^{jk} + {U^k}_{;j} P^{ji}\right ) - \frac{1}{3} {U^j}_{;j} P^{ik},
\end{equation}

with the projection operator 
\begin{equation}
\label{proj_ten} 
P^{ik} = \eta^{ik} - U^i U^k.
\end{equation}

The divergence of four-velocity is 
\begin{equation} 
\Theta = {U^i}_{;i}.
\end{equation}

The relativistic Euler equation is written as 
\begin{equation}
\label{euler}
P_{is} \, {T^{sk}}_{;k} \, = \,0.
\end{equation}

The energy conservation law has the form
\begin{equation}
\label{energy}
U_{s}\, {T^{sk}}_{;k} \, = \,0,
\end{equation}

the rest-energy conservation law reads
\begin{equation}
\label{continuity}
{(\rho \, U^k)}_{;k}  = 0.
\end{equation}

The covariant derivative in a non-coordinate basis is 
\begin{equation}\nonumber
{A^i}_{;j} = {\bf e}_j (A^i) + {\Gamma^i}_{kj} A^k,
\end{equation}

while the divergence of a rank-2 contravariant tensor is 
\begin{equation} \nonumber
{A^{ij}}_{;j} = {\bf e}_j (A^{ij}) + {\Gamma^i}_{kj} A^{kj} + {\Gamma^j}_{kj} A^{ik}.
\end{equation}

The energy flux vector and the the shear tensor (the deformation tensor free from pure scaling) are purely space-like objects:
\begin{equation} \label{sigma_orth}
U_i q^i = 0, \quad U_i \sigma^{ik} = 0, \quad {\sigma^i}_i = 0.  
\end{equation}

\section{Construction of the standard accretion disk model}

\subsection{Basic assumptions and the vertical balance equation}
\label{sec_assumpt}
Thus, we consider a disk from the standpoint of local observers rotating around a black hole near its equatorial plane
with a relativistic Keplerian velocity.  Before writing the dynamic equations in the projection onto tetrad
(\ref{e_1}-\ref{e_4}), 
we discuss basic assumptions of the model and their consequences. 
In addition to obvious assumptions about axial symmetry and stationarity of the flow (meaning that the derivatives
$\partial_t$ and $\partial_\phi$ 
are zero) the main assumption, which we have already used, is the small disk thickness, $\delta = h(r)/r \ll 1$, where $h(r)$ is the characteristic height of the disk along the
$z$-axis (more precisely, the disk half-thickness).

The disk symmetry with respect to the plane $z=0$ implies that 
$U^t, U^\phi, U^r, q^t, q^\phi, q^r, \rho, p, \eta, \zeta, \epsilon$ are even functions of $z$, and $U^z$ and $q^z$ are odd functions of $z$.

We also assume that the characteristic scale of variations of these quantities in the radial direction is much larger than that in the vertical direction,
that is, their ratio is greater than $\sim\delta^{-1}$\footnote{We note that the assumption about the velocity components in the disk plane, $U^r$, $U^\phi$, 
should also be made that in vertical direction they can substantially change only on scales $\sim r$, 
otherwise, terms in the shear tensor could arise that strongly dynamically contribute to the vertical balance condition, which would lead to a disk totally different from the basic case of interest here.}.

Next, kinematic arguments suggest that 
\begin{equation}
\label{U_zr}
U^z \sim \delta U^r.
\end{equation}
If the energy flux determined by vector ${\bf q}$ is proportional to the internal energy gradient $\epsilon$,
then, for the local comoving observer, $q_{loc}^t=0$ and $q_{loc}^\phi, q_{loc}^r \sim \delta q_{loc}^z$. 
Taking (\ref{U_zr}) into account, implies that the projection of ${\bf q}$ onto the four-velocity of circular equatorial motion is also small, i.e. of the order of $\sim \delta q_{loc}^z$. 
From the standard Lorentz transformations, we obtain that $q^t,q^\phi,q^r \sim \delta q^z$, i.e. the energy flux relative to the tetrad should be directed mostly normally to the disk plane. 

Now, taking all the above into account, we consider the projection of the relativistic analog of Euler equation (\ref{euler}) onto the ort ${\bf e}_z$
in more detail:

\begin{equation}
\label{euler_z}
{T^{zk}}_{;k} + U_z U_s  \, {T^{sk}}_{;k} \, = \,0.
\end{equation}

Using the symmetry of physical quantities discussed above and symmetry properties of tetrad orts and connection coefficients (which become odd functions of $z$ 
if they have at least one index $z$), discussed in Section \ref{sec_tetrad}, it is easy to check that equation
(\ref{euler_z}) is indeed an odd function of $z$. 
Further, we see that the first term in (\ref{euler_z}) separately yields the term $\partial_z p$ and other terms containing $p$ are smaller due to the smallness of $U^z$. 
All other terms together can always be written as $\sim z \rho f(r) (1 \, + \, g(r,z)) $ with the function $g(r,z) \sim O(\delta^0)$.

Thus, we arrive at the important conclusion that necessarily 
\begin{equation}
\label{smallness_p}
\frac{1}{\rho}\frac{\partial p}{\partial z} \sim \delta \ll 1.
\end{equation}

This means that in a thin disk the variables $p,\, \partial_r p \sim \delta^2$,
i.e. these variables are small relative to the dominant action of the gravitational force in this direction. Therefore, particles of the disk must move in trajectories close to geodesic ones. 
Clearly, in a steady-state and axially symmetric flow, this can be realized only in two cases: when the matter moves almost radially towards the gravitating center 
(and the specific angular momentum in the disk is everywhere close to zero) or when the matter moves in almost circular orbits (and the specific angular momentum, oppositely, is maximal). 
We note that both cases are consistent with the general assumptions discussed above and the result (\ref{smallness_p}). However, in the last case, the strict vertical hydrostatic equilibrium 
holds in the disk in the first order in $\delta$; in other words,
(\ref{euler_z}) can be rewritten in the form
\begin{equation}
\label{hydrostatics}
\frac{1}{\rho} \frac{\partial p}{\partial z} \sim z f(r) ( 1 + \delta^2 ... ).
\end{equation}

When the flow is almost radial, the corrections in the parentheses in  (\ref{hydrostatics})
are not small, and their value is determined by the contribution from the prevailing radial motion, 
when, due to the change in the disk thickness at each radius, the particles are accelerated in the $z$ direction. 

Thus, the standard disk model includes one more independent assumption on the closeness of the fluid particle trajectories to equatorial circular orbits around the central black hole. 
Therefore, we will additionally suppose that in our reference frame $U^\phi, U^r \sim s U^t$ with $s \ll 1$ and later we can see how this second small parameter is related to $\delta$. 

Consequently, we write equations first not only in the leading order in $\delta$ but also by assuming $s=0$, i.e. that the flow moves along geodesic orbits and $U^i = (1,0,0,0)$. 
Wherever needed, we then additionally evaluate the contribution from the terms in the leading order in $s$.

\subsubsection{Deformation of the velocity field}

We first find the non-zero components of the shear tensor in the leading order.  First, the velocity divergence vanishes:
\begin{equation}
\Theta = {U^j}_{;j} \, = \, {\Gamma^j}_{kj}\, U^k = {\Gamma^j}_{0j} = 0
\end{equation}

Next, we have
$$
U^i_{;j} P^{jk} = \Gamma^i_{tk} \eta^{kk} - \Gamma^i_{tt}, 
$$
and, in view of the symmetry in $i$ and $k$, we see that the only non-zero components of the shear tensor are
$\sigma^{ik}$ are 
\begin{equation} \label{sigma_NT}
\sigma^{r\phi} = -\frac{1}{2} \left ( {\Gamma^\phi}_{tr} + {\Gamma^r}_{t\phi} \right ) = 
\frac{1}{2} \left ( \frac{1}{2} \frac{H}{r^{3/2} C} + r^{3/2} \right ) = \frac{3}{4} \, \frac{D}{r^{3/2} C},  
\end{equation}

\begin{equation}
\sigma^{r z} = -\frac{1}{2} {\Gamma^z}_{t\phi} = O(z).
\end{equation}

\subsubsection{Equation of hydrostatic equilibrium}

Substituting $U^i = (1,0,0,0)$ in (\ref{euler_z}) and taking the smallness (due to the low sound velocity in the flow) of several non-zero terms containing $\eta$ and components of ${\bf q}$ into account, we obtain
\begin{equation}
\label{hydrostatics_rel}
\frac{\partial p}{\partial z} = \rho\, \Gamma^z_{tt} = - \rho \frac{z}{r^3} \frac{H}{C}.
\end{equation}

\subsubsection{Radial direction}

The radial projection of the relativistic Euler equation for $s=0$ reads
\begin{equation}
\label{euler_rad}
{T^{rk}}_{;k} = 0,
\end{equation}
and excluding terms $\sim \delta^4$ containing the connection coefficients and components of ${\bf q}$, 
we have only one non-zero term of the order of $\delta^2$ which has the form\footnote{The order of components $q^i$ 
can be estimated as follows. In the stationary case, the divergence of the energy flux must be of the order of the power generated due to viscous dissipation, which 
is, in turn, proportional to some scalar characterizing the degree of the velocity shear and the viscosity coefficient
$\eta$. In our case, the viscosity coefficient  $\eta < \rho h c_s \sim \delta^2$.
The divergence is mainly due to the term $\partial_z q^z$. This immediately implies that $q^z\sim \delta^3$ and $q^{t,\phi,r} \sim \delta^4$.}

$$
-[p \eta^{rk}]_{;k} = D^{1/2} \, \frac{\partial p}{\partial r}
$$

Clearly, this term should be balanced by the leading terms $\sim s$.
Evidently, the contribution from
$$
[\rho U^r U^k]_{;k}
$$
should be considered first, and here it can be only due to terms containing one of the connection coefficients of zeroth order in
$z$ 
and the time velocity component. There is only one such term:
$2\Gamma^r_{t\phi}U^t U^\phi = 2 r^{-3/2} U^\phi$. 

Hence, we reach an important conclusion that $s \sim \delta^2$, i.e. the velocity components in the disk plane are
\begin{equation}
U^r, U^\phi \sim \delta^2,
\end{equation}
which is used when determining the force balance in the azimuthal direction.

\subsection{Azimuthal direction}

We consider the last projection of the relativistic Euler equation, its component along the azimuthal ort. 
Let us proceed in the same way as above and first write terms that are present in the case $s=0$.
Again, we take $U^i = (1,0,0,0)$ and see that
$$[(\rho + \epsilon + p) U^\phi U^k]_{;k}=0,$$ because $\Gamma^\phi_{tt}=0$ through the order $\sim\delta^2$ 
(see the discussion at the end of Section \ref{sec_tetrad}). 
Next, the term with pressure is absent by virtue of the axial symmetry, and terms with  $q^i$ cannot contribute to the order higher than $\sim \delta^4$.  

It thus remains to consider the contribution

$$
[2\eta \sigma^{\phi k}]_{;k} = D^{1/2} (2\eta \sigma^{r\phi})_{,r} + (2\eta \sigma^{\phi z})_{,z} + 
4\eta \Gamma^{\phi}_{r\phi} \sigma^{r\phi} + \eta \, O(\delta^2) =  
$$
\begin{equation}
\label{part_1}
= -\frac{3}{2} D^{1/2} \left ( \eta \frac{D}{r^{3/2} C} \right )_{,r} + (\eta \Gamma^t_{z\phi})_{,z} + 
3\eta \, (r D^{1/2})_{,r} \, \frac{D}{r^{5/2} C} + \eta\, O(\delta^2).
\end{equation}

Here, we are also dealing with terms of the second order in $\delta^2$; therefore, it is necessary to find the leading contribution from terms $\sim s$. 
Again, we consider only the prevailing part due to ideal fluid term:
$$
(\eta_{\phi i} - U_\phi U_i) [\rho U^i U^k]_{;k}.
$$
The second part, which is proportional to $U_\phi$, can be neglected because the term in square brackets cannot contribute to the zeroth order in $\delta$, 
since there are no connection coefficients of the form $\Gamma^i_{tt} \sim \delta^0$, as was discussed at the end of Section \ref{sec_tetrad}. 

As a result, we obtain
\begin{multline}  \label{part_2}
[\rho U^\phi U^k]_{;k} = \rho\Gamma^{\phi}_{lk} U^l U^k + \rho\Gamma^{k}_{lk} U^\phi U^l =
\rho( \Gamma^\phi_{tr} + \Gamma^\phi_{rt} ) U^r = \\- \rho\frac{U^r}{r^{3/2}} \left ( \frac{1}{2} \frac{H}{C} -1 \right ) \equiv
\rho\frac{U^r}{2r^{3/2}} \frac{E}{C},
\end{multline}
where
\begin{equation}
\label{E}
E = 1 - \frac{6}{r} + \frac{8a}{r^{3/2}} - \frac{3a^2}{r^2}.
\end{equation}

We now introduce the notation 
\begin{equation}
\label{T}
T_\nu \equiv \int_{-h}^{+h} T^{r\phi}_\nu = 2 \sigma^{r\phi} \int_{-h}^{+h} \eta \, dz,
\end{equation}
where $T_\nu$ is the vertically integrated density of the flux of the $\phi$-component of momentum in the radial direction.
Then, by integrating (\ref{part_1}) and (\ref{part_2}) over the disk thickness and combining them in one equation, we have
\begin{equation}
\label{T_eq}
\frac{\partial T_\nu}{\partial r} + \frac{2 T_\nu}{rD} \left ( 1 - \frac{1}{r} \right ) + 
\frac{\Sigma U^r}{2r^{3/2}} \frac{E}{CD^{1/2}} = 0,
\end{equation}
where the contribution from $\sigma^{\phi z}$ vanishes due to its being an odd function of $z$, and we neglect the dependence of $U^r$ on $z$, 
which gives rise to a higher-order correction (see footnote 2).
In formula (\ref{T_eq}) we have introduced the surface density of the disk
\begin{equation}
\label{Sigma}
\Sigma \equiv \int_{-h}^{+h} \rho \,dz.
\end{equation}

Important equation (\ref{T_eq}) with known boundary conditions at the inner disk radius allows us to calculate the profile
$T_\nu(r)$ for disk provided that the radial velocity distribution is known.

We note that equation for $T_\nu$ 
can also be derived from the angular momentum conservation law, which was used in the original paper
\cite{NT} (see equations  (5.6.3)-(5.6.6) therein).

\subsection{Rest energy conservation law. Radial momentum transfer}

To solve equation (\ref{T_eq}),
the radial velocity profile should be specified. 
It can be obtained from the rest energy conservation law (\ref{continuity}):
\begin{equation}
\label{cont_1}
{\bf e}_r (\rho U^r) + {\bf e}_z (\rho U^z) + \Gamma^i_{ki} \rho U^k = 0.
\end{equation}

Clearly, the substitution $U^i=(1,0,0,0)$ does not yield non-zero terms up to the order $\sim \delta^2$ (see the discussion at the end of Section \ref{sec_tetrad}). 
In our reference frame, this fact can be easily understood: the circular axially symmetric motion corresponds to zero velocity divergence. 
It is straightforward to check that the following terms $\sim s$ will appear in the continuity equation: 
\begin{equation}
\label{cont_2}
D^{1/2} (\rho U^r)_{,r} + (\rho U^z)_{,z} - \frac{(r \,D^{1/2})_{,r}}{r} \, \rho U^r = 0,
\end{equation}
where the last term arises due to the contribution from $\Gamma^\phi_{r\phi} \rho U^r$,
and similar terms with other velocity components, even if they appear, have an order higher than $\sim \delta^4$. 

After integrating over $z$, the contribution from the second term in (\ref{cont_2}) vanishes because
$\rho\to 0$ far from the equatorial disk plane, and we obtain
\begin{equation}
\label{cont_3}
(\Sigma U^r r D^{1/2} )_{,r} = 0.
\end{equation}

The combination whose derivative is found in (\ref{cont_3}) is a constant, which must be identified with the radial flux of matter. 
After additional integration over $\phi$ we obtain that 
\begin{equation}
\label{cont_4}
2\pi \, \Sigma U^r r D^{1/2} = -\dot M,
\end{equation}
where $\dot M > 0$ is the rate of the matter inflow into the disk at infinity, i.e. the mass accretion rate.

After substituting (\ref{cont_4}) in (\ref{T_eq}), we finally obtain  
\begin{equation}
\label{T_eq_AB}
\frac{d T_\nu}{dr} + P_1 T_\nu + P_2 = 0, 
\end{equation}
where
$$
P_1 = \frac{2}{r D} \left ( 1 - \frac{1}{r} \right ),
$$
$$
P_2 = -\frac{\dot M}{4\pi} \frac{E}{r^{5/2} CD}.
$$

The solution to  (\ref{T_eq_AB}) with the boundary condition $T|_{r_{ms}}=0$ is written in the form

\begin{equation}
\label{T_sol}
T_\nu =   \frac{1}{F(r)} \int_{r_{ms}}^r P_2(x) F(x) \, dx,
\end{equation}

\begin{equation}
\label{F_int}
F(r) = exp \left ( \int_{r_{ms}}^r P_1(x)dx \right ).
\end{equation}

The integral (\ref{F_int}) is elementary, and as a result we obtain
\begin{equation}
\label{T_sol_fin}
T_\nu = \frac{\dot M} {4\pi \, r^2 D} \int_{r_{ms}}^r  \frac{E}{r^{1/2} C} \, dr.
\end{equation}

\subsection{Energy balance}

Here, we consider equation (\ref{energy}). As above, let us set $U^i=(1,0,0,0)$ and find terms of the leading order in $\delta$.
As in the case of the azimuthal projection of the relativistic Euler equation, ‘ideal’ terms $[(\rho+\epsilon+p) U^t U^k]_{;k}$ and ${p\eta^{0k}}_{;k}$ 
do not contribute here. From the shear term, we have

$$
[2\eta \sigma^{tk}]_{;k} = \Gamma^t_{lk} \sigma^{lk} = 
2\eta [(\Gamma_{t\phi r}+ \Gamma_{tr\phi})\sigma^{r\phi} + O(\delta^2)] = 2\eta [4\sigma^{r\phi} + O(\delta^2)].
$$

Terms with $q^i$ contribute due to rapid change in the energy flux component normal to the disk with $z$:
$$
(U^t q^k)_{;k} = \frac{\partial q^z}{\partial z} + O(\delta^4).
$$

Summing all terms, we obtain from the energy balance equation 
\begin{equation}
\label{energy_bal}
\frac{\partial q^z}{\partial z} = 4\eta \left ( \sigma^{r\phi} \right )^2 = \frac{3}{2} T^{r\phi}_\nu \frac{D}{r^{3/2}C},
\end{equation}
whence, after integrating over the disk thickness, we derive an important relation
\begin{equation}
\label{F}
Q = \frac{3}{4} \frac{D}{r^{3/2} C} T_\nu, 
\end{equation}
where $Q=q^z(z=h)$ is the vertical energy flux escaping from the disk. After specifying $Q$, 
we can  calculate the radial profile of the effective temperature of the disk surface, because by definition
$Q =\sigma T_{eff}^4$.   
This is the universal result of the standard accretion disk theory: $T_{eff}$ does not depend on the specific nature of the dissipation of the 
kinetic energy of matter or on the mechanism of thermal energy transfer toward the disk surface, and is proportional to the value of $\dot M$, 
times some universal known function of $r$.

Thus, we have obtained the explicit form of the viscous stress integrated over the disk thickness, $T_\nu$, and the explicit 
form of the radiation energy flux from its surface, $Q$. 
At the same time, we know only the combination $\Sigma U^r$, 
and not each of these variables separately. In addition, we should determine the disk half-thickness profile, $h(r)$,
and the temperature, pressure and density distributions, $T(r,z)$, $p(r,z)$ and $\rho(r,z)$, inside it. 
To do this, the vertical disk structure should be calculated.

\subsection{Energy transfer equation and the vertical disk structure}
\label{sec_ener_str}

The vertical disk structure is defined by three equations. Two of them have already been obtained above: the vertical hydrostatic balance equation
(\ref{hydrostatics_rel}) and thermal energy generation equation (\ref{energy_bal}). 

The remaining equation is the transfer equation for energy dissipating in the disk. In the simplest case, the energy transfer is due to the photon diffusion in heated matter.
Strictly speaking, we should writ a relativistic analog of the radiation heat conductivity equation, which is
a variant of the kinetic Boltzmann equation for photons when their mean free path length 
is much smaller than the characteristic spatial length of the problem. 
Boltzmann equation is relativistically generalized in section 2.6 in \cite{NT}.
The standard transition to the diffusion approximation yields the following equation (see expression 2.6.43 in \cite{NT}):
\begin{equation}
\label{ph_diff}
q^i = \frac{1}{\tilde\kappa \rho} \frac{4}{3}b T^3 P^{ik} ( {\bf e}_k( T ) + a_k T), 
\end{equation}
where $\tilde\kappa$ is the Rosseland mean opacity of matter, $T$ is the temperature, $b$ is the radiation constant and
$a_k \equiv U_{k;j} U^j$ is the four-acceleration. The discussion of equation 
(\ref{ph_diff}) can be also found on p. 165 of \cite{Mihalas}.

As regards (\ref{ph_diff}), we first note that the four-acceleration never exceeds the order $\sim\delta^2$, 
because the four-velocity itself differs from the geodesic value (free circular equatorial motion) only in the second order in $\delta$. 
In contrast, the derivative in the first term in parentheses in the right-hand side of (\ref{ph_diff}) for $k=z$ raises the order in $\delta$, 
since $T$, as well as $\epsilon$, , vary significantly across  the disk thickness. As a result, as already discussed in Section \ref{sec_assumpt},
we see that $q^z$ is the leading component of vector ${\bf q}$ and is determined by the equation
\begin{equation}
\label{ph_diff_2}
q^z = - \frac{1}{3\tilde\kappa \rho} \frac{\partial (b T^4)}{\partial z},
\end{equation}
which is identical to the Newtonian form for a thin disk.

Equations (\ref{hydrostatics_rel}), (\ref{energy_bal}) and (\ref{ph_diff_2}) must be supplemented with the equation of state of matter 
$$
p(\rho,T),
$$ 
the opacity law 
$$
\tilde\kappa(\rho,T),
$$ 
and the explicit form of
$$
\eta(\rho,T), \mbox{or} \quad T^{r\phi}_\nu(\rho,T)
$$ 
depending on the type of parametrization of the turbulent viscosity in the disk. 

In addition, it is necessary to set boundary conditions at the integration interval
$z\in [0,h]$. 
In the simplest case, we assume that the disk has no atmosphere and 
$$
\rho|_{z=h}=T|_{z=h}=0.
$$ 
Furthermore, the energy flux vanishes in the disk equatorial plane:
$$
q^z|_{z=0}=0.
$$ 
Finally, we denote 
$$
2\int_0^h T^{r\phi}_\nu dz = T_\nu.
$$ 
Note that the above equations and boundary conditions for the vertical disk structure automatically guarantee the validity of equations
(\ref{cont_4}), (\ref{T_sol_fin}) and (\ref{F}) for the radial disk structure. 

After calculating the vertical structure, we can specify the surface density distribution using (\ref{Sigma}) and then 
$U^r$ using (\ref{cont_4}).

\subsection{Parametrization of turbulent viscosity and the explicit disk structure}
\label{sec_expl_disk}

Estimates carried out in \cite{SS_1973} and \cite{NT} according to the algorithm described in Section \ref{sec_ener_str}, 
show that at sufficiently high accretion rate $\dot M$, which is the free parameter of the problem, the radiation energy becomes dominant in the inner parts of the disk. 
The estimate of the threshold value of $\dot M$ can be found, for example, in \cite{SS_1973} (see formula 2.18 therein).  
It turns out that the disk thickness far away from its inner radius is independent on $r$ and for 
$\dot M$ of the order of and above the critical value, $\dot M_{cr}$ (when the disk luminosity reaches the Eddington value in the inner parts of the disk), $\delta >1$, 
corresponding to the spherization of the flow (see expression 7.1 and its discussion in \cite{SS_1973}). 
In addition, later studies showed that the radiation-dominated region is thermally unstable \cite{SS_1976} and convectively unstable \cite{BK_1977}.

This means that for the correct description of the inner parts of accretion disks at high accretion rates, when $\delta$ increases, terms of higher order in 
$\delta$ should be taken into account. These include the radial pressure gradient $\sim\delta^2$ in the radial force balance and the advection term,
$U^r T \partial S/\partial r \sim \delta^4$, which arises in the energy balance and accounts for the radial heat transfer. 
The latter, in fact, implies that the heat diffusion time in the vertical direction 
is comparable to its radial advection due to radial transfer of matter. 
In other words, the main property of the standard accretion disk model considered here is violated: the local energy balance in the disk, 
when the heat generated due to turbulent energy dissipation is locally released from the disk surface. 
It was found that the account for the new terms also allows one to correctly describe the region near
$r_{ms}$, where in the standard model $U^r\to\infty$, and to construct a stationary solution with $\delta <1$ for $\dot M$ of the order and above $\dot M_{cr}$
that is stable under thermal perturbations (so-called ‘slim-disks’, see \cite{BK_1981} and \cite{Abram_88} and their citations list, 
and, e.g., \cite{BK_2010}). 
Later, these results were confirmed by numerical simulations (see, e.g., \cite{Okuda_98} and \cite{Agol_01}). 
We add that the transition from a standard disk to a slim disk with increasing $\dot M$ 
in the relativistic model around a rotating black hole should occur even earlier due to a higher accretion efficiency (which is, in turn, due to both decreasing $r_{ms}$
and the additional angular momentum loss from the disk surface by radiation).

Now, assuming that $\dot M \ll \dot M_{cr}$,
let us calculate the disk vertical profile, which is to be useful in the next part of the paper, 
in the simplest case where the pressure is mainly determined by fully ionized hydrogen plasma, i.e.
\begin{equation}
\label{p_eq}
p = 2\rho k T/m_p,
\end{equation}
where $m_p$ is the mass of a proton, $k_B$ is the Boltzmann constant, and the opacity is determined by Thomson scattering, $\tilde\kappa= \kappa_T = 0.4 cm^2/g$.

Let us also assume that the kinematic viscosity $\nu$ is independent of $z$ and can be parametrized in the form 
\begin{equation}
\label{nu_par}
\nu = \alpha c_s h,
\end{equation}
where $0<\alpha<1$ is the Shakura parameter determining the turbulent viscosity in the disk (see \cite{Sh_1972} and \cite{SS_1973}), 
and $c_s$ is the speed of sound in the equatorial disk plane.  
Here, due to (\ref{p_eq}), 
\begin{equation}
\label{c_s_T}
c_s^2 = 2 k T_c/m_p, 
\end{equation} 
where $T_c = T(z=0)$.

Equation (\ref{ph_diff_2}) yields 
$$
\int_0^h dz q^z \rho = \left . - \frac{1}{3\kappa_T} b T^4 \right |_0^h = \frac{1}{3\kappa_T} b T_c^4.
$$
On the other hand, 
$$
\int_0^h dz q^z \rho = C_q F \int_0^h \rho dz = \frac{1}{2} C_q \Sigma F,
$$
where $C_q$ is some correction factor of the order of unity corresponding to the difference between the escaping radiation flux, $Q$, 
and its mean value along the disk thickness. 
As a result, we get

\begin{equation}
\label{T_centr}
T_c = \left ( \frac{3\kappa_T}{2} \frac{C_q}{b} \Sigma F \right )^{1/4}, 
\end{equation}

Next, for simplicity we assume that the entropy is constant along $z$ and,
dividing the left-hand side of (\ref{hydrostatics_rel}) by $\rho$, we
introduce the enthalpy, $dw = dp/\rho$, integrate 
(\ref{hydrostatics_rel}) over $z$ and obtain the central value of  $w$, $w_c \equiv w(z=0)$:
$$
w_c = -\int_0^h dw = \int_0^h \frac{z}{r^3} \frac{H}{C} = \frac{h^2}{2r^3}\frac{H}{C}, 
$$ 
Hence, using that $w_c = n c_s^2$, where $n$ is the polytrope index, we obtain
\begin{equation}
\label{c_s_h}
c_s^2 = \frac{h^2}{2nr^3} \frac{H}{C}.
\end{equation}

Finally, due to definition  (\ref{T}), parametrization (\ref{nu_par}) and equation (\ref{T_sol_fin}) we find 
\begin{equation}
\label{T_equat}
T_\nu =  \frac{3}{2} \frac{D}{r^{3/2}C} \alpha \Sigma c_s h = \frac{\dot M}{2\pi} \frac{Y}{r^{3/2} D}, 
\end{equation}
where in the second equality we introduce the new variable
\begin{equation}
\label{Y}
Y \equiv (2r)^{-1/2} \int_{r_{ms}}^r  \frac{E}{r^{1/2} C} \, dr,
\end{equation}
which in the Newtonian limit, far away from the inner edge of the disk, tends to unity. 

Equations (\ref{F}), (\ref{c_s_T}), (\ref{T_centr}) and (\ref{c_s_h}) are sufficient to exclude
all unknowns except $\Sigma$ and free parameters $\dot M$ and $\alpha$ from (\ref{T_equat}). 
We thus obtain the following surface density profile $\Sigma$:
\begin{equation}
\label{Sigma_expl}
\Sigma = \Sigma_0 \alpha^{-4/5}\dot M^{3/5} r^{-3/5} C^{3/5} D^{-8/5} H^{2/5} Y^{3/5},
\end{equation}
where the dimensional constant $\Sigma_0$ combines all relevant physical constants and numerical coefficients. 
Its explicit form and numerical value (which depends on the black hole mass to which we normalize all quantities) can be found by the reader. 

Now, using formulas (\ref{c_s_h}), (\ref{c_s_T}), (\ref{T_centr}) and (\ref{Sigma_expl}), it is possible to derive the profile $h(r)$.
The resulting disk aspect ratio reads
$\delta(r) = h(r)/r$:
\begin{equation}
\label{delta_expl}
\delta(r) = \delta_* r^{1/20} C^{9/20} D^{-1/5} H^{-9/20} Y^{1/5}, 
\end{equation}
where $\delta_*$ is a constant that determines the characteristic disk thickness $\delta$.

\chapter{Relativistic twisted accretion disk}

\section{Introductory remarks}
\label{sec_intro}

In Chapter 1 we described a flat disk in the equatorial disk around a rotating black hole. 
Its axially symmetric structure was evident and consistent with the symmetry of space near the black hole. 
If now we relax the main assumption that the flow of matter at all distances coincides with the equatorial plane,
the question arises: what can the dynamics of this more complicated flow, both stationary and non-stationary, be? 
Is this flow similar to a disk in any way? For thin disks considered here, the answer to this question proves to be positive under some restrictions.     

The main reason for the deformation of (for example, initially flat) disk is that the black hole spin gives rise 
to an additional off-center gravitational interaction with the gas elements of the flow.  
It can be shown that far away from the event horizon but close to the equatorial plane of the black hole this interaction 
is presented by an axially symmetric field of force directed to the black hole spin axis in the planes parallel to the equatorial one (see 
\cite{Thorne} chapter 3, paragraph A).

This force is called gravitomagnetic force and is given in this case by the expression
\begin{equation}
\label{GM_force}
F_{GM}= \frac{4a\Omega}{r^2}\frac{\partial}{\partial r},
\end{equation}
where $\Omega$ is the Keplerian frequency and $\partial/\partial r$ 
is the radial coordinate ort of the cylinder reference frame. Clearly, this external force can change the proper angular momentum of the disk elements 
(and hence deform the disk) moving outside the equatorial plane of the black hole.  
Here only the projection of the gravitomagnetic force onto the angular momentum direction matters, which is proportional 
to the sine of the angle between the angular momentum vector and the black hole spin axis. 
As we will see shortly, the restriction that allows us to treat the new configuration as a disk 
(both stationary and non-stationary) requires that the gravitomagnetic force be smaller than the central gravitational attraction force, 
i.e. requires the smallness of the parameter $a\ll1$.  
In addition, one more restriction can be formulated that the non-complanarity of the disk with the equatorial plane of the black hole, 
as well as the degree of its deviation from the planar form (i.e. twist, warp) 
should not exceed some small values for the disk to be hydrodynamically stable (see \cite{II_1997} paragraph 7 and \cite{ZhI} paragraph 4.2.4).

Let us split a thin planar disk into rings of small widths. In each ring, the motion of gas elements is mainly due to gravitational attraction force from the central body. 
The characteristic time of this motion is $t_d\sim\Omega^{-1}$. In addition, $t_d$ 
determines the time it takes for the disk to restore the hydrodynamic equilibrium across the ring, since the disk aspect ratio 
(the ratio of the the disk thickness to the radial distance) 
is of the order of the ratio of the sound velocity to the orbital velocity. This conclusion can be also arrived at by noticing 
that vertical pressure gradient is $\delta^{-1}$ times smaller than the unit mass gas element acceleration, 
i.e. exactly as small as the ratio of the radial size of the ring to its vertical scale. Thus, we can conclude that if other forces
acting on a given ring from the adjacent rings or from the black hole lead to the dynamics with the characteristic time $t_{ev}$ 
much greater than the dynamical one, $t_{ev}\gg t_d$, the hydrostatic equilibrium is maintained in the ring, in other words, the ring remains flat, 
and the entire flow preserves the disk-like form. This is undoubtedly so in a flat disk, because in this case equally oriented rings 
interact by the viscous force acting in the azimuthal direction and 
the angular momentum changes due to inflow and outflow of the matter accreting through the ring, with the both processes occurring on the diffusion time scale,
$t_{\nu}\sim\Omega^{-1}\delta^{-2}\gg t_d$. 

Now let the disk be tilted with respect to the equatorial plane of the black hole by a small angle $\beta\ll 1$. 
If in a flat disk the gravitomagnetic force contributes only to the modulus of acceleration of gas elements 
moving in circular orbits, but now, due to a non-zero projection of this force ($\propto \beta$) 
onto the angular momentum of gas elements, this force makes the orbits to precess around the black hole spin axis. For free particles, this effect is described 
in detail in the second part of the next Section in terms of the difference between the frequencies of circular and vertical motions. 
We also show that the precession frequency is much smaller than the circular frequency for 
$a\ll 1$ (see formula (\ref{lense_tirring})), 
which is equivalent to the condition $t_{ev}\gg t_d$ for whole rings composed of gas elements.

Equation (\ref{lense_tirring}) suggests that the precession of the rings is differential, i.e. depends on the distance to the center. 
As a result, the relative orientation of initially coaxial rings changes and the disk is no longer flat. 
However, we keep in mind that under the condition $t_{ev}\gg t_d$ each of the rings behaves 'rigidly' in its vertical direction, which is now also a function of $r$.
The new configuration is similar to a twisted (or warped) disk, i.e. a flow symmetric relative to some (now not planar) surface, 
which can be called the equatorial surface of the twisted disk.  Here, the cross-section of the equatorial surface by a plane passing through the center is a circle -- the instantaneous shape of orbits of gas elements rotating with a given radial distance $r$.
The disk turns into a set of rings tilted to the black hole equatorial plane by a constant angle $\beta$  
but with depending on $r$ node lines (the line formed by the intersection of the ring planes with the black hole equatorial plane). 
The node line is now determined by the position angle $\gamma(r)$
measured in the equatorial plane in the positive direction from a fixed direction to the ascending node of a given ring. 
The key point here is that the pressure gradient in the twisted disk, directed (as in any thin disk in general) 
almost normal to its warped surface, is not normal to the planes of the rings composing the disk. 
Therefore, we conclude that the pressure gradient acquires two projections. 
The main projection is coaxial with the rotational axis of each ring. Let us conventionally denote it as $(\nabla p)_{\xi}$, 
where $\xi$ is the distance from the equatorial surface of the twisted disk measured along the direction of rotation of the ring 
($\xi$ reduces to $z$ in the case of a flat disk).  We note from the beginning that $(\nabla p)_{\xi}\propto \xi$ 
due to the hydrostatic equilibrium across the ring.  The second projection of the pressure gradient, conventionally denoted as
$(\nabla p)_{r}$, lies in the ring's plane along the radial direction connecting the disk center and a given gas element of the ring. 
The ratio of these two projections is a small value proportional to the rate of change of 
orientations of rings in the disk, which, in turn, depends on the radial direction chosen in the given ring's plane. 
From purely geometrical considerations, we rigorously show in what follows that for a disk with
$\beta=const$ the ratio 
$(\nabla p)_{r}/(\nabla p)_{\xi}\propto \beta d\gamma/dr \cos\psi$, where $\psi$ is the angle measured in the azimuthal direction 
for the given ring from its ascending node to the given gas element. 
Note that the normal to the twisted disk surface is orthogonal to the ring's plane only in two diametrically opposite points -- 
where the ring's plane intersects the planes of the adjacent rings. At $\beta=const$, these points are characterized by $\psi=\pm \pi/2$. 
At the same time, in other pair of points with $\psi = 0,\pi$ the value of $(\nabla p)_{r}$ reaches positive and negative maxima.

Thus, in a flat disk, the dynamics in the radial direction is controlled in the leading order in $\delta\ll 1$
by the gravitation force and the corrections $\sim\delta^2$ 
are neglected, whereas in a twisted disk the radial projection of the pressure gradient starts additionally contributing to the radial balance. 
This addition, on the one hand, depends on the degree of the twist, and on the other hand, increases proportionally to the distance from the equatorial disk surface, $\xi$. 
Next, because it also depends harmonically on the azimuthal direction, the gas elements (for $\xi\neq 0$) are subjected to periodic disturbance 
by this force with the orbital period, and their orbits become ellipses with small eccentricity. 
As is well known, the eigenfrequency of the small oscillations of free particles in eccentric orbits is  equal to the epicyclic frequency, $\kappa$. 
Because the pressure gradient projection considered here excites exactly such oscillations, the radial profile of the epicyclic frequency, $\kappa(r)$, 
is an important characteristic that determines the shape of both stationary and non-stationary twisted configurations. 
In the next Section, we derive the required relativistic profile $\kappa(r)$ for equatorial circular orbits in the Kerr metric (see equation (\ref{epic_freq_Kerr})). 
Note from the beginning that in the special case of Newtonian gravitation $\kappa=\Omega$, 
and hence the action of the external exciting force on gas elements with the same frequency results in a resonance: 
the amplitude of the perturbed motion, characterized by perturbation of the orbital velocity, ${\bf v}$, 
must increase without a bound. This growth, however, is always limited by turbulent viscosity in the disk. Indeed, since the exciting force amplitude $\propto \xi$, 
so is the amplitude of ${\bf v}$. But this would mean the presence of the vertical velocity shear, $\partial_\xi {\bf v}$, 
in each ring. Together with the vertical density gradient (and hence the vertical gradient of the dynamic viscosity) in the disk,
this gives rise to a volume viscous force that damps the driving of individual layers of each disk's ring by the resonance force. 
Note that near the black hole, where the frequency $\kappa$ deviates from $\Omega$, the amplitude ${\bf v}$ 
remains bounded even in the absence of viscous forces. This allows the existence of stationary twisted disks with low viscosity around the black holes, 
in which $\beta(r)$ takes an oscillatory form (see \cite{II_1997}).

Thus, we see that the twist of the disk caused by the gravitomagnetic force necessarily results in a perturbation of the circular motion of gas elements in disk’s rings. 
The velocity field of this perturbation, ${\bf v}$, depends on $r$ (in addition to its being proportional to $\propto \xi$,
as explained above) and is determined by the current shape of the disk. By virtue of the continuity of the flow, this gives rise to density inhomogeneities outside the disk 
equatorial surface, $\rho_1\propto\xi$. Because $(\nabla p)_{r}\propto\cos\psi$,
these inhomogeneities take the opposite signs in the diametrically opposite points of any given ring. 
But this implies that the ring is subjected to the total torque of the central gravitational force 
acting on the density excesses out of the equatorial plane of the ring (i.e. outside $\xi=0$). We denote this torque as ${\bf T}_{g}$.
Because the disk is thin and the gravitational acceleration along ring’s axis is itself $\propto\xi$, 
the corresponding component of the gravitational force, and ${\bf T}_g$ as well, are quadratic in $\xi$. In addition, we remind that the torque ${\bf T}_g$ 
is proportional to the small warp magnitude, ${\bf T}_g \propto \beta d\gamma/dr$. 
Thus, we arrive at the conclusion that the dynamics of the twisted disk rings is controlled by ${\bf T}_g$, 
together with the torque due to the gravitomagnetic force discussed earlier in this introductory Section. Note that in the case 
$\beta=const$ considered here, $(\nabla p)_r$ and, correspondingly, $\rho_1$
take the maximum absolute values (but with the opposite signs) at $\psi=0,\pi$, i.e. at the node line of each ring
\footnote{To make the description as rigorous as possible, it is important also to add that the noted coincidence of azimuthal location of maxima of
$(\nabla p)_r$ and $\rho_1$ occurs only when the action of viscosity  on the gas elements of the ring is neglected.}. 
But this implies that ${\bf T}_g$ lies in the plane made by the angular momentum  of each ring and the black hole spin axis. 
By virtue of the symmetry of the problem, the total contribution to ${\bf T}_g$ from other azimuths does not alter its direction. Therefore, immediately after the gravitomagnetic force 
turns an imaginary tilted planar disk into a twisted configuration with  $\beta=const$, 
the gravitational force acting on the asymmetrically located matter of the disk relative to the surface $\xi=0$ 
tends to change the disk rings tilt angles: either to align the rings with the equatorial plane of the black hole or, conversely, to remove them from it. On the other hand, once
$\beta$ becomes dependent on $r$, the maxima of absolute values of $(\nabla p)_r$ are shifted from the node line of each ring to some new  $\psi$, 
which gives rise to a component in ${\bf T}_g$ that also contributes to the precession motion of the disk rings, as the gravitomagnetic torque does. 

The dynamics of twisted disks sketched above is complicated by the presence of non-zero viscosity in the disk. 
First of all, each ring of the disk is subjected to the action of the viscous force arising due to the difference between the direction of the
tangential velocity of the ring and that of the adjacent rings. This difference is maximal in the directions where the ring planes intersect,
i.e. exactly where $(\nabla p)_r$ vanishes. In the above example of configuration with $\beta=const$ this corresponds to $\psi=\pm \pi/2$, 
i.e. perpendicular to the node line of the rings. The viscous force, being proportional to the difference in tangential velocities, is directed at these points perpendicular 
to the ring plane and has different signs on different sides from the node line. Therefore, the corresponding torque, ${\bf T}_\nu$ is perpendicular to the plane made by the ring’s angular momentum and the black hole spin axis. 
In other words, the viscous interaction between the disk rings leads only to their precession around the black hole spin. 
Note also that the viscous torque ${\bf T}_\nu \propto \beta d\gamma/dr$, which appears due to the difference between the tangential velocities of adjacent rings,
and ${\bf T}_\nu \propto \xi^2$ due to the viscosity coefficient.
It is important to note that as soon as the profile $\beta(r)$ is formed due to the gravitational torque ${\bf T}_g$, the viscous torque ${\bf T}_\nu$
starts causing alignment/misalignment of the ring with the equatorial plane of the black hole. This happens for the same reasons by which ${\bf T}_g$ 
also starts contributing to the precession motion, as discussed above.

In addition to causing the appearance of ${\bf T}_\nu$,
the viscosity in a twisted disk, like in a flat accretion disk, leads to the radial diffusion transfer of the angular momentum component parallel 
to the equatorial plane of the black hole (which is non zero exactly for a tilted/twisted disk) toward the disk center due to simple transport of the accreting matter, 
and toward its periphery due to the corresponding angular momentum outflow. In the case of a relativistic disk, 
an additional loss of this angular momentum component occurs due to the thermal energy outflow by radiation from the disk surface (see equation (C6) in \cite{ZhI}).

All forces participating in the dynamics of twisted disks appear in the so-called ‘twist’ equation – the principal equation of the twisted disk theory. 
This equation is derived and analyzed in the subsequent Sections.

\subsection{Weakly perturbed circular equatorial motion: epicyclic frequency and frequency of vertical oscillations}
\label{sec_freq}
In a twisted disk, the motion of matter outside the equatorial plane of the Kerr metric is assumed; this motion is not necessarily circular in the projection onto that plane. 
Therefore, we first analyze the properties of free particles moving in orbits slightly different from circular ones. 

We first assume that particles move exactly in the equatorial plane but in slightly non-circular orbits. 
The problem can be solved using relativistic hydrodynamic equations with zero pressure and by assuming that there is a small addition to the purely circular velocity. 
Then, instead of equations (\ref{euler}) and (\ref{energy}), it is better to use the original equations in the form 
\begin{equation}
\label{epic_Tik}
{T^{ik}}_{;k} = 0,
\end{equation}
where in the considered case of free motion, ${T^{ik}} = \rho U^i U^k$ and $\rho=const$. 
Under the last assumption, the velocity field, as follows from the rest-energy conservation law  (\ref{continuity}), 
is divergence-free, and (\ref{epic_Tik}) is equivalent to the following equation:
\begin{equation}
\label{epic_Uik}
{U^i}_{;k} U^k = 0.
\end{equation}
We now single out from the four-velocity field a small addition to the main circular equatorial motion and denote it as $v^i$. 
The unperturbed motion corresponds to rest in the projection onto tetrad (\ref{e_1}-\ref{e_4}) used to construct the flat accretion disk model, i.e. is given by the four-velocity $U^i_0 = \{1,0,0,0\}$. 
Substituting the sum $U^i_0 + v_i$ in (\ref{epic_Uik}), we obtain linear equations for small perturbations of the four-velocity, $v_i$, 
which is assumed to be a function of $t$ only:
\begin{equation}
\label{epic_lin_eq}
{v^i}_{;k} U^k_0 + {U^i_0}_{;k} v^k = 0. 
\end{equation}
Taking into account that ${U^i_0}_{;k} = \Gamma^i_{tk}$, for $i=1,2$ we obtain the system of equations
\begin{equation}
\label{epc_eqs_1}
{v^r}_{;t} + \Gamma^r_{t\phi} v^\phi = C^{-1/2} B \frac{dv^r}{dt} - 2 r^{-3/2} v^\phi = 0,
\end{equation}
\begin{equation}
\label{epic_eqs_2}
{v^\phi}_{;t} + \Gamma^\phi_{tr} v^r = C^{-1/2} B \frac{dv^\phi}{dt} + 
r^{-3/2}\left (1-\frac{1}{2}\frac{H}{C}\right ) v^r = 0.
\end{equation}
It follows that small perturbations of the four-velocity components in the equatorial plane of the rotating black hole oscillate in time. For example, $v^r$ 
satisfies the equation
\begin{equation}
\label{epic_v_r}
\frac{d^2v^r}{dt^2} + \frac{2C}{r^3 B^2} \left ( 1 - \frac{H}{2C} \right ) v^r = 0, 
\end{equation}
which implies that the square of the frequency of these oscillations, which is the epicyclic frequency by definition, has the form
\begin{equation}
\label{epic_freq_Kerr}
\kappa^2 = r^{-3} B^{-2} ( 2C-H ) = 
r^{-3}\left ( 1 + \frac{a}{r^{3/2}} \right )^{-2} 
\left ( 1-\frac{6}{r} + \frac{8a}{r^{3/2}} - \frac{3a^2}{r^2}\right ).  
\end{equation}

A somewhat different derivation of $\kappa$ can be found in the Appendix in \cite{Okazaki87}. 
It is important to note that (\ref{epic_freq_Kerr}) contains a derivative with respect to the coordinate time, and therefore the epicyclic frequency 
is determined by the clock of an infinitely remote observer, similarly to circular frequency (\ref{kepl_freq}) introduced above. By comparing equation (\ref{eq_r_ms}),
which defines the location of the innermost stable circular equatorial orbit in the Kerr metric, $r_{ms}$, with 
(\ref{epic_freq_Kerr}), we infer that $\kappa^2(r_{ms})=0$. For $r<r_{ms}$ the epicyclic frequency becomes imaginary, and equation (\ref{epic_v_r})
has exponentially growing solutions. It must be so because in this region the free circular motion around a rotating black hole becomes unstable. In Section
\ref{sec_rms} this result was obtained from the analysis of the form of the effective centrifugal potential 
in which a test particle moves in an equatorial circular orbit. Nevertheless, we see that $r_{ms}$
can be determined alternatively from the calculation of the profile $\kappa^2(r)$ in the Kerr metric. 

It is well known that for Newtonian motion so-called Keplerian degeneration occurs when 
$\kappa=\Omega$ for a non-circular motion, which causes non-relativistic orbits to be closed. 
However, this symmetry is broken for relativistic free motion, and the epicyclic frequency $\kappa$ differs from $\Omega$ 
already near a non-rotating ($a=0$) black hole, where its square is
\begin{equation}
\label{epic_freq_Schw}
\kappa^2 = r^{-3} \left ( 1 - \frac{6}{r} \right ) = \Omega^2 \left ( 1 - \frac{6}{r} \right ) < \Omega^2.
\end{equation}
The difference between the epicyclic and circular frequencies results in the well known effect of the precession of an elliptical orbit. 
Far away from the horizon of a Schwarzschild black hole, i.e. for $r\gg 1$, 
the frequency of the orbit rotation, called the Einstein precession frequency, is 
$\Omega_p\approx 3 / r^{5/2}$.

We now suppose that we rotate together with the test particle at some radius. When considering the problem in the projection onto tetrad (\ref{e_1}-\ref{e_4}),
this particle remains at rest. We now impart to the particle a small velocity in the direction perpendicular to the equatorial plane. 
Equation (\ref{hydrostatics_rel}) of hydrostatic equilibrium for a flat disk 
implies that in our reference frame the particle, being in free motion, is subjected to acceleration that is proportional to $z$ and tends to return 
the particle to $z=0$. 
As a result, the test particle will harmonically oscillates with a frequency whose square is

\begin{equation}
\label{vert_freq_loc}
{\Omega_v^{l}}^2 = \frac{H}{r^3 C},
\end{equation}
where the superscript `l'
reminds us that the frequency is measured in the reference frame comoving with the particle in its main circular equatorial motion. 
To re-define this frequency as measured by the clock of an infinite observer, as has been done for both circular and epicyclic frequencies, the frequency
$\Omega_v^l$
must be divided by the time dilation factor (the difference between the proper time of the particle and the time at infinity), i.e. by the
$t-$component of the four-velocity  (\ref{geod_vel}).
Thus, the square of the frequency of vertical oscillations is 
\begin{equation}
\label{vert_freq}
{\Omega_v}^2 = r^{-3}B^{-2} H = r^{-3}\left ( 1 + \frac{a}{r^{3/2}} \right )^{-2} 
\left ( 1 - \frac{4a}{r^{3/2}} + \frac{3a^2}{r^2} \right ),
\end{equation}
which coincides, for example, with the expression presented in \cite{Kato90} (see also \cite{Ipser96}).
Eq. (\ref{vert_freq}) implies that around a non-rotating black hole  $\Omega_v = \Omega$.
This means that the vertical and circular motions have the same period, and the total motion of the particle 
is again the circular motion in a closed orbit whose plane, however, 
is now slightly tilted toward the initial equatorial plane. The situation changes for $a\neq 0$, 
because for $\Omega_v \neq \Omega$ the orbit is not closed any more, and the orbital plane starts precessing around the spin axis of the black hole. 
The frequency of the orbital precession is equal 
to the difference between the circular and vertical frequencies. For a slowly rotating black hole with
$a\ll 1$ the precession frequency of a slightly tilted orbit is
\begin{equation}
\label{lense_tirring}
\Omega_{LT} = \Omega - \Omega_v \approx r^{3/2} \left ( 1 - \frac{a}{r^{3/2}} \right ) - 
r^{3/2} \left ( 1 - \frac{3a}{r^{3/2}} \right ) = \frac{2a}{r^3} \ll \Omega.
\end{equation}
This is simply the angular velocity of the frame dragging by the rotating black hole  (see equation
(\ref{LT})) in the limit $a\ll 1$. The frequency $\Omega_{LT}$ is also referred to as the Lense-Thirring frequency. 

In the most general case, where the test particle deviates from circular motion simultaneously in the vertical and horizontal directions, 
the particle's motion in space can be described by a slightly elliptical orbit, 
with both plane and apse line turning with an angular velocity proportional to the difference between 
the circular and vertical frequency and the difference between the circular and  epicyclic frequency, respectively. For
$a\ll 1$, the precession of the orbital plane occurs on a timescale much longer than the dynamical time, $t_{LT} \gg t_d$, 
where $t_{LT}\sim\Omega_{LT}^{-1}$ (see the discussion in the previous Section).

\section{Choice of the reference frame}

\subsection{The metric}

Taking the general conclusions in Section 2.1 into account, we consider slowly rotating black holes, $a\ll 1$. 
In this case, the linear expansion of the Kerr metric in the parameter $a$ is sufficient. Then formula (\ref{Kerr}) takes the form

\begin{equation}
ds^2 =  (1-2/ R ) dt^2 -(1-2/R)^{-1} dR^2 -
R^2(d\theta^2 + sin^2\theta d\phi^2) + 4 \frac{a}{R}
sin^2\theta\, d\phi\, dt. \label{Kerr_LT}
\end{equation}
Metric (\ref{Kerr_LT})
is identical to that of a non-rotating black hole written in the Schwarzschild coordinates, except for one non-diagonal term responsible for the Lense-Thirring precession. 

Our main purpose in this Section is to introduce the relativistic reference frame that follows the disk twist. 
The symmetry of the problem implies that the equations of motion should have the simplest form in such a frame. As for a flat disk, 
it is convenient to use some orthonormal non-coordinate basis. For this basis to follow the disk shape, 
its two spatial orts should be tangent to the disk symmetry plane. At each spatial point we take the orts of the 'flat' basis, 
which are determined, say, by the equatorial plane of the black hole, and turn them by the angles
$\beta$ and $\gamma$ defining the disk shape. This is done in the simplest way by using a Cartesian coordinate system with the axis  $z$ 
parallel to the black hole spin. However, we should first understand which four-dimensional basis (whose dual tetrad must transform the metric
(\ref{Kerr_LT}) into the Minkowski metric) in the flat-space limit would produce the spatial part described by the Cartesian reference frame.

This can be done by changing the radial variable in (\ref{Kerr_LT}), namely, by passing from $R$ to the so-called 'isotropic' radial coordinate, $R_I$:

\begin{equation}
R = R_I\left (1+\frac{1}{2R_I} \right )^2 \label{R_I}.
\end{equation}

Substituting (\ref{R_I}) in (\ref{Kerr_LT}) yields
\begin{equation}
ds^2 = \left ( \frac{1-\frac{1}{2R_I}}{1+\frac{1}{2R_I}} \right )^2
dt^2 - \left (1+\frac{1}{2R_I} \right )^4 (dR_I^2 + R_I^2d\theta^2 + R_I^2
sin^2\theta d\phi^2) + 4\frac{a\,sin^2\theta}{R_I \left ( 1
+\frac{1}{2R_I} \right )^2} dt\, d\phi \label{Kerr_R_I},
\end{equation}
where the second term represents the elementary spherical volume. Now, it is easy to transform to the Cartesian coordinates by the change $\{x=R_I\cos\phi\sin
\theta,y=R_I\sin\phi\sin\theta, z=R_I\cos \theta\}$. With account for $R^2_I\sin^2\theta d\phi = xdy - ydx$ we have

\begin{equation}
ds^2 = K_1^2 dt^2 + 2a K_1 K_3 (xdy-ydx) dt - K_2^2 (dx^2 + dy^2+ dz^2) \label{Kerr_car},
\end{equation}
where
\begin{equation}
K_1 = \frac{1-\frac{1}{2R_I}}{1+\frac{1}{2R_I}}, \quad K_2 = \left (
1+\frac{1}{2R_I} \right )^2, \quad  K_3 =
\frac{2}{R_I^3}\frac{1}{1-\left (\frac{1}{2R_I} \right )^2},
\label{KKK}
\end{equation}
are functions of $R_I = (x^2+y^2+z^2)^{1/2}$ only.

Metric (\ref{Kerr_car}) generates the following dual basis
\begin{equation}
\mbox{\boldmath$e$}^t = K_1 dt + a K_3 (xdy-ydx),
\quad \mbox{\boldmath$e$}^x = K_2 dx, \quad
\mbox{\boldmath$e$}^y = K_2 dy, \quad
\mbox{\boldmath$e$}^z = K_2 dz. \label{dual_tw_1}
\end{equation}

Note that basis  (\ref{dual_tw_1}) corresponds to the observers at rest in the Schwarzschild coordinates, because their world lines defined by the condition 
$U^i = {\bf e}^i/ds=\{1,0,0,0\}$, correspond to the equalities $dx=dy=dz=0$. 
Their identical clocks are synchronized in such a way that 
in equal time intervals determined by the ort
${\bf e}^t$,
light travels an equal distance in any direction defined by the combination of the ${\bf e}^{x,y,z}$. 
If the observers used the coordinate time, $t$, they would discover, for example, that the light signal in the azimuthal direction prograde with the 
black hole spin travels a larger distance than in the opposite (retrograde) direction. 
This follows from the frame-dragging effect of a rotating black hole 
and is equivalent to the well-known tilt of light cones in the azimuthal direction. 
Finally, we note that another choice of the orthonormal basis is possible in principle, which also compensates the space-dragging effect. 
Such a basis is called the frame of locally non-rotating observers, which is moving with the azimuthal angular velocity equal to (\ref{LT}); 
mathematically, this corresponds to the correction of the azimuthal ort instead of the time one (see \cite{BPT_1972}).

Below, we need to rotate the spatial part of (\ref{dual_tw_1}), so as to obtain the dual twisted basis and then the original basis, 
which, as we recall, is needed to write down the projection of hydrodynamic equations. 
For this, let us first introduce the twisted cylindrical coordinates.

\subsection{Twisted coordinates}

We define the twisted cylindrical coordinates $\{\tau, r,\,\psi,\,\xi\}$ such that the condition $\xi=0$
determines a coordinate surface coincident with the equatorial surface of a twisted disk. Here,
$\tau$, $r$, $\psi$ and $\xi$ are the new time variable and twisted analogs of the radial, azimuthal and vertical cylindrical coordinates, respectively\footnote{Here and
hereafter, $r$ denotes the twisted radial coordinate.} These coordinates were first introduced in \cite{Pet_1977} and \cite{Pet_1978}. 
At each fixed $r=const$, angle $\psi$ is measured in the positive direction from the ascending node of the circle $\xi=0$ 
crossing the equatorial plane of the black hole. The relation between $\{\tau, r,\,\psi\,\xi\}$ and $\{t,\,x,\,y\,\,z\}$ 
can be obtained by a sequence of rotations at each radial distance by the angles $\beta(r,\tau)$ and $\gamma(r,\tau)$.

Let us take the radius vector with coordinates

\begin{equation}
\label{vector}
\left[ 
\begin {array}{c} 
\tau\\
 r\cos\psi\\
r\sin\psi\\
\xi
\end {array} 
\right],  
\end{equation}
where three spatial Cartesian coordinates are defined in a frame with the $z$-axis tilted by the angle $\beta(r,\tau)$ toward the black hole spin and the $x$-axis 
lying in the black hole equatorial plane and turned by the angle $\gamma(r)$ relative to some direction common for all $r$.

Next, we consecutively rotate this frame by the angle $\beta(r,\tau)$ about its $x$-axis in the negative direction and then by the angle $\gamma(r,\tau)$ 
about its $z$-axis in the negative direction. After these two rotations, this frame transforms into a 'flat' Cartesian frame common at all $r$ with 
the $xy$-plane coinciding with the equatorial plane of the black hole. 
Herewith, the new coordinates of the radius-vector are obtained by multiplying (\ref{vector}) first by the matrix  
\begin{equation}
\label{A1}
{\it A_1}(\beta)\, = \, \left[ 
\begin {array}{cccc} 
1&0&0&0\\
0&1&0&0\\
0&0&\cos \beta  & -\sin \beta  \\
0&0&\sin \beta  &  \cos \beta 
 \end {array} \right], 
\end{equation}
and then by the matrix
\begin{equation}
\label{A2}
{\it A_2}(\gamma)\, = \, \left[ 
\begin {array}{cccc} 
1&0&0&0\\
0&\cos \gamma & -\sin \gamma &0\\
0&\sin \gamma &  \cos \gamma &0\\
0&0&0&1
\end {array} \right]. 
\end{equation}

As a result, we obtain the following relation between the twisted cylindrical and 'flat' Cartesian coordinates taken in the linear approximation in small $\beta$:

\begin{equation}
\label{transform}
\begin {array}{lll} 
 t &= &\tau\\
 x& =& r \cos \gamma \cos \psi  - \sin \gamma  \left(r \sin \psi - \xi \beta \right) \\
 y& =& r \sin \gamma \cos \psi  + \cos \gamma  \left(r \sin \psi - \xi \beta \right) \\
 z& =& r \beta  \sin \psi  + \xi. 
\end {array} 
\end{equation}

\subsection{Tetrad transported by the twist-following observers}

We now pass from the 'flat' basis (\ref{dual_tw_1})
to the twisted one by rotating its spatial orts by the twisting angles at each spatial point. First, we need to perform the rotation strictly opposite to what we did in 
the previous paragraph. This means that we should take basis (\ref{dual_tw_1}) as a column and first multiply it by the matrix $A_2(-\gamma)$ and then by the matrix 
$A_1(-\beta)$. After that, because we wish to obtain the basis corresponding to the (twisted) cylindrical frame, 
it is necessary to additionally 'advance' the three spatial orts by azimuthal angle $\psi$, what is achieved by additional multiplication of the basis by the matrix $A_2(-\psi)$. 

As a result, we obtain the twisted dual basis that contains some linear combinations of the 'flat' coordinate orts, $\{dt,\,dx,\,dy,\,dz\}$. 
It remains to express it as linear combinations of coordinate orts of the twisted coordinate frame, $\{d\tau,\,dr,\,d\psi,\,d\xi\}$.
For this, it suffices to take differentials of the coordinate transformation 
(given by (\ref{transform}) in the linear approximation in $\beta$) 
and to substitute them in the twisted dual basis obtained after the rotations.  It can be checked that in the approximation linear in $\beta$ and $a$, we have

\begin{equation} \mbox{\boldmath$e$}^\tau = (K_1-a r \xi K_3
\partial_\varphi U) d\tau +
                      a\xi K_3 \partial_\varphi (Z - r W )d r + ar K_3 (r-\xi Z) d\varphi -
                       a r K_3 \partial_\varphi Z d\xi,
\label{dual_tw_21}
\end{equation}
\begin{equation}
\mbox{\boldmath$e$}^r = -\xi K_2 U d\tau +
K_2(1-\xi W) dr, 
\label{dual_tw_22}
\end{equation}
\begin{equation}
 \mbox{\boldmath$e$}^\varphi = -\xi K_2
\partial_\varphi U d\tau -
                        \xi K_2 \partial_\varphi W dr + r K_2
                        d\varphi,
\label{dual_tw_23}
\end{equation}
\begin{equation}
\mbox{\boldmath$e$}^\xi = r K_2 U d\tau + r K_2 W
dr + K_2 d\xi, 
\label{dual_tw_24}
\end{equation}
where we introduce the new azimuthal variable $\varphi = \psi + \gamma(r)$ 
and pass to partial derivatives with respect to the corresponding new coordinates.

We also introduce new variables characterizing the disk geometry:
\begin{equation} 
\Psi_1 = \beta \cos \gamma, \,\, \Psi_2 = \beta \sin \gamma \label{Psi}
\end{equation}
and from now on use them instead of the angles
$\beta $ and $\gamma $. Additionally,
\begin{equation} 
Z = \beta \sin \psi = \Psi_1 \sin \varphi -
\Psi_2 \cos \varphi, \quad  U=\dot Z, \quad W=
Z^\prime,
\label{ZUW}
\end{equation}
where partial derivatives with respect to $\tau$ and $r$ are denoted by the dot and the prime.

It follows that for $\beta=\gamma=0$ and with the additional transition to the Cartesian coordinates, basis (\ref{dual_tw_21}-\ref{dual_tw_24}) 
is transformed into the 'flat' basis (\ref{dual_tw_1}).

As discussed above, observers transporting basis (\ref{dual_tw_1}) are at rest in the Schwarzschild coordinates. 
On the contrary, observers corresponding to basis (\ref{dual_tw_21}-\ref{dual_tw_24})
move in space by following the changing shape of the twisted disk 
(in the non-stationary dynamics).

As we have seen in Chapter 1, the original basis onto which hydrodynamic equations are projected is obtained by inverting the dual basis matrix. 
Using (\ref{dual_tw_21}-\ref{dual_tw_24}), in the approximation linear in $\beta$ and $a$, we have

\begin{equation}
{\bf e}_\tau = \frac{1}{K_1} \left ( \partial_\tau + \xi U
\partial_r + \frac{\xi}{r} \partial_\varphi U \partial_\varphi - r
U \partial_\xi \right ), \label{dual_tw_31}
\end{equation}
\begin{equation}
{\bf e}_r = \frac{1}{K_2} \left ( -a\xi\frac{K_3}{K_1}
\partial_\varphi Z \partial_\tau + (1+\xi W) \partial_r +
\frac{\xi}{r} \partial_\varphi W \partial_\varphi - r W
\partial_\xi  \right ),
\label{dual_tw_32}
\end{equation}
\begin{equation}
 {\bf e}_\varphi = \frac{1}{K_2} \left ( -a\frac{K_3}{K_1} (r
- \xi Z) \partial_\tau - a\xi \frac{ K_3}{K_1} r U \partial_r +
\left ( \frac{1}{r} - a\xi \frac{K_3}{K_1} \partial_\varphi U
\right ) \partial_\varphi + ar \frac{K_3}{K_1} r U \partial_\xi
\right ),\label{dual_tw_33}
\end{equation}
\begin{equation}
{\bf e}_\xi = \frac{1}{K_2} \left (  ar\frac{K_3}{K_1}
\partial_\varphi Z  \partial_\tau + \partial_\xi \right ).
\label{dual_tw_34}
\end{equation}

With the original and dual bases in hands, using the algorithm presented in Section \ref{sec_bases}, 
we can calculate the connection coefficients. This very cumbersome but straightforward procedure yields the following non-zero connection coefficients
in the linear approximation in $\beta$ and $a$: 

 \begin{equation}
\begin{array}{ll}

\Gamma_{\tau r \tau} = \frac{K_1^\prime}{K_1 K_2}, & \Gamma_{\tau
r \varphi} = a\frac{K_3}{K_2^2}\,\left ( 1 - \frac{1}{2} \left (r
- \xi Z \right ) K_4 \right ), 
\\&\\ 
\Gamma_{\tau r \xi} = - a\frac{K_3}{K_2^2}
\partial_\varphi Z \left ( 1 - \frac{1}{2r} \left (r^2+\xi^2
\right ) K_4
\right ) , &
\Gamma_{\tau \varphi r} = -\Gamma_{\tau r \varphi}, 
\\&\\
\Gamma_{\tau
\varphi \xi} = a\frac{K_3}{K_2^2} \left ( Z + \frac{\xi}{2r} \left (r-\xi Z
\right ) K_4 \right ), & \Gamma_{\tau \xi \tau} =
\frac{\xi}{r}\frac{K_1^\prime}{K_1 K_2},
\\&\\
\Gamma_{\tau \xi r} = -\Gamma_{\tau r \xi}, & \Gamma_{\tau \xi
\varphi} = -\Gamma_{\tau \varphi \xi}, 
\\&\\
 \Gamma_{r \varphi \tau} =
\frac{\xi}{r} \frac{1}{K_1}  \partial_\varphi U - \Gamma_{\tau r
\varphi}, &
\Gamma_{r \varphi r} = \frac{\xi}{r} \frac{1}{K_2}
\partial_\varphi W, 
\\&\\ 
\Gamma_{r \varphi \varphi} = \frac{(r
K_2)^\prime}{r K_2^2} - a\xi \frac{K_3}{K_1 K_2} \partial_\varphi
U, & \Gamma_{r \xi \tau} = \frac{U}{K_1} - \Gamma_{\tau r \xi},
\\&\\
\Gamma_{r \xi r} = \frac{W}{K_2} -
\frac{\xi}{r}\frac{K_2^\prime}{K_2^2}, & \Gamma_{r \xi \varphi} =
-ar\frac{K_3}{K_1 K_2} U, 
\\&\\ 
\Gamma_{r \xi \xi} = \frac{K_2^\prime}{K_2^2}, &
\Gamma_{\varphi \xi \tau} = \frac{1}{K_1} \partial_\varphi U -
\Gamma_{\tau \varphi \xi}, 
\\&\\ 
\Gamma_{\varphi \xi r} =
\frac{1}{K_2} \partial_\varphi W, & \Gamma_{\varphi \xi \varphi} =
-\frac{\xi}{r}\frac{K_2^\prime}{K_2^2} - ar\frac{K_3}{K_1 K_2}
\partial_\varphi U,
\\&\\
\\
\end{array}
\label{connect}
\end{equation}
where $K_4 \equiv K_3/K_1 (K_1/K_3)^\prime$.
The other non-zero $\Gamma_{ijk}$, as usual, can be obtained by taking their asymmetry in the first two indices into account.

Thus, the basis (\ref{dual_tw_31}-\ref{dual_tw_34}) together with the connection coefficients (\ref{connect}) are the sum of two parts: the main part that persists at 
$\beta=0$ and a small additional part $\propto\beta$. In what follows, we conventionally denote these parts ``${\rm B_0}$'' and ``${\rm B_1}$'', respectively.

\section{System of twist equations}

\subsection{Projection of dynamical equations onto the twisted basis for a thin disk}

\subsubsection{Separation of equations into two systems describing a flat disk and a twisted disk}

We take the relativistic hydrodynamic equations in the original form:

\begin{equation}
\label{Tik_2}
T^{ik}_{\,\,;k} = 0,
\end{equation}
where the stress-energy tensor and its components are presented in Section \ref{sec_rel_eq}.
Equations (\ref{Tik_2}) should now be projected onto the twisted basis (\ref{dual_tw_31}-\ref{dual_tw_34}).
To perform this, we assume that $\beta\ll 1$.
In other words, mathematically we consider the twist of the disk as a small perturbation to its 'ground' state, 
i.e. to the model of a flat disk, also referred to as the background. 
It is important that the appearance of a twist gives rise to new terms in equations 
not only due to the bending of the basis, but also due to the appearance of additional perturbations 
of physical quantities themselves that enter the stress-energy tensor, including the density, pressure and four-velocity.

For a twisted disk, instead of (\ref{Tik_2}) we can write 
\begin{equation}
\label{Tik_split}
( {({T_0}^{ik} + {T_1}^{ik})}_{;k} )_0 + ( {({T_0}^{ik} + {T_1}^{ik})}_{;k} )_1 = 0, 
\end{equation}
where ${T_0}^{ik}$ corresponds to the background state and ${T_1}^{ik}$ 
is a small {\it Eulerian} perturbation of the stress-energy tensor.  The indices 0 and 1 that follow the notation of 
the covariant derivative mean that the derivative is taken in bases ${\rm B_0}$ and ${\rm B_1}$, respectively.

The action of the covariant derivative with index 0 on ${T_0}^{ik}$, evidently, yields 0, because these are equations for the background:
\begin{equation}
 \label{Tik_0}
( {{T_0}^{ik}}_{;k} )_0 = 0.
\end{equation}

Then, in the approximation linear in $\beta$, we find the twist equations:
\begin{equation}
\label{Tik_tw}
( {{T_1}^{ik}}_{;k} )_0 + ( {{T_0}^{ik}}_{;k} )_1 = 0.
\end{equation}

We assume that in a twisted disk the four-velocity, pressure, rest-mass energy density, internal energy, 
viscosity coefficient and energy flux density, as defined in their standard sense (see Section 
\ref{sec_rel_eq}), are given by
$$
U^i = U_0^i + v^i, \, p = p_0+p_1, \, \rho = \rho_0+\rho_1, \, \epsilon =\epsilon_0+\epsilon_1,\, 
$$
$$
\eta=\eta_0+\eta_1,\, q^i=q^i_0+q^i_1, 
$$
respectively. Here, the indices 0 and 1 denote values related to the background and
perturbations, respectively; 
$v^i$ are perturbations of the four-velocity\footnote{To shorten the equations, we  omit the term with the second viscosity $\zeta$: 
using the analysis be given below, it can be shown that this term does not contribute to the final equations in the leading order in the small parameters of the problem.}

Thus, $T_0^{ik}$ is the stress-energy tensor that contains only unperturbed quantities in accordance with definition (\ref{Tik}) and its perturbation has the form
$$
T_1^{ik}=w_1 U_0^i U_0^k + w_0 (v^i U_0^k + U_0^i v^k) - p_1 \eta^{ik} +
2\eta_1 \sigma_0^{ik} + 2\eta_0 \sigma_1^{ik} - 
$$
\begin{equation}
\label{T_1_ik}
U_0^i q_1^k - U_0^k q_1^i - v^i q_0^k - v^k q_0^i,
\end{equation}
where $w_0=\rho_0+\epsilon_0+p_0$ is the background enthalpy and $w_1 = \rho_1+p_1+\epsilon_1$ is its perturbation.

In addition, $\sigma_0^{ik}$ is the shear tensor which contains only unperturbed quantities in accordance with definition (\ref{shear}), and $\sigma_1^{ik}$
is its perturbed part of the form
$$
\sigma_1^{ik} = {1\over 2}[ (v^{i}_{;j})_0 P_0^{jk}+(v^{k}_{;j})_0 P_0^{ji}]-{1\over
3} ( v^{j}_{;j})_0 P_0^{ik} + 
$$
$$
{1\over 2}[({U_0}^{i}_{;j})_0 P_1^{jk}+({U_0}^{k}_{;j})_0 P_1^{ji}] - {1\over
3}({U_0}^{j}_{;j})_0 P_1^{ik} +
$$
\begin{equation}
\label{sigma_1_ik}
{1\over 2}[({U_0}^{i}_{;j})_1P_0^{jk}+({U_0}^{k}_{;j})_1P_0^{ji}] - {1\over
3}({U_0}^{j}_{;j})_1 P_0^{ik},
\end{equation}
where $P_0^{ik}$ is the projection tensor that contains only unperturbed quantities in accordance with definition (\ref{proj_ten}), 
and its perturbation is written as $P_1^{ik} = - U_0^i v^k - U_0^k v^i$.

Everywhere below we omit the index 0 for the unperturbed quantities. In addition, the viscous part of the stress-energy tensor in the disk is marked with
``$\nu$''  wherever necessary: $T^{ik}_\nu \equiv 2\eta \sigma^{ik}$.

\subsubsection{Additional relations used to write the equations}

The relations given below are valid through terms of the order of $\propto\delta^2$, which is sufficient for the theory of twisted disks in the leading order 
in the small parameter $\delta$. 
In deriving these relations, such a simplification enables us to consider that in the background solution, only
$U^\tau$ and $U^\varphi$ are non-zero, while $U^r\propto\delta^2$ and 
$U^r$ can be temporarily set equal to zero.

We first note that the following relation between the components $U^\tau$ and $U^\varphi$ is used below:
\begin{equation}
\label{U_norm}
(U^\tau)^2 = (U^\varphi)^2 + 1, 
\end{equation}
which follows from the expression for the norm of the four-velocity in an orthonormal basis. Constraint (\ref{U_norm}) 
is also useful in the differential form: 
\begin{equation}
\label{U_norn_diff}
U^\tau dU^\tau = U^\varphi dU^\varphi.
\end{equation}

Next, because the normalization of the four-velocity is also valid in the twisted disk, and the four-velocity perturbations are small, in the linear approximation we have
$$
(U^\tau + v^\tau)^2 - (U^\varphi+v^\varphi)^2 = (U^\tau)^2 + 2U^\tau v^\tau - (U^\varphi)^2 - 2U^\varphi v^\varphi = 1, 
$$ 
and hence, with account for (\ref{U_norm}), $v^i$ is ``orthogonal'' to $U^i$:
\begin{equation}
\label{v_orth}
U^\tau v^\tau = U^\varphi v^\varphi.
\end{equation}

Finally, from the condition that $\sigma^{ik}$ is space-like, we have
$$
\sigma^{r\tau} U^\tau= \sigma^{r\varphi} U^\varphi,
$$
and thus, in the basis ${\rm B_0}$ used in this Section, in the flat disk model, 
not only $T^{r\varphi}_\nu$, but also $T^{r\tau}_\nu$
is non-zero in the order of $\delta$ that is 
of interest for us here:
\begin{equation}
\label{T_rt}
T^{r\tau}_\nu = \frac{U^\varphi}{U^\tau} T^{r\varphi}_\nu.
\end{equation}
Note that in basis  (\ref{e_1}-\ref{e_4}) co-moving with the azimuthal motion, only the component $T^{r\varphi}_\nu$ was non-zero 
(see (\ref{sigma_NT})).

\subsubsection{Equation of free azimuthal motion}

The quantities corresponding to the background model and entering twist equations (\ref{Tik_tw}) should be obtained separately from equations (\ref{Tik_0}). 
For this, it suffices to use the results of Chapter 1 taking only the transition from basis (\ref{e_1}-\ref{e_4}) 
to the basis ${\rm B_0}$ into account.

Nevertheless, when deriving the twist equations, it is also necessary to use some of equations (\ref{Tik_0}) written exactly in the basis
${\rm B_0}$.  We mean the $r$- and $\xi$-projections of these equations in the leading order in the small disk thickness which, as we know, describe 
its azimuthal rotation in the equatorial plane of the black hole 
and its vertical hydrostatic equilibrium. We emphasize that these relations 
are valid for both stationary and non-stationary accretion flow for any viscosity parametrization, 
as well as for any specific vertical and radial structure of the flow. 
Only the condition $\delta\ll 1$ is important.

At the first stage of deriving the twist equation we will need only the $r$-projection of (\ref{Tik_0}). 
Setting  $T^{ik}=\rho U^i U^k$ we find that ${T^{rk}}_{;k}=0$ yields
\begin{equation}
\label{tw_geod}
\frac{K_1^\prime}{K_1} (U^\tau)^2 + a\frac{K_3}{K_2} (2-rK_4) U^\tau U^\varphi - \frac{(rK_2)^\prime}{rK_2} (U^\varphi)^2 = 0.
\end{equation}

Exactly this combination (\ref{tw_geod}) is used in the derivation; however, it can be checked that together with (\ref{U_norm}) in the approximation linear in $a$ 
it gives the solution 
\begin{equation}
\label{U_phi_tw}
U^\varphi = (r_S-3)^{-1/2} \left ( 1 - a r_S^{-1/2}(r_S-3)^{-1} \right ),
\end{equation}
where we have transited to the Schwarzschild radial coordinate $r_S$ which is equivalent to $r$,
that we used in Chapter 1 in the expression for $U_g^\varphi$ (see formula (\ref{geod_vel})). 
It is easy to check that $U_g^\varphi = U^\varphi/r_S$, as must be the case with the transition from the coordinate basis to ${\rm B_0}$ taken into account.

\subsubsection{'Gauge' condition of the twisted frame}

The principal kinematic constraint for the twisted reference frame requires a constant vertical position of fluid particles:
\begin{equation}
\label{xi_0}
\frac{d\xi}{d\tau} = 0,
\end{equation}
which is provided by fast establishment of hydrostatic equilibrium across the disk compared to the dynamical time of the twist change, as discussed in Section 2.1.
However, as has been already noted in \cite{Hatchett}, an important point is that this does not mean that the projection of the four-velocity of the fluid onto
${\bf e}_\xi$ is also zero, because our basis is non-coordinate and its orts are not tangent to the coordinate lines. 

By definition,

$$
v^\xi = \frac{{\bf e}^\xi}{ds}.
$$

Using (\ref{dual_tw_24}) we have: 
$$
v^\xi = r K_2 U \frac{d\tau}{ds} + r K_2 W \frac{dr}{ds};
$$
we should substitute $d\tau/ds$ and $dr/ds$ in this relation in the zeroth order in $\beta$,
in other words, as values corresponding to the flat disk dynamics. Expressions for 
${\bf e}^\tau$, ${\bf e}^\varphi$ and ${\bf e}^r$ at $\beta=0$ give

\begin{equation} \label{for_v_xi}
\frac{d\tau}{ds} = \frac{1}{K_1}\left ( U^\tau - a r^2 K_3\frac{d\varphi}{ds} \right ), \quad
\frac{dr}{ds} = \frac{U^r}{K_2}, \frac{d\varphi}{ds} = \frac{U^\varphi}{rK_2},
\end{equation}
where by definition $U^i \equiv {\bf e}^i/ds$.
As a result, we obtain
\begin{equation}
\label{v_xi}
v^\xi = r U^\tau K \frac{K_2}{K_1} U + r U^r W,
\end{equation}
where
$$
K= \left ( 1 - ar\frac{K_3}{K_2}\frac{U^\varphi}{U^\tau} \right )
$$
In (\ref{v_xi}) the velocity components $U^{\tau}$ and $U^r$ should be taken from the corresponding 
background solution for a flat disk.

\subsubsection{Explicit form of the system of equations for a twisted disk}

Now, using (\ref{U_norm}-\ref{v_xi}), we write equations (\ref{Tik_tw}) in 
explicit form by keeping only the terms in the leading order in the two small parameters $\delta$ and $u\equiv t_{d}/t_{ev}$\footnote{As we discussed above, the smallness of 
$t_d/t_{ev}$ is necessary to ensure that the accretion flow outside the equatorial plane of the black hole can be considered as a 'disk'. 
In turn, this is jointly ensured by the smallness of both $\delta$ and $t_d/t_{LT} \ll 1$ (see Section \ref{sec_freq}).}.
Here, we take into account that quantities of 'thermal' origin in the background solution are small, i.e.
$p,\,e,\,\eta \propto \delta^2 \rho$ and $q^\xi\propto \delta^3 \rho,\, q^{r,\varphi}\propto \delta^4$ (see Chapter 1).

We postpone discussing the effects of the fluid non-ideality for a while. Note that this assumption not only corresponds to vanishing terms with 
viscosity coefficient and energy flux density, or their perturbations, but also means the absence of contributions $\propto U^r$.
To select the leading-order terms in the ideal fluid approximation, we start by considering second terms in the
$\tau$-, $r$- and $\varphi$-projections of (\ref{Tik_tw}). It turns out that such terms are proportional to $\delta \beta$ here, and in the $r$-projection of (\ref{Tik_tw})
this contribution is due to the projection of the vertical pressure gradient onto the orbital plane of motion of matter in the twisted disk (see the analysis in Section
2.1 where this quantity was denoted by ``$(\nabla p)_r$''). In addition, in the $\tau$- and $\varphi$-projections
of (\ref{Tik_tw}) involve terms $\propto \delta^{-1} u \beta$, which should be also kept. On the other hand, the first terms in the
$\tau$-, $r$- and $\varphi$-projections of (\ref{Tik_tw}) give rise to terms containing Eulerian velocity perturbations, $v^{\tau,r,\varphi}$, 
as well as the Eulerian rest-mass energy density perturbation, $\rho_1$.  Hence, we conclude that 
\begin{equation}
\label{v_tw}
v^{\tau,r,\varphi}\propto \max\{ \delta, \delta^{-1} u \}\, \beta, \quad \mbox{and} \quad \rho_1\propto \max\{\delta, \delta^{-1} u\} \,\rho\beta.
\end{equation}
In addition, for reasons that become clear below, we temporarily keep partial derivatives of $v^i$ and $\rho_1$ with respect to time, despite their being 
$u^{-1}$ times smaller than the quantities themselves. Finally, the first terms of the
$\tau$- and $\varphi$-projections of
(\ref{Tik_tw}) also contain terms with the combination $\partial_\xi \rho v^\xi$, whose amplitudes are restricted by the order
$\propto\max\{\delta, \delta^{-1} u\}\,\beta$ 
by equation (\ref{v_xi}). 

Now, using the result (\ref{v_tw}), it is easy to select the leading terms entering in the $\tau$-, $r$- and $\varphi$-projections of (\ref{Tik_tw}) 
due to the fluid non-ideality. The most troublesome here is the contribution due to the shear tensor perturbation, $2\eta\sigma^{ik}_1$, which appears in $T^{ik}_1$ 
(see (\ref{T_1_ik}) and (\ref{sigma_1_ik})). However, most of the terms from this contribution contain jointly
$\eta\propto \delta^2$ and $v^i\propto \delta\beta$; therefore, it is clear that it is necessary to include only the terms 
in which the derivative with respect to $\xi$ (lowering the order in $\delta$) occurs twice. This fact strongly reduces the number of 'viscous' terms to be kept. 
Besides, by similar considerations, the final expressions will not contain terms with ${\bf q}$, ${\bf q}_1$ and $\eta_1$.
Finally, we stress once again that in addition to the purely 'viscous' terms mentioned above, the contribution due to the radial advection that 
appears in the background solution with non-zero viscosity should not be forgotten. 
We are concerned with the terms that can appear in the 'non-viscous' part of the stress-energy tensor (see the first term in (\ref{Tik})) due to the non-zero value of  $U^r\propto \delta^2$.

Taking all the above into account and using relations derived in three preceding Sections, we obtain the $\tau$-, $r$- and $\varphi$-projections of (\ref{Tik_tw}) in the form
\begin{multline}
K\frac{K_2}{K_1} (U^\tau)^2 \dot \rho_1 +
\left ( 2U^\varphi - ar\frac{K_3}{K_2}\frac{(U^\varphi)^2 +
(U^\tau)^2}{U^\tau} \right ) \frac{K_2}{K_1} \rho \dot v^\varphi +
\frac{1}{r} U^\tau U^\varphi \partial_\varphi \rho_1 + \\
\frac{1}{r}
\frac{(U^\varphi)^2 +  
(U^\tau)^2}{U^\tau} \rho \partial_\varphi v^\varphi + 
\partial_r (\rho U^\tau v^r) + \partial_\xi \rho U^\tau v^\xi +
\frac{(rK_1^2 K_2^2)^\prime}{rK_1^2K_2^2} \rho U^\tau v^r +
F_{\nu}^{\tau} = \\ 
r\partial_\xi \rho (U^\tau)^2 K \frac{K_2}{K_1} U
+ \frac{\xi}{r} \rho U^\tau U^\varphi \partial_\varphi W,
\label{tau_tw}
\end{multline}

\begin{multline}
 K\frac{K_2}{K_1} U^\tau \dot v^r +
\frac{U^\varphi}{r} \partial_\varphi v^r - \left
[2\frac{K_1^\prime}{K_1 U^\varphi}  + a\frac{K_1}{rK_2 U^\tau}
\left ( \frac{r^2 K_3}{K_1} \right )^\prime \right ] v^\varphi
+{1\over \rho}F^{r}_{\nu} =\\
 W r \frac{\partial_\xi p}{\rho} -
a\xi\frac{K_3^2}{K_1 K_2} \left ( \frac{K_1}{K_3} \right )^\prime
Z U^\tau U^\varphi, \label{r_tw}
\end{multline}

\begin{multline}
 K\frac{K_2}{K_1} U^\tau U^\varphi \dot \rho_1 +
\left ( \frac{(U^\varphi)^2+(U^\tau)^2}{U^\tau} -
2ar\frac{K_3}{K_2}U^\varphi \right ) \frac{K_2}{K_1} \rho \dot
v^\varphi + \frac{(U^\varphi)^2}{r} \partial_\varphi \rho_1 + 2
\frac{U^\varphi}{r} \rho \partial_\varphi v^\varphi +\\
\partial_r (\rho U^\varphi v^r) + \partial_\xi\rho U^\varphi v^\xi +
\frac{(r^2 K_1 K_2^3)^\prime}{r^2 K_1 K_2^3} U^\varphi \rho v^r -
a\frac{K_1}{r K_2} \left ( \frac{r^2 K_3}{K_1} \right )^\prime
U^\tau \rho v^r + F^{\varphi}_{\nu} = \\
K\frac{K_2}{K_1} r \partial_\xi\rho U^\tau U^\varphi U + \frac{\xi}{r} \rho
(U^\varphi)^2 \partial_\varphi W, \label{phi_tw}
\end{multline}
where
$$
K = \left ( 1 - ar\frac{K_3}{K_2} \frac{U^\varphi}{U^\tau} \right
),
$$
and $F_{\nu}^{\tau,r,\varphi}$ is the total contribution due to non-zero viscous forces and the radial advection of matter in the background solution $\propto U^r$. 

Explicitly,
\begin{multline}
F^{\tau}_{\nu} = {U^{\varphi}\over U^{\tau}}(\partial_\xi
T_{\nu}^{\varphi \xi}-rW \partial_\xi T_{\nu}^{r\varphi }) - r
\partial_\xi \rho U^\tau U^r W, \quad
F_{\nu }^{r}=\partial_\xi T_{\nu}^{r\xi}, \\
F^{\varphi}_{\nu}=(\partial_\xi T_{\nu}^{\varphi \xi}-rW
\partial_\xi T_{\nu}^{r\varphi }) - r \partial_\xi \rho U^\varphi U^r W,
\label{F_nu_hor}
\end{multline}
where
\begin{multline}
T_{\nu}^{r\xi}=-{\eta \over K_{2}}(\partial_\xi v^{r} +
U^{\varphi}\partial_\varphi W), \quad
T_{\nu}^{\varphi \xi}=-{\eta
\over K_{2}} \left (\partial_\xi v^{\varphi} - 2a{K_{3}\over
K_{2}}U^{\tau} (U^\varphi)^2 Z \right ), \\ 
T_{\nu}^{r\varphi}=-\eta r
\left ({U^{\varphi}\over rK_{2}}\right )^\prime. \label{T_nu}
\end{multline}

We note that $T_{\nu}^{r\xi}$ and $T_{\nu}^{\varphi \xi}$  have the meaning of perturbations of the viscous stress tensor. 
In these expressions, the terms $\propto\beta$ contributing to the shear tensor perturbations appear due to the twisted basis. Conversely,
$T_{\nu}^{r\varphi}$ relates to the background. Nevertheless, for the sake of brevity, we use the same notation with the index ``$\nu$'' for these 
three quantities.

Finally, we assume in (\ref{tau_tw}-\ref{T_nu}) that in the relativistic coefficients $K_1$, $K_2$ and $K_3$, the argument $R_I$ is replaced by $r$, 
because $R_I^2=r^2+\xi^2$ and accounting for the dependence on $\xi$ here always gives rise to a small correction $\propto\delta^2$ only.

It remains to write the explicit form of the $\xi$-projection of (\ref{Tik_tw}).
Similarly, we start with the contribution of terms in the ideal fluid approximation, and first rearrange the first term in (\ref{Tik_tw}).
The leading-order terms in $\delta$ here are, in particular, $\rho v^\varphi$ and $\rho_1$, but additionally multiplied by $\xi$. 
This means that their amplitudes are restricted by the order $\max\{\delta^2, u\}\, \beta$. 
Moreover, $v^\xi$ now enters the term ``$U^\varphi \rho \partial_\varphi v^\xi$'' which also implies the raising of the order of smallness by $\delta$ 
compared to (\ref{tau_tw}-\ref{phi_tw}) (it can be seen that in formulas (\ref{tau_tw}) and (\ref{phi_tw}) $v^\xi$ entered in combination with $\partial_\xi\rho$).
Besides, of all terms of a ‘thermal’ origin we must now keep the term with $\partial_\xi p_1$, since it also is of the order of $\delta^2$
due to $p_1 \sim \delta^2\rho_1 \propto \rho\delta^3\beta$.

Turning now to the second term in the $\xi$-projection of (\ref{Tik_tw}), we write all terms through the order $\propto\max\{\delta^2, u\}\,\beta$.
From similar considerations, the terms due to fluid non-ideality (including ‘advective’ terms proportional to $\propto U^r$) 
are also kept here, with their smallness increased by the coefficient $\delta$ compared to what we did in (\ref{tau_tw}-\ref{phi_tw}).

We thus obtain the following equation:
\begin{multline}
U^\varphi
\partial_\varphi v^\xi + r\frac{\partial_\xi p_1}{\rho} + \xi
\frac{(U^\varphi)^2}{r} \left ( 1 - 2
ar\frac{K_3}{K_2}\frac{U^\tau}{U^\varphi} \right )
\frac{\rho_1}{\rho} +\\
 2\xi U^\varphi v^\varphi \left [
\frac{K_1^\prime}{K_1} - \frac{K_2^\prime}{K_2} - \frac{ar}{2}
\frac{K_3^2}{K_1 K_2} \left ( \frac{K_1}{K_3} \right )^\prime
\left ( \frac{U^\tau}{U^\varphi} + \frac{U^\varphi}{U^\tau} \right
) \right ] +{r\over \rho}F^{\xi}_{\nu} = \\
-\left [ \frac{K_2}{K_1} \partial_\varphi U - 2a\frac{K_3 Z}{K_2}
+ a\frac{\xi^2}{r} \frac{K_3^2 Z}{K_1 K_2} \left ( \frac{K_1}{K_3}
\right )^\prime \right ] r U^\tau U^\varphi + ar^2\frac{K_3}{K_1}
(U^\varphi)^2
\partial_\varphi U, \label{xi_tw}
\end{multline}
where
\begin{multline}
F^{\xi}_{\nu}={1\over rK_{1}K_{2}^{3}} \partial_r
(rK_{1}K_{2}^{3}T_{\nu}^{r\xi}) + \partial_\xi T_{\nu}^{\xi \xi} +
{1\over r}\partial_\varphi T_{\nu}^{\varphi \xi} + \\
\partial_\varphi W ( T_{\nu}^{r\varphi} + T^{r\varphi}_{adv}  )
 + a \frac{K_1}{r K_2} \left (\frac {r^2 K_3}
{K_{1}}\right )^{'} \partial_\varphi Z \left (\frac{U^\varphi}{U^\tau}
T_{\nu}^{r\varphi}+ \frac{U^\tau}{U^\varphi} T^{r\varphi}_{adv} \right ),
\label{F_xi}
\end{multline}
and $T^{r\varphi}_{adv}=\rho U^{\varphi}U^{r}$. We do not provide the explicit form of $T^{\xi\xi}_\nu$ here, because it is not required 
in the final form of the twist equations. 

Everywhere in (\ref{xi_tw}-\ref{F_xi}), except in the second term in square brackets in the right-hand side of (\ref{xi_tw}), the argument $R_I$  
in the relativistic coefficients $K_1$, $K_2$ and $K_3$ is replaced by $r$. 
The mentioned term is an exception because this term alone has the zeroth order in small parameters $\delta$ and $u$ in equation (\ref{xi_tw}). 
But because we have kept the terms $\propto \max\{\delta^2, u\}$ in (\ref{xi_tw}),
in the term under discussion it would be necessary to take corrections $\propto\delta^2$ into account due to the dependence of the relativistic coefficients $K_2$ and $K_3$
on $\xi$. We did not do that for the reason discussed in the next paragraph.

\subsection{Completing the derivation of twist equations}
\label{sec_completion}

Thus, we have written the twist equations in the leading orders in small parameters $\delta$ and $u$. 
All corrections linear in the Kerr parameter $a$ have been taken into account. If we temporarily set $a=0$ and consider equation (\ref{xi_tw}),
we see that, on the one hand, it contains the terms proportional to the rate of change of the disk twist,
$U$, 
and on the other hand, it has terms containing perturbations of the physical quantities of the order $\propto\delta^2$. 
Thus, we can say that due to the internal forces only, does a thin twisted disk evolve on a long timescale such that $u\sim \delta^2$. 
Then it becomes totally clear that equations (\ref{tau_tw}-\ref{phi_tw}) are restricted by the order $\propto\delta$, and equation (\ref{xi_tw}) is restricted
by the order $\propto\delta^2$. 

At the same time, when the parameter $a$ is non-zero, a ‘large’ term of the zeroth order in $\delta$
and $\propto a Z$ arises in the right-hand side of equation (\ref{xi_tw}). This term describes the gravitomagnetic interaction of the rotating black hole with the tilted/twisted disk. In order that all terms in 
(\ref{xi_tw}) be balanced with each other, we must assume that $a\sim \delta^2$. 
But it then becomes clear that all additional corrections $\sim a$ in equations (\ref{tau_tw}-\ref{phi_tw}) have the next order in $\delta$ 
and can be omitted. The same relates to all terms $\propto a\delta^2$ in equation (\ref{xi_tw}), including the correction $\propto\delta^2$ 
due to the dependence of the relativistic coefficients on $R_I$ in the gravitomagnetic term itself.

In fact, this means that when considering the dynamics of a twisted thin accretion disk near a rotating black hole, 
it suffices to use the background model, i.e. the corresponding flat disk, in the Schwarzschild metric with $a=0$.
The assumption of the slow black hole rotation itself was needed because otherwise the accretion flow 
(including non-stationary one) could not be regarded as a disk, since 
the vertical hydrostatic equilibrium there would be violated (see Section 2.1). 
Of course, these conclusions relate to only slightly tilted/twisted and geometrically thin disks with $\beta\ll 1$, $\delta\ll 1$.  

In what follows, we therefore set $a=0$ in all terms except the gravitomagnetic one. This significantly simplifies further calculations that 
are required for obtaining the twist equations in the final form. Let us first analyze equations (\ref{tau_tw}) and (\ref{phi_tw}). 
It is convenient to consider their combinations which contain either $\dot\rho$ or $\dot v^\varphi$.

Eliminating $\dot v^\varphi$ for $a=0$ we obtain the equation
\begin{multline}
U^\varphi
\partial_\varphi \rho_1 + \frac{1}{(U^\tau)^2} \rho
\partial_\varphi v^\varphi + \frac{U^\tau}{K_2^2}
\frac{\partial}{\partial r} \left ( r K_2^2 \frac{\rho
v^r}{U^\tau} \right ) =  \\ 
\xi  U^\varphi \rho
\partial_\varphi W + \frac{U^\varphi}{(U^\tau)^2} (\partial_\xi
T_{\nu}^{\varphi \xi}-rW
\partial_\xi T_{\nu}^{r\varphi }),
\label{tw_cont}
\end{multline}
where we have omitted the term $\dot\rho_1$, which is of the next order in $\delta$ compared to the other terms. In the Newtonian limit as
$r\rightarrow \infty$, equation (\ref{tw_cont}) reduces to the continuity equation for perturbations. 

Next, eliminating $\dot\rho_1$ for $a=0$\footnote{$a=0$ also in the expression for $T^{\varphi\xi}_\nu$.}, we obtain the equation
\begin{equation}
\frac{K_2}{K_1} \dot v^\varphi +
\frac{1}{r}\frac{U^\varphi}{U^\tau} \partial_\varphi v^\varphi +
\left ( \frac{\partial_r U^\varphi}{U^\tau} +
\frac{K_1^\prime}{K_1} \frac{U^\tau}{U^\varphi} \right ) v^r +
\frac{1}{\rho U^\tau}(\partial_\xi T_{\nu}^{\varphi \xi}-rW
\partial_\xi T_{\nu}^{r\varphi }) = 0. \label{tw_phi_new}
\end{equation} 
In the Newtonian limit, (\ref{tw_phi_new}) reduces to the azimuthal component of the Navier-Stokes equation for perturbations.

Finally, (\ref{r_tw}) with $a=0$ takes the form
\begin{equation}
 \frac{K_2}{K_1} U^\tau \dot v^r +
\frac{U^\varphi}{r} \partial_\varphi v^r - 2\frac{K_1^\prime}{K_1
U^\varphi} v^\varphi +{1\over \rho}\partial_\xi T_{\nu}^{r\xi} = W r
\frac{\partial_\xi p}{\rho}. \label{tw_r_new}
\end{equation}
In the Newtonian limit, (\ref{tw_r_new}) reduces to the radial component of the Navier-Stokes equation for perturbations.

It is important to explain why we have kept terms with $\dot v^r$ and $\dot v^\varphi$ in equations (\ref{tw_phi_new}) and (\ref{tw_r_new})
although they are of the next order in $\delta$. 
As mentioned in Section 2.1, in the Newtonian limit the epicyclic frequency becomes equal to the Keplerian circular frequency, 
which results in a resonance growth of the amplitude of velocity perturbations 
of gas elements in the disk under the action of the radial projection of the vertical pressure gradient,
$(\nabla p)_r$, which is limited only by the viscosity. Mathematically expressed, in the limit of an inviscid Keplerian disk, equation (\ref{tw_phi_new}) 
yields in the leading order in the parameter $u$ (with the term with $\propto \dot v^\varphi$ omitted) 
such a relation between $v^r$ and $v^\varphi$ that the sum of the second and the third terms in (\ref{tw_r_new}) vanishes. 
But, because there is a term $\propto \delta\beta$ in the right-hand side of (\ref{tw_r_new}), it follows that $\dot v^r$ (and hence $\dot v^\varphi$ as well) 
acquires the first order in $\delta$ in the considered case. Either viscosity or relativistic corrections eliminate the Keplerian resonance, and the amplitudes of
$\dot v^r$ and $\dot v^\varphi$ decrease again to the third order in $\delta$.

Now, from equation (\ref{xi_tw}) we need to derive the so-called twist equation that plays the principal role in the twisted disk theory. 
For this, we need to explicitly determine the value $\partial_\xi p/\rho$, 
which is done in the next Section. Although the Schwarzschild approximation is sufficient, we also take linear corrections in $a$ into account. 
This is required below in obtaining an additional expression for the Lense-Thirring frequency in terms of the relativistic coefficients used in the twisted basis.

\subsubsection{Equation of the vertical hydrostatic equilibrium}

Let us write the $\xi$-projection of (\ref{Tik_0}) in the basis $B_0$  to the leading order in $\delta$, 
as we did in Chapter 1 employing basis (\ref{e_1}-\ref{e_4}) (see equation (\ref{hydrostatics_rel})). 
Taking into account that the four-velocity of the flow is $\{U^\tau,\,0,\,U^\varphi,\,0\}$ in the leading order in $\delta$, we obtain the following equation

\begin{equation}
\label{hydrostat_tw}
\frac{\partial_\xi p}{\rho} = \frac{\xi}{r} (U^\varphi)^2 
\left [ \frac{K_2^\prime}{K_2} - \left ( \frac{U^\tau}{U^\varphi}\right )^2 + 
ar\frac{K_3 K_4}{K_2} \frac{U^\tau}{U^\varphi}\right ],
\end{equation}
where $U^\tau$ and $U^\varphi$ satisfy the normalization condition (\ref{U_norm})  and the geodesic equation (\ref{tw_geod}).
With this in mind, we arrive at the final form of the hydrostatic equilibrium equation
\begin{equation}
\label{hydrostat_tw_fin}
\frac{\partial_\xi p}{\rho} = - \frac{\xi}{r} \frac{(U^\varphi)^2}{r} \left ( 1 - 2 a r\frac{K_3}{K_2}
\frac{U^\tau}{U^\varphi} \right ),
\end{equation}
where the Schwarzschild profiles of $U^\tau$ and $U^\varphi$ are used in the term with the parameter $a$.

It can be checked that with the substitution $\xi \to z/K_2$, equation (\ref{hydrostat_tw_fin}) is equivalent to (\ref{hydrostatics_rel})  in the linear approximation in $a$.
Here, we should only take into account that $r_S=K_2 r$, where $r_S$ is the Schwarzschild coordinate equivalent to the coordinate $r$ in (\ref{hydrostatics_rel}).

\subsubsection{Twist equation}

Our goal is to rewrite (\ref{xi_tw}) in divergent form. Without accounting for the gravitomagnetic term, equation (\ref{xi_tw}), in which we also set $a=0$, 
must respect the conservation law of the angular momentum projection of the twisted disk onto the equatorial plane of the black hole 
(the conservation of the disk angular momentum projection onto the black hole spin in our problem, linear in $\beta$, 
follows from equations for the background, since the corrections due to the small tilt are proportional to $\propto 1-\cos\beta \sim \beta^2$), 
which reflects spherical symmetry of the Schwarzschild metric. 

It turns out that to do this it is necessary to eliminate from (\ref{xi_tw}) $v^\varphi$ and $\rho_1$, 
in the left-hand side of this equation. Therefore, we will use equations (\ref{tw_cont}-\ref{tw_r_new}) with $\dot v^r = \dot v^\varphi=0$
for our purposes, because nowhere we will deal with resonance combinations of $v^r$ and $v^\varphi$ that vanishes in the main order in $u$ 
in the Keplerian inviscid limit (see the comment to equations (\ref{tw_phi_new}) and (\ref{tw_r_new}) above).

First, in the right-hand side of (\ref{tw_cont}) we rewrite the term with $\partial_\varphi W$ through $v^r$ and $v^\varphi$ using (\ref{tw_r_new}) 
and (\ref{hydrostat_tw_fin}) with $a=0$. In the resulting expression for $\rho_1$ we replace $v^\varphi$ using (\ref{tw_phi_new}).
Here, the derivative with respect to $\varphi$ can be eliminated using the harmonic dependence on $\varphi$ (see (\ref{ZUW}) ).
In other words, $\partial_{\varphi\varphi}=-1$.
Substituting the obtained expressions for $\rho_1$ and $v^\varphi$ in (\ref{xi_tw}),  integrating over $\xi$ and 
performing integration by parts wherever necessary, using the fact that the corresponding surface terms vanish as $\rho\to 0$,
we arrive at the compact equation

\begin{multline}
\Sigma U^\tau U^\varphi \left \{\partial_\varphi U - a\frac{K_1
K_3}{K_2^2} Z \right \} + \partial_\varphi W \frac{K_1}{K_2} \left
\{\Sigma U^\varphi U^r + \bar T_{\nu}^{r\varphi} \right \}=\\
-\frac{1}{ 2r^2  K_2^4} \int d\xi\, \{\, \partial_r (\xi r K_1
K_2^3 U^\varphi \rho \partial_\varphi v^r + r^2 K_1 K_2^3 T^{r\xi}
), \label{tw_eq}
\end{multline}
where, as usual, $\Sigma= \int \rho\, d\xi $ is the surface density of the disk, and the bar over $T^{r\varphi}$ 
means that it is integrated over $\xi$.
In the Appendix of \cite{ZhI}, it is shown that (\ref{tw_eq}) can be used to obtain the angular momentum conservation for the twisted disk. 

Equations (\ref{tw_phi_new}), (\ref{tw_r_new}) and (\ref{tw_eq}) represent a closed system of equations describing the dynamics 
of twisted configurations provided that the corresponding model background is specified. 
Unknown variables in this system include the velocity perturbations $v^r$ and $v^\varphi$ and the quantity $Z$ characterizing the disk geometry. 
We emphasize that in deriving these equations we essentially used only three main assumptions: 
$a\ll 1$, $\delta\ll 1$ and $\beta\ll 1$. 
This means that the equations describe the dynamics of any geometrically thin accretion flow (disk) with any parametrization of viscosity, 
any radial and vertical structure in both stationary and non-stationary case. In the latter case, we mean the non-stationary background: 
the equations determine not only the dynamics of twist perturbations propagating in a stationary flat disk, 
but also the dynamics of twisted rings/tori, when the evolution of the geometrical form occurs in parallel with its expansion 
in the radial direction due to turbulent viscosity, which also results in the evolution of the background itself.

\subsubsection{Once again about the characteristic frequencies of the problem}

In Section \ref{sec_freq},  we already obtained relativistic expressions for the characteristic frequencies of the problem. These include the circular and 
epicyclic frequencies of free equatorial motion, as well as the frequency of vertical oscillations and the precession frequency of tilted orbits.
Here, we wish to obtain expressions for these frequencies, but now in terms of the values used above to construct the theory of twisted disks, i.e. in the basis
$B_0$. These expressions are required to write the twist equations in a more compact form. 

The circular frequency of the free equatorial motion as measured by the clock
of an infinitely remote observer, which we already presented in equation (\ref{kepl_freq}), 
can be obtained simply by dividing $d\varphi/ds$ by $d\tau/ds$ given in (\ref{for_v_xi}). We obtain 

\begin{equation}
\label{Om_circ}
\Omega = \frac{K_1}{K K_2} \frac{U^\varphi}{rU^\tau}.
\end{equation}
Using (\ref{U_phi_tw}) and (\ref{U_norm}), and also remembering that $r_S = r K_2$, 
we can check that (\ref{Om_circ}) coincides with (\ref{kepl_freq}) in the linear approximation in $a$.

We now consider small vertical deviations from the circular equatorial motion. In Section \ref{sec_freq}, 
we discussed that the frequency of vertical oscillations as measured by an infinitely remote observer, $\Omega_v$, is the locally measured frequency, 
$\Omega_l$, divided by the $t$-component of the four-velocity of circular motion, $U^t_g$. 
The frequency $\Omega_l$ explicitly enters the equation of hydrostatic equilibrium (see equation (\ref{hydrostatics_rel})  or equivalent equation (\ref{hydrostat_tw_fin})
with the substitution $\xi\to z/K_2$). 
Using relations (\ref{for_v_xi}),  we express 
$U^t_g \equiv d\tau/ds$ in terms of $U^\tau$:
$$
U^t_g = K K_1^{-1} U^\tau,
$$
whence 
\begin{equation}
\label{Om_vert}
\Omega_l = \Omega_v \frac{KU^\tau}{K_1} = \frac{U^\varphi}{rK_2} \frac{\Omega_v}{\Omega},
\end{equation}
where the final expression was obtained using (\ref{Om_circ}).

But then, from a comparison of (\ref{Om_vert}) with (\ref{hydrostat_tw_fin}),  we see that 
\begin{equation}
\label{Om_vert_fin}
\Omega_v = \Omega \left (1 - ar\frac{K_3}{K_2}\frac{U^\tau}{U^\varphi} \right ),
\end{equation}
where the Schwarzschild profiles for $U^\tau$ and $U^\varphi$ are used in the term with the parameter $a$.

Then, using (\ref{lense_tirring}), we obtain the Lense-Thirring frequency
\begin{equation}
\label{lense_tirring_fin}
\Omega_{LT} = a\frac{K_1 K_3}{K_2^2}.
\end{equation}

It suffices for our purposes to know the epicyclic frequency in the Schwarzschild case with $a=0$.
This expression can be the most easily derived directly from the twist equations, 
more precisely, from their part that describes the dynamics in the plane of disk rings, i.e. from
(\ref{tw_phi_new}) and (\ref{tw_r_new}). 
Setting the ‘viscous’ terms and radial projection of the pressure gradient in the right-hand side of
(\ref{tw_r_new}) equal to zero, as well as omitting the dependence of $v^r$ and $v^\varphi$ on 
$\varphi$, we obtain equations for the Eulerian perturbations that describe a free motion of gas elements slightly deviating from the circular motion. 
Clearly, these equations are equivalent to (\ref{epc_eqs_1}-\ref{epic_eqs_2}) which were written in the basis (\ref{e_1}-\ref{e_4}).
From these equations, we obtain the following equation for $v^r$:
\begin{equation}
\label{epic_v_r_tw}
\ddot v^r + 2\frac{K_1 K_1^\prime}{K_2^2 (U^\tau)^2} 
\left ( \frac{\partial_r U^\varphi}{U^\varphi} + \frac{K_1^\prime}{K_1}\frac{(U^\tau)^2}{(U^\varphi)^2} \right ) v^r  = 0,
\end{equation}
where the expression before $v^r$ is equal to $\kappa^2$. 
It can be rewritten in a more compact form
\begin{equation}
\label{kappa_tw}
\kappa^2 = 2 \frac{K_1^\prime (K_1 U^\tau)^\prime}{K_2^2 U^\tau (U^\varphi)^2}
\end{equation}
to ensure that it coincides with (\ref{epic_freq_Schw}),
considering that the radial Schwarzschild coordinate $r_S=r K_2$ enters the last equation.

Finally, for convenience, we introduce the following quantity with the dimension of frequency that appears in our problem. In the Schwarzschild case, $a=0$,
\begin{equation}
\label{tilde_freq}
\tilde\Omega = \frac{K_1^\prime}{K_2} \frac{1}{U^\tau U^\varphi} = \frac{r_S-3}{r_S^2 (r_S-2)^{1/2}},
\end{equation}
which tends to the Keplerian value in the Newtonian limit. 

Using (\ref{Om_circ}), (\ref{kappa_tw}) and (\ref{tilde_freq}) allows us to write equations (\ref{tw_phi_new}) and (\ref{tw_r_new}) 
in a more compact form. Lense-Thirring frequency (\ref{lense_tirring_fin}), evidently, enters the gravitomagnetic term in (\ref{tw_eq}).
However, we deal with this rewriting in the next Section when considering a specific background model.

\subsection{Twisted equations in the particular case of a stationary vertically isothermal $\alpha$-disk}

We now consider the form the twist equations take in the specific background of a stationary
$\alpha$-disk which we have discussed in Chapter 1. This does not mean, however, that only stationary twisted solutions are to be considered. 
In other words, the equations we obtain are also applicable to arbitrary non-stationary dynamics of the corresponding twisted disk. 
For example, they enable us to calculate the evolution of the shape of an (infinite) initially flat disk instantly tilted 
to the equatorial plane of a rotating black hole.  The initial stage of the evolution of such a disk was qualitatively described in Section \ref{sec_intro}.
In addition, these equations describe the wave-like (in the case of a disk with  sufficiently small
$\alpha < \delta$; see also \cite{Pap_Lin})
or diffusion-like (in the case of a disk with sufficiently large $\alpha > \delta$; see also \cite{PP_1983}) 
dynamics of some twisted perturbation imposed on the disk lying initially in the equatorial plane of the black hole.

\subsubsection{Explicit form of the necessary background profiles}

The twist equations contain the quantity $\bar T^{r\varphi}_\nu$ (as well as $\bar \eta$), 
related to the corresponding flat disk model. We could obtain the explicit form of these quantities by integrating the $\tau-$ and $\varphi-$projections
of equations (\ref{Tik_0}). However, it is simpler to use the results of Chapter 1, where we have already obtained this quantity, denoted by
$T_\nu$ there (see equation (\ref{T_sol_fin}) ). 
We should only take into account that now we are working in another basis than that used for the flat disk, and therefore the transition from
$T_\nu$ to $\bar T^{r\varphi}_\nu$ should be specified. First, using the orthogonality condition for the shear tensor and, hence, for the viscous stress tensor, (\ref{sigma_orth}), 
we see that
only one component of the viscous stress tensor, ${T^{r\varphi}_\nu}^\prime$, is non-zero in basis (\ref{e_1}-\ref{e_4}), because the four-velocity there has only non-zero time component 
up to the terms $\propto\delta^2$. The prime here marks basis (\ref{e_1}-\ref{e_4}).
Further, the (orthonormal) bases are different only in that the observers associated with basis (\ref{e_1}-\ref{e_4}) 
move in the azimuthal direction with the velocity of the free equatorial circular motion, whereas the basis ${\rm B_0}$ corresponds to the observers at rest. 
Therefore, the transformations of vectors and tensors must be equivalent to the usual Lorentz transformations. 
Using \cite{LL2} (see exercise 1, paragraph 6 therein) we see that ${T^{r\varphi}_\nu} = U^\tau {T^{r\varphi}_\nu}^\prime$ where $U^\tau$
is the Lorentz factor of azimuthal motion.
Finally, it should be additionally taken into account that integration over $\xi$ differs from that over $z$ by the coefficient $K_2$
As a result, we obtain
\begin{equation}
\label{T_relation}
\bar T^{r\varphi}_\nu = \frac{U^\tau}{K_2} T_\nu.
\end{equation}
We note that it is possible to change from ${T^{r\varphi}_\nu}^\prime$ to $T^{r\varphi}_\nu$ using the relation (\ref{tetrade_ten}) 
by writing it for two bases, equating the right-hand sides and then multiplying 
one of the sides of the obtained equalities by matrices inverse to the basis matrices there. 
Here, we should only take into account that in the basis ${\rm B_0}$
the radial coordinate was changed, (\ref{R_I}), i.e. that $r_S=r K_2$ in the notation of this part of the paper. 

Next, for the case $a=0$, which is sufficient here, it is easy to express $T_\nu$ in terms of elementary functions. 
Indeed, the integral in (\ref{T_sol_fin}) can be taken by the substitution $y\equiv\sqrt{r_S}$:
$$
\int \frac{E}{r_S^{1/2} C} dr_S = \int \frac{y^2-6}{y^2-3} dy = y + \frac{\sqrt{3}}{2} \ln { \frac{y+\sqrt{3}}{y-\sqrt{3}} }.
$$ 

For $\bar T^{r\varphi}$ with account for (\ref{T_relation}) we finally obtain 
\begin{equation}
\bar  T^{r\varphi}={\dot M \over 2\pi}U^{\tau}r^{-3/2}{L(r)\over
K_2^{5/2}K_{1}^2},\label{T_tw1}
\end{equation}
where
\begin{equation}
L=1-{\sqrt 6\over y}-{\sqrt 3\over 2y}\ln{{(y-\sqrt 3)(3+2\sqrt
2)\over (y+\sqrt 3)}}. \label{L_expr}
\end{equation}
As it must be, $L=0$ at $r_S=6$. We note that $L = Y(a=0)$ where $Y$ was defined in (\ref{Y}).

On the other hand, the expression for $\sigma^{r\varphi}$ in (\ref{sigma_NT}), in our case $a=0$ in the basis $B_0$, can be rewritten in the form
$$
\sigma^{r\varphi} = \frac{3}{4} \frac{D}{r_S^{3/2}C} U^\tau = \frac{3}{4} K_1^2 U^\varphi_g U^\tau_g U^\tau = \frac{3}{4} \frac{K_1}{r K_2} U^\varphi (U^\tau)^2, 
$$
where, as usual, we use relations (\ref{for_v_xi}).

Then
\begin{equation}
\label{T_tw2}
\bar  T^{r\varphi} = \frac{3}{2} \bar \eta \frac{K_1}{r K_2} (U^\tau)^2 U^\varphi.
\end{equation}

As in Section \ref{sec_expl_disk}, equating expressions (\ref{T_tw1}) and (\ref{T_tw2}) we obtain
\begin{equation}
\label{eta_tw}
\bar \eta ={\dot M \over 3\pi} \left ( {r^{-1/2}\over
U^{\tau}U^{\varphi}}{L\over K_1^3K_2^{3/2}} \right ).
\end{equation}
In the Newtonian limit, far away from the inner edge of the disk, equation (\ref{eta_tw}) gives a well-known result $\bar \eta = \dot M/ (3\pi)$.

We assume that the kinematic viscosity is proportional to the characteristic disk half-thickness times the sound velocity in the disk:
\begin{equation}
\nu \sim \alpha c_s h_{p},  \label{nu_prescr}
\end{equation}
where $h_{p}$ is the proper characteristic disk half-thickness, which in our coordinate system is 
$h_{proper}=K_2h$ and $\alpha$ is the Shakura parameter, which is assumed to be constant. Because (\ref{hydrostat_tw_fin}) implies that
$c_{s} \sim \sqrt{P/\rho} \sim U^{\varphi} h/r$, 
we finally define $\alpha$ by the equality
\begin{equation}
\nu = \alpha K_2 U^{\varphi} h^{2}/r.  \label{alpha_prescr}
\end{equation}

Using (\ref{eta_tw}) and (\ref{alpha_prescr}), we obtain the relation
\begin{equation}
\Sigma h^{2} ={\dot M \over 3\pi \alpha}\left( {r^{1/2}\over
U^{\tau}(U^{\varphi})^{2}}{L\over K_1^3K_2^{5/2}} \right ). \label{sigma_h2}
\end{equation}

To find $U^r$ in the advective term in (\ref{tw_eq}), 
we use the rest-energy conservation law in the basis $B_0$ for the stationary disk.  Again, we use result
(\ref{cont_4}). Recalling the transition to the isotropic radial coordinate, the relation between the coordinate and physical velocities
(\ref{e_dual_3}) and (\ref{for_v_xi}), and the difference in the definitions of $\Sigma$, we obtain
\begin{equation}
-{\dot M\over 2\pi}=\Sigma K_{1}K_{2}^{2}rU^{r}. \label{tw_cont_bg}
\end{equation}

Then $U^{r}$ can be derived from (\ref{tw_cont_bg}) and (\ref{sigma_h2}) as
\begin{equation}
U^{r}= - {3\alpha \over 2}{\delta^{2}\over L} K_1^2
U^{\tau}(U^{\varphi})^{2}\sqrt{K_2 r}. \label{tw_Ur}
\end{equation}

Finally, we need to know the profile $\delta(r)$. 
Note that this value is invariant under the transition between the bases
(\ref{e_1}-\ref{e_4}) and $B_0$, since the change from $h_p$ to $h$ and from $r_S$ to $r$ is scaled with the same coefficient $K_2$.

In the gas-pressure-dominated disk with the Thomson scattering opacity, it follows from (\ref{delta_expl}) with $a=0$ that
\begin{equation}
\delta (r)= \delta_{*}
K_1^{1/2}K_2^{1/20}(U^{\tau})^{-9/10}L^{1/5}r^{1/20}. \label{delta_1}
\end{equation}

In order to derive a simpler form of twist equations, we need to specify the vertical profile of the rest-energy density. 
Here we use its simplest form in an isothermal disk:
\begin{equation}
\rho = \rho_c \,{\rm exp} \left (-\frac{\xi^2}{2h^2} \right ), \label{rho_xi}
\end{equation}
where $\rho_{c}(r)$ is the equatorial density.

\subsubsection{Transition to complex amplitudes}

In the case of an isothermal disk, the velocity perturbations $v^r$ and $v^\varphi$ taken in the form 
\begin{equation}
v^\varphi = \xi (A_1 \sin \varphi + A_2 \cos \varphi)\quad v^r =
\xi (B_1 \sin \varphi + B_2 \cos \varphi) \label{vAB}
\end{equation}
satisfy equations (\ref{tw_phi_new}) and (\ref{tw_r_new}) provided that $\nu$ 
does not change with the height and the amplitudes $A_1$, $A_2$, $B_1$ and $B_2$ are functions of $r$ and $\tau$.
Indeed, in this case, all ‘thermal’ terms are $\propto\xi$, and the dependence on $\xi$ with the ansatz (\ref{vAB}) is identically satisfied in the considered equations. 

Let us introduce the complex amplitudes
\begin{equation}
{\bf A} = A_2 + i A_1, \quad {\bf B} = B_2 + i B_1 \quad
\mbox{and} \quad {\bf W} = \Psi_1 + i \Psi_2 = \beta e^{i\gamma}
\label{compl_ampl}
\end{equation}

By composing two combinations, $(\ref{tw_phi_new}) +{\rm i} \, \partial_\varphi (\ref{tw_phi_new})$ and  
$(\ref{tw_r_new}) +{\rm i} \, \partial_\varphi (\ref{tw_r_new})$,  
we see that all terms in these combinations are $\propto {\rm e}^{-i\varphi}$. 
In particular, the terms containing $W$ and $\partial_\varphi W$ pass into the terms respectively containing
$-{\rm i} {\bf W}^\prime {\rm e}^{-i\varphi}$ and ${\bf W}^\prime {\rm e}^{-i\varphi}$. 

As a result, we obtain the following complex equations
\begin{equation}
\dot {\bf A} - (i-\alpha)\Omega {\bf A} +
\frac{\kappa^2}{2\tilde\Omega} {\bf B} = -\frac{3}{2} i \alpha K_1
(U^\tau)^2 U^\varphi \Omega {\bf W}^\prime \label{compl_A},
\end{equation}
\begin{equation}
\dot {\bf B} -  (i-\alpha)\Omega {\bf B} - 2\tilde \Omega {\bf A}
= - (i+\alpha) U^\varphi\Omega\, {\bf W}^\prime \label{compl_B},
\end{equation}
where we have used equation (\ref{alpha_prescr}), as well as expressions for frequencies (\ref{Om_circ}), (\ref{kappa_tw}) and (\ref{tilde_freq})
obtained in Section \ref{sec_completion}.

In a similar way, by using (\ref{vAB}) and (\ref{compl_ampl})  and composing the combination
$(\ref{tw_eq}) +{\rm i} \, \partial_\varphi (\ref{tw_eq})$,  we derive an equation for complex amplitudes. 
In the right-hand side of this equation, the integration over $\xi$ should be performed under the derivative with respect to $r$.
For an isothermal disk with density distribution (\ref{rho_xi}), the equality
$\int \rho \xi^2 d\xi = \Sigma h^2$ holds. Thus, the derivative with respect to $r$ 
acts on terms proportional to $\Sigma h^2$ or $\bar \eta$. 
Instead of these combinations, we substitute equations (\ref{sigma_h2}) and (\ref{eta_tw}) there
and group common constant factors before the derivative with respect to $r$. 
Additionally, instead of $U^r$ and $\bar T^{r\varphi}$ we substitute
expressions (\ref{tw_Ur}) and (\ref{T_tw1})
in the left-hand side of the discussed equation and then divide the whole equation by $\Sigma$.
The obtained equation contains $\dot M$ and $\Sigma$ only in the combination $\dot M/\Sigma$,
which we express through $\delta^2$ and other known quantities using (\ref{sigma_h2}).
Also using the expression for Lense-Thirring frequency (\ref{lense_tirring_fin}), we finally arrive at the following equation
\begin{multline}
\dot {\bf W} - i\Omega_{LT} {\bf W} + \frac{3}{2} \alpha \delta^2
\frac{K_1^2}{K_2} U^\varphi \left ( U^\tau - K_1 (r K_2)^{1/2}
\frac{U^\varphi}{L} \right ) {\bf W}^\prime = \\
\frac{\delta^2 K_1^3 U^\varphi}{2 r^{1/2} K_2^{3/2} L} \frac
{\partial}{\partial r} \left \{ r^{3/2} K_2^{1/2} \frac{L}{K_1^2
U^\tau U^\varphi} (\,\,(i+\alpha){\bf B} +\alpha U^\varphi {\bf
W}^\prime\,) \right \}. \label{compl_W}
\end{multline}

Equations (\ref{compl_A}-\ref{compl_W}) form a closed system of equations for the quantities ${\bf A}$, ${\bf B}$ and ${\bf W}$ 
as functions of $r$ and $\tau$.
In the weak gravity limit they reduce to equations (30), (31) and (33) in \cite{Dem_Iv}.

\section{Stationary twisted disk}

\subsection{Main equation and boundary condition}

We now consider stationary solutions of the system of equations (\ref{compl_A}-\ref{compl_W}). 
The main goal of this Section is to calculate the shape of a stationary twisted disk. 

We set $\dot {\bf A} = \dot {\bf B} = \dot {\bf W} = 0$.
After eliminating ${\bf A}$ from (\ref{compl_A}-\ref{compl_B}) we obtain
\begin{equation}
\label{B_W}
\left [ 1 +
\frac{\kappa^2}{(i-\alpha)^2 \Omega^2} \right ] (i-\alpha)\Omega
{\bf B} = \left [ (i+\alpha)U^\varphi \Omega -
\frac{3i\alpha}{i-\alpha} K_1 (U^\tau)^2U^\varphi \tilde \Omega
\right ] {\bf W}^\prime,
\end{equation}
whence we express ${\bf B}$ through ${\bf W}^\prime$ and substitute it in equation (\ref{compl_W}).  
We thus obtain the equation
\begin{equation}
\frac{K_1}{r_S^{1/2} L} \frac{d}{d r_S} \left ( \frac{r_S^{3/2} L}{K_1
U^\tau} f^{*}(\alpha, r_S) \frac{d {\bf W}}{d r_S} \right ) - 3\alpha
U^\tau (1-L^{-1}) \frac{d {\bf W}}{d r_S} + \frac{4ia}{\delta^2
K_1^3 r_S^3 U^\varphi} {\bf W} = 0, \label{tw_stat}
\end{equation}
where the asterisk denotes complex conjugation and
\begin{equation}
f(\alpha, r_S) = (1 + \alpha^2 - 3i\alpha K_1^2)\,\,
\frac{r_S(i-\alpha)} {\alpha r_S(\alpha+2i)-6} + \alpha. \label{f_alpha}
\end{equation}

We note that equation (\ref{tw_stat}) was written after changing to the Schwarzschild radial coordinate $r_S$. 
In what follows, we wish to consider only the case $a>0$, i.e. a prograde disk. It can be seen that the problem has two free parameters. 
First of all, this is the combination $\tilde\delta \equiv \delta_*/\sqrt{|a|}$. 
Clearly, $\tilde\delta$ ranges from $0$ to $\infty$ 
and characterizes the relative role of the hydrodynamic and gravitomagnetic forces acting on the disk rings. Second, (\ref{tw_stat})
contains the disk viscosity parameter $0<\alpha<1$.
Equation (\ref{tw_stat})  in the rigorous Newtonian limit with non-zero viscosity reproduces the corresponding equation (2.10)
from \cite{KP_1985} and, additionally, with post-Newtonian corrections, reproduces equation (33) from \cite{II_1997}, 
what is checked in \cite{ZhI} (see paragraph 4.1 therein).

The coefficients of equation (\ref{tw_stat}) have a singular point at the inner edge of the disk at $r_S = \bar r_S \equiv 6$, where $L$ 
vanishes. The regularity of the solution at $\bar r_S$  must yield a condition for the function ${\bf W}$. 
Using this condition as the initial one, we can integrate (\ref{tw_stat}) from $\bar r_S$
to infinity and to obtain the form of the stationary twisted disk. We expand equation (\ref{tw_stat}) in series in orders of the small $x_0=r_S - \bar r_S\ll 1$.
In practice, to do this, all quantities that take non-zero values at $\bar r_S$ should be set exactly equal to these values and $L$ should be
expanded to the main order in $x_0$. From (\ref{L_expr}), we find
\begin{equation}
\label{L_appr}
L\approx \frac{x_0^2}{72},
\end{equation}
whence we see that another quantity in (\ref{tw_stat}) that vanishes at the inner disk edge, $\delta$, can be written as
$$
\delta = \delta_{ms} x_0^{2\epsilon}, 
$$
where $\epsilon$ is the power-law exponent $L$ in equation (\ref{delta_1}). 
Accordingly, $\delta_{ms}$ is also given by equation (\ref{delta_1}), 
which is taken at $\bar r_S$ and into which we now substitute the coefficient $72^{-1}$ from (\ref{L_appr}) instead of $L$.

After that, it is easy to obtain the equation valid for $x_0\ll 1$,
\begin{equation}
\label{bound_eq}
\frac{d}{dx_0} \left ( x_0^2 \frac{d{\bf W}}{dx_0} \right ) + C_1 x_0^{2-4\epsilon} {\bf W} + C_2 \frac{d{\bf W}}{dx_0} = 0,
\end{equation}
where 
$$
C_1 = - \frac{2{\rm i}}{f(\alpha,r_S)} \frac{U^\tau}{U^\varphi} \frac{\Omega_{LT}}{K_1^3 r_S\delta_{ms}^2}
$$
and
$$
C_2 = - \frac{216\alpha}{f(\alpha,r_S)} \frac{(U^\tau)^2}{r_S}
$$
are taken at $\bar r_S$.
We see that for any finite viscosity the last term in (\ref{bound_eq}) becomes dominant sufficiently close to the disk edge; 
therefore, the boundary condition can be straightforwardly written as
\begin{equation}
\label{bound_cond1}
\frac{d{\bf W}}{dx_0} \Biggr |_{\bar r_S} = 0. 
\end{equation}

On the other hand, from (\ref{bound_eq}) with $\alpha=0$, we obtain a simpler equation whose solution is a Bessel function:
\begin{equation}
\label{bound_cond2}
{\bf W} = C x_0^{-1/2} J_\frac{1}{2-4\epsilon} \left ( z \right ),
\end{equation}
where
\begin{equation}
\label{z}
z = \sqrt{C_1} \frac{x_0^{1-2\epsilon}}{1-2\epsilon}.
\end{equation}
As $x_0\to 0$ (\ref{bound_cond2}) tends to a non-zero constant but with a zero derivative with respect to $x_0$. 
Therefore, in this case we return to condition (\ref{bound_cond1}). 

Due to the linearity of the problem, it suffices to take an arbitrary non-zero value of ${\bf W}$ in $\bar r_S$, to set the first derivative of ${\bf W}$ at $\bar r_S$
equal to zero, and with these boundary conditions to integrate (\ref{tw_stat}) up to infinity. 
The modulus and phase of ${\bf W}$ give the profiles $\beta(r_S)$ and $\gamma(r_S)$ for a stationary twisted disk. 
In what follows, we normalize the profile $\beta$ to unity at infinity.

\subsection{Disk with a marginally small viscosity}

We consider a disk with a very low viscosity separately. 
Clearly, it is possible to analytically treat the accretion disk setting formally $\alpha\to 0$, if simultaneously $\dot M \to 0$. 
In such a disk, $U^r\to 0$, however, it then has definite profiles of $\Sigma$ and $h$. 

In addition, to obtain an analytical solution, we consider the case $\tilde\delta\ll 1$; 
in other words, we assume a sufficiently thin disk around a rapidly rotating black hole. 

Setting $\alpha=0$ in (\ref{tw_stat}) yields
\begin{equation}
{d\over dr_S} \left (b {d\over dr_S} {\bf W}\right )+ \lambda {\bf W} =0,
\label{eq_invisc}
\end{equation}
where
\begin{equation}
b= {r_S^{5/2}L\over K_1 U^{\tau}}, \quad \lambda = {24 a L\over
\delta^2 K_1^4 U^{\varphi} r_S^{5/2}}. \label{b_lam}
\end{equation}

The coefficients in (\ref{eq_invisc})
take real values; therefore, there are real solutions of this equation. This means that in the absence of viscosity in the stationary twisted disk
$\gamma=const$ which can be set equal to zero by the corresponding choice of the reference frame. Therefore, the variable ${\bf W}$ is identical to the angle  $\beta$ in this Section.

\subsubsection{The form of the disk near its inner edge}

In the foregoing, we have already presented the solution near the inner edge of the inviscid disk (see equation (\ref{bound_cond2}) ). 
The constant $C_1$  in this case has the explicit form:
\begin{equation}
\label{C_1}
C_1 = {24 a U^{\tau}\over r_S^5K_1^3U^{\varphi}\delta^{2}_{ms}},
\end{equation}
and is taken at $r_S=\bar r_S$.

Using the known approximation for the Bessel function of small argument, we obtain a relation between the constant $C$ in (\ref{bound_cond2})
and the value of ${\bf W}$ at $\bar r_S$, 
${\bf W}(r_S) \equiv {\bf W}_0$:
\begin{equation}
C=\Gamma \left ({3-4\epsilon \over 2(1-2\epsilon)} \right ) \left ({\sqrt \chi \over
2(1-2\epsilon )}\right )^{-1/2(1-2\epsilon)}{\bf W}_{0}, \label{C_W0}
\end{equation}
where $\Gamma (x)$ is the gamma-function. 

In addition, we need the asymptotic of (\ref{bound_cond2})
for $z\gg 1$. Clearly, $z$ can be large, even for $x\ll 1$, because $\sqrt{C_1} \sim \tilde\delta^{-1}\gg 1$. 
Hence, for  $z\gg 1$ we obtain
\begin{equation}
{\bf W}\approx  C\sqrt{2\over \pi xz}\cos \left (z -{\pi \over
2}{1-\epsilon \over 1- 2\epsilon }\right ). \label{large_z}
\end{equation}

\subsubsection{The form of the disk at large distance}

We consider equation (\ref{tw_stat}) for $r_S\gg 1$ and $\alpha\to 0$. 
Importantly, we cannot set all variables in (\ref{tw_stat}) to      
their Newtonian values and have the viscosity simultaneously vanished. This already follows from the fact that then
$f(\alpha,R) \to 1/(2\alpha) \to\infty$. 
Physically, this reflects the fact that, as we mentioned above, in the absence of viscosity in the strictly Newtonian potential, 
a Keplerian resonance occurs when the circular and epicyclic frequencies coincide, and perturbations in the twisted disk grow 
infinitely due to the action of the radial projection of the vertical pressure gradient. 
Therefore, a stationary twist is impossible in this case. Taking the next-order term in the expansion of
$f(\alpha,r_S)$ in small $r_S^{-1}$ into account, we obtain
\begin{equation}
\label{f_appr}
f(\alpha,r_S) \approx \frac{1}{2\alpha \left ( 1 + \frac{3{\rm i}}{\alpha r_S} \right )}.
\end{equation}
As $\alpha\to 0$, $f(\alpha,r_S)$ now remains finite at any finite $r_S$.
Nevertheless, it makes the leading contribution due to relativistic effects, and all other variables in
(\ref{tw_stat}) can now be set equal to their Newtonian values $U^\tau=1$, $U^\varphi = r_S^{-1/2}$, $L=1$ and $K_1=1$.
Moreover, we neglect the weak dependence of $\delta$ on $r_S$ far from the black hole and set $\delta=\delta_*$.

After that, by introducing the new independent variable $x_1 \equiv r_S^{-1/2}\ll 1$, we obtain the equation
\begin{equation}
\label{eq_largeR}
x_1{d^2\over dx_1^2}{\bf W}-2{d\over dx_1}{\bf W}+96{\tilde\delta}^{-2}x_1^4{\bf W}=0.
\end{equation}

The solution of (\ref{eq_largeR}) can again be expressed in terms of a Bessel function:
\begin{equation}
{\bf W}=x_1^{3/2}(A_1J_{-3/5}(z_1)+A_2J_{3/5}(z_1)), \label{sol_largeR}
\end{equation}
where
\begin{equation}
z_1={8\over 5}\sqrt 6{\tilde \delta}^{-1}x_1^{5/2},\label{z_1},
\end{equation}
$A_1$ and $A_2$ are constants.

When $r_S$ is so large that $z_1\ll 1$, the first and second terms in
(\ref{sol_largeR}), multiplied by $x_1^{3/2}$, respectively tend to a non-zero constant and to zero.
This allows us to express the constant $A_1$ in terms of the value of ${\bf W}$ at infinity, ${\bf W}_\infty$:
\begin{equation}
{\bf W}_\infty ={\left ({5\over 4\sqrt 6}\right )}^{3/5}{{\tilde \delta
}^{3/5}\over \Gamma (2/5)}A_1. \label{W_infty}
\end{equation}

In the opposite case, at $z_1 \gg 1$, i.e. closer to the black hole, we obtain another asymptotic form:
\begin{equation}
{\bf W}\approx \sqrt {{5\tilde \delta \over 2\pi
\sqrt{24}}}r_S^{-1/8}\left [A_1\cos \left (z_1 +{\pi \over 20}\right ) + A_2\sin \left (z_1
-{\pi \over 20}\right )\right ]. \label{large_z1}
\end{equation}

\subsubsection{WKB-solution for the shape of the disk}

Throughout the disk the asymptotic solutions (\ref{large_z}) and (\ref{large_z1}) can be matched by a WKB-solution of equation (\ref{eq_invisc}). 
Indeed, because we are considering the case $\tilde\delta \ll 1$, 
the ratio of $\lambda $ and $b$ in (\ref{eq_invisc}), $\tilde \lambda=\lambda/b$, is large at all $r_S$ such that $z$ and $z_1$ are large. 

The WKB-solution has the form
\begin{equation}
{\bf W}\approx {C_3\over (\lambda b )^{1/4}}\cos
\left (\int^{r_S}_{\bar r_S} \sqrt {\tilde \lambda } dr_S +\phi_{WKBJ}\right ),
\label{wkbj_sol}
\end{equation}
where the constants $C_3$ and $\phi_{WKBJ}$ should be chosen such that (\ref{wkbj_sol}) 
is smoothly matched with formula (\ref{large_z}) in the corresponding region. 
It can be checked that this yields
\begin{equation}
\phi_{WKBJ}= -{\pi \over 2} {1-\epsilon \over 1-2\epsilon }
\label{phi_wkbj}
\end{equation}
and
\begin{equation}
C_3=6^{1/4}\sqrt {{1-2\epsilon \over \pi K_1 U^{\tau}}}C,
\label{C_3}
\end{equation}
where we assume that $K_1$ and $U^{\tau}$ 
are taken at $r_S=\bar r_S=6$ and $L\approx x^2/72$ near $\bar r_S$.

Next, in the limit $r_S\rightarrow \infty $ 
we can set $\lambda $ and $b$ before the cosine in (\ref{wkbj_sol}) equal to their Newtonian values. In addition, the integral in (\ref{wkbj_sol}) 
can be represented as $I(r_S)\equiv\int^{r_S}_{\bar r_S} \sqrt {\tilde \lambda
}dr_S = I- \int^{\infty}_{r_S} \sqrt {\tilde \lambda }  dr_S $, where
$I=\int^{\infty}_{\bar r_S}  \sqrt {\tilde \lambda } dr_S $. 
Taking into account that the Newtonian value $\tilde \lambda
=24{\tilde \delta }^{-2}R^{-9/4}$, we have that  $\int^{\infty}_{r_S}\sqrt
{\tilde \lambda } dr_S  \approx {8\sqrt 6\over 5}{\tilde \delta
}^{-1}r_S^{-5/4},$ and therefore
\begin{equation}
{\bf W}\approx C_3 {{\tilde \delta }^{1/2}\over 24^{1/4}}\cos
\left ({8\sqrt 6\over 5}{\tilde \delta }^{-1}r_S^{-5/4} - I-\phi_{WKBJ}\right ).
\label{sol_wkbj_largeR}
\end{equation}
Solution (\ref{sol_wkbj_largeR}) must be smoothly matched with expression (\ref{large_z1}) in the corresponding region, which yields the values of constants $A_1$ and $A_2$. 
It can be checked that they are
\begin{eqnarray}
A_1=\sqrt {2\pi \over 5}C_3 \cos \left.\left (I+\phi_{WKBJ}-{\pi \over
20}\right )\right /\cos {\pi \over 10},\nonumber\\ \quad A_2=\sqrt {2\pi \over 5}C_3 \sin
\left.\left (I+\phi_{WKBJ}+{\pi \over 20}\right )\right /\cos {\pi \over 10}. \label{A_1_2}
\end{eqnarray}

Thus, equations (\ref{bound_cond2}), (\ref{wkbj_sol}) and (\ref{sol_largeR})  jointly with coefficients (\ref{C_3}), (\ref{A_1_2}) 
and phase (\ref{phi_wkbj}), determine the shape of an inviscid stationary relativistic twisted disk at all distances in the range from $r_S=\bar r_S$ to $r_S=\infty$.

\subsubsection{Resonance solutions in the low-viscosity disk}

We note that equations (\ref{C_W0}), (\ref{C_3}), (\ref{A_1_2}) and (\ref{W_infty}) provide a relation between ${\bf W}_0$ and ${\bf W}_\infty$:
\begin{equation}
{\bf W}_\infty = C_{tot}(\tilde
\delta) {\bf W}_0, \label{W_0_infty} 
\end{equation} 
where the explicit form of $C_{tot}(\tilde \delta )$ follows from these formulas. In particular, as follows from (\ref{W_infty}) and (\ref{A_1_2}), 
$C_{tot}(\tilde \delta )\propto \cos
(I+\phi_{WKBJ}-{\pi \over 20})$. 

We hence conclude that for some discrete set of
$\tilde \delta$ for which
$\cos (I+\phi_{WKBJ}-{\pi \over 20})=0$ so 
${\bf W}_\infty =0$ despite that ${\bf W}_0\ne 0$. 

From equations (\ref{b_lam}), it is possible to represent the integral
$I$ in the form
$I=\tilde \delta^{-1} \tilde I$, where $\tilde I$ does not depend on
$\tilde \delta $. 
This allows us to write the singular values of $\tilde\delta$ explicitly:
\begin{equation}
\tilde \delta_k = {\tilde I\over {\pi \over 2}{\left ({11\over 10}+
{1-\epsilon \over 1 -2\epsilon}+2k\right)}}, \label{delta_res}
\end{equation}
where $k$ is an integer number.

\begin{figure}
\begin{center}
\vspace{1cm}
\includegraphics[width=10cm,angle=-0]{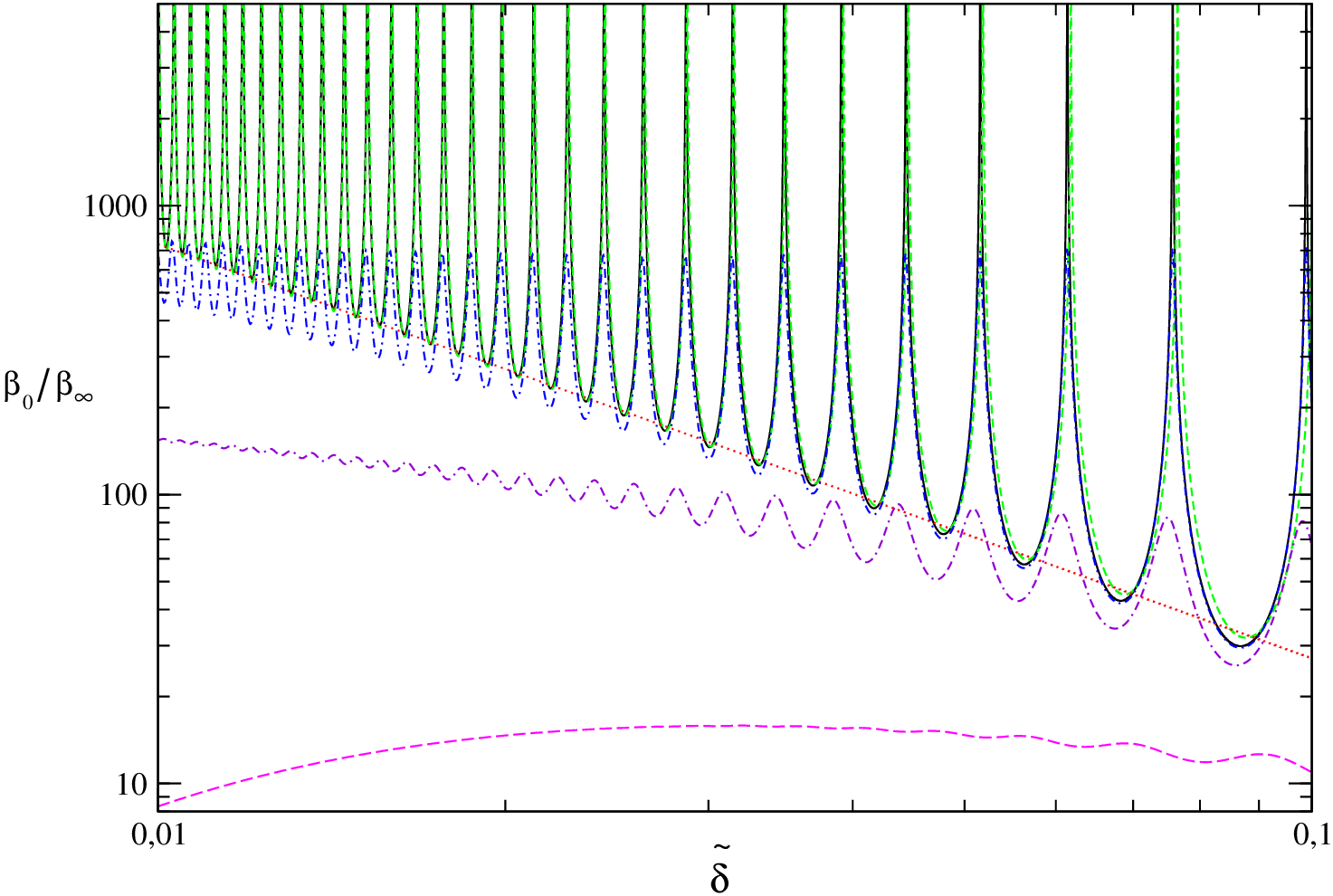}
\end{center}
\caption{
Ratio of the tilt angle of the inner disk edge to the tilt at infinity, $\beta_{0}/\beta_{\infty}$, 
as a function of the parameter $\tilde \delta$. 
The solid curve shows the numerical solution of equation (\ref{tw_stat}) with $\alpha=0$, 
the dotted curve represents the analytical dependence $C_{tot}^{-1}(\tilde \delta )$, where $C_{tot}$ is given by equation (\ref{W_0_infty}). 
The dashed, dash-dotted and dash-dot-dashed curves are obtained by numerical integration of equation (\ref{tw_stat}) with
$\alpha=10^{-4}$, $10^{-3}$ and $10^{-2}$, respectively.}
\label{fig_res}
\end{figure}

The values of $\tilde\delta_k$ 
correspond to such a balance between the external gravitomagnetic force and the internal pressure gradient in the disk 
that leads to the disk twist even if the matter flowing into the disk at infinity moves in the equatorial plane of the black hole. 
We note that the solution in the form of a flat disk lying entirely in the black hole equatorial plane, of course, also exists for these
$\tilde\delta_k$. 
This non-uniqueness of the solution disappears for any small viscosity in the disk, for which 
${\bf W}_\infty=0$ always implies ${\bf W}_0=0$. For small $\alpha\ll 1$, 
the disk 'feels' these 'resonance' solutions, and its inner parts 
deviate significantly from the equatorial plane of the black hole, 
even when the outer parts of the disk lie almost in the equatorial plane. Figure (\ref{fig_res})
shows the curve corresponding to the analytical solution (\ref{W_0_infty}), 
as well as several curves for a viscous twisted disk obtained by integrating the original equation (\ref{tw_stat}).  
We see that already for $\alpha=10^{-3}$ the discussed resonances are almost entirely suppressed.

\subsection{The disk behavior on the plane of parameters $\alpha$ and $\tilde\delta$}

\begin{figure}
\begin{center}
\vspace{1cm}
\includegraphics[width=10cm,angle=-0]{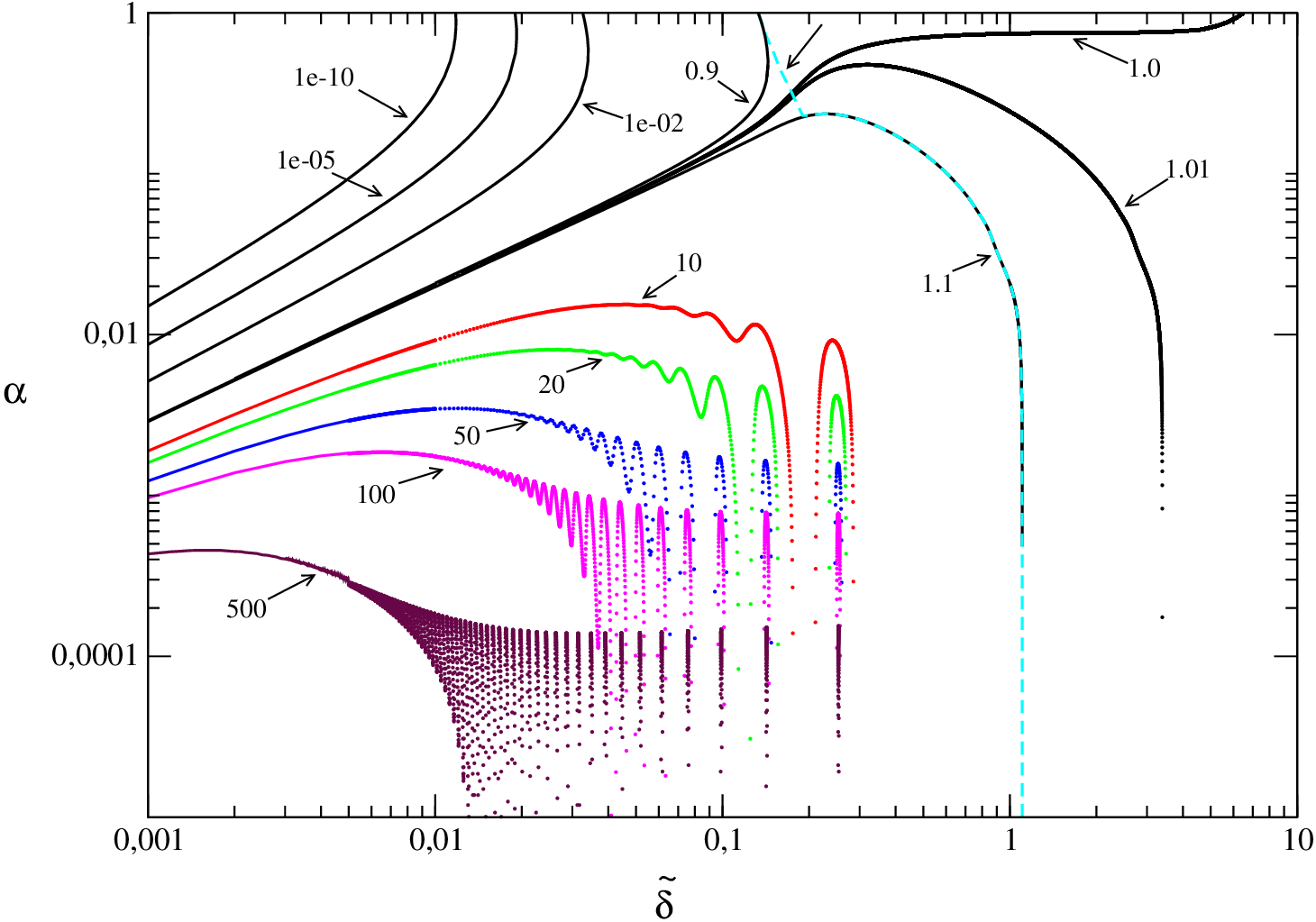}
\end{center}
\caption{
Contours of constant ratio $\beta_{0}/\beta_{\infty}$
on the parameter plane $(\tilde \delta$, $\alpha )$. 
The numbers show the value of $\beta_{0}/\beta_{\infty}$
for each curve. The dashed curve in the right part of the Figure separates the region where the change in
$\beta$ with $r_S$ is more than 10\% of $\beta_{\infty}$ 
(to the left) from the region where the disk twist is insignificant, and $\beta$ deviates by less than
10\% from $\beta_\infty$ (to the right).} \label{fig_plane}
\end{figure}

In conclusion, we present the full study of regimes of behavior of a stationary twisted relativistic disc near a rotating black hole.
It is convenient to show the results of numerical integration of equation (\ref{tw_stat}) on the plane of free parameters of the problem, 
$\tilde\delta$ and $\alpha$. The first parameter varies in the range $10^{-3}<\tilde\delta<10$ and the second parameter varies in the range  $0<\alpha<1$. 
As follows from Fig. (\ref{fig_plane}), at small $\tilde\delta$, i.e. when the gravitomagnetic force exceeds the internal forces in a twisted disk, 
it either lies in the equatorial plane of the black hole, i.e. $\beta_0/\beta_\infty\to 0$, 
or, conversely, the tilt of its rings strongly increases in the inner parts of the disk, with oscillations of
$\beta(r_S)$ along the radial coordinate. Note that for small viscosity, these oscillations become so strong that the corresponding gradient of the tilt angle, 
$\beta^\prime$, leads to supersonic perturbations of the velocity components, $v_r$ and $v^\varphi$, at heights of the order of the disk thickness,
$\xi \sim h$. 
This, in turn, must lead to the generation of various hydrodynamic instabilities and sound waves, which cause additional disk heating (and hence also an increase in
$\tilde\delta$), as well as the growth of $\alpha$. These processes should partially suppress the oscillations of $\beta$ discussed above. 

\begin{figure}
\begin{center}
\vspace{1cm}
\includegraphics[width=10cm,angle=-0]{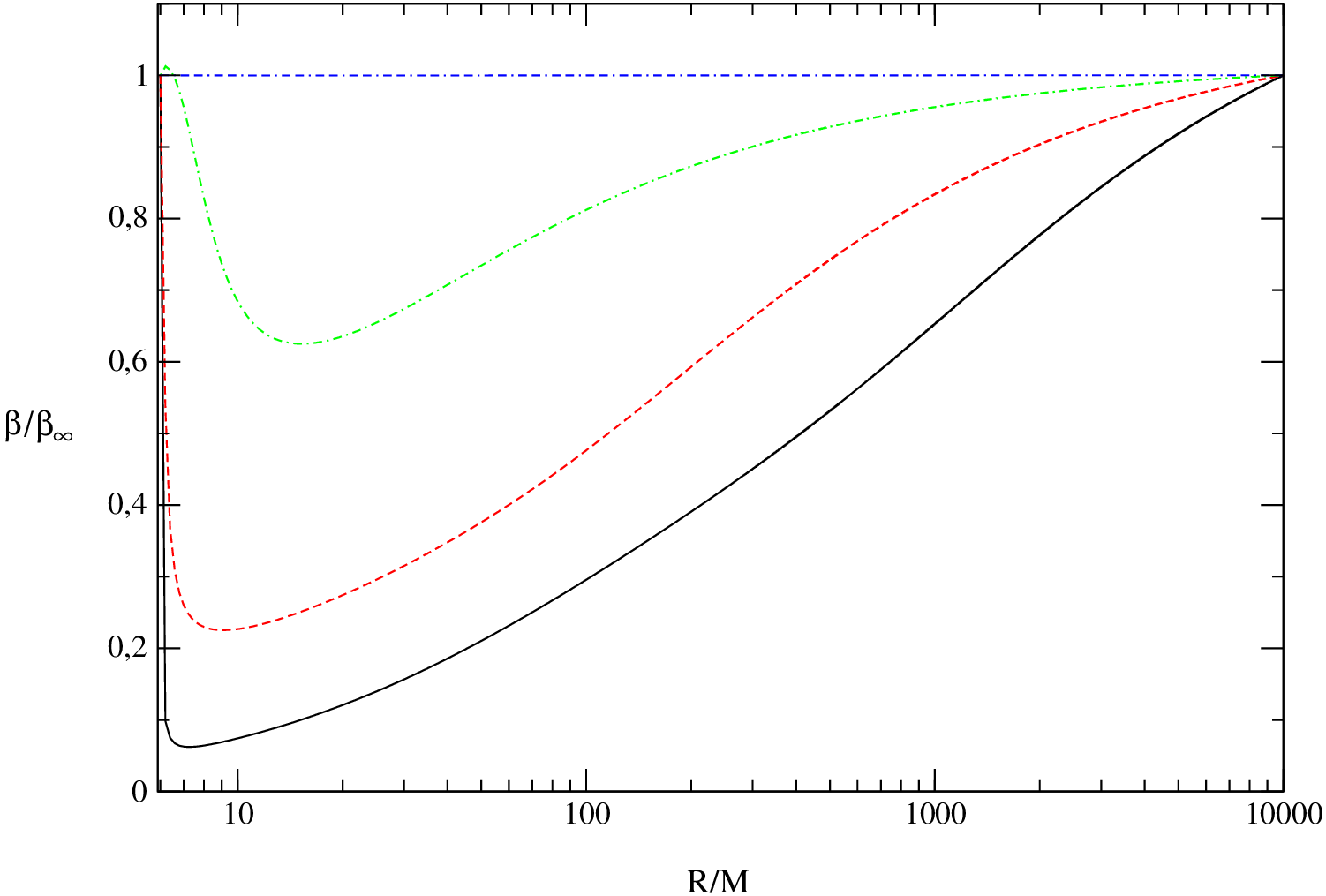}
\end{center}
\caption{
The dependence of $\beta $ on $r_S$ along the curve in Fig. (\ref{fig_plane}) 
for which $\beta_{0}/\beta_{\infty}=1$. 
$\tilde \delta $ takes the values  $10^{-3}$, $10^{-2}$, $10^{-1}$ and $1$ 
for the solid, dashed, dash-dotted and dash-dot-dashed curves, respectively. } 
\label{fig_beta}
\end{figure}

The disk alignment into the equatorial plane of the black hole occurs at sufficiently high viscosity, when the condition
$\alpha > \tilde\delta$ is satisfied with a large margin, and is referred to as the Bardeen-Petterson effect \cite{BP_1975}. 
It is seen from Fig. (\ref{fig_plane}) that this effect occurs only in sufficiently viscous and thin disks. 
But already for $\tilde\delta \sim \alpha$ the ratio $\beta_0/\beta_\infty$ becomes of the order of unity, which means the absence of the disk alignment. 
At the same time, oscillations of $\beta$ disappear. Figure (\ref{fig_beta}) shows the profiles of $\beta(r_S)$ when $\beta_0/\beta_\infty=1$ 
for several $\tilde\delta$. It is seen that for not very small $\tilde\delta$, 
the twisted disk has a sufficiently smooth shape, which suggests the possibility of the existence of such configurations in nature. We note that
$\beta$ behaves non-monotonically: it first decreases and then increases with the decrease of $r_S$. 
The latter can have important implications both for the disk structure itself and and for its observational manifestations. For example, the hot inner regions of such a disk should 
illuminate its outer parts much stronger compared to the flat disk case. Clearly, this is due to the disk inner parts being tilted with respect to the outer parts.

In the region where $\tilde\delta$ is of the order of or greater than unity, the action of the gravitomagnetic force becomes insignificant, and the disk is weakly twisted. 
In Fig. (\ref{fig_plane}), the area to the right of the dashed line is where $\beta(r_S)$ deviates from $\beta_\infty$ by less than 10\%.
It is also worth noting that for $\tilde\delta > 0.1$ the Bardeen-Petterson effect is completely absent for any $\alpha$.

\section{Conclusion}

We have presented a detailed technical derivation of the governing equations for the evolution of the shape of a relativistic twisted disk, 
as well as for perturbations of the velocity and density inside it. Only three simplifying assumptions
have been used: the smallness of the disk aspect ratio, $\delta\ll 1$, the slowness of the black hole rotation, $a\ll 1$, 
and the smallness of the disk's rings tilt to the equatorial plane of the black hole, $\beta\ll 1$.
This allowed us to formulate equations (\ref{tw_phi_new}), (\ref{tw_r_new}) and (\ref{tw_eq}) 
for three variables describing Eulerian perturbations of the azimuthal velocity,
$v^r$ and $v^\varphi$ and the geometrical form of the disk, $Z$. In general, the dependence of 
$v^r$ and $v^\varphi$ on the twisted coordinates $r$, $\xi$ and $\tau$, 
and the dependence of $Z$ on $r$ and $\tau$ should be found. In accordance with equation (\ref{ZUW}), all these variables depend harmonically on the azimuthal coordinate. 
The governing equations contain the profiles of the background solution, representing an accretion disk
with similar radial and vertical structure but lying in the equatorial plane of the black hole. 
We note once again that not only the twisted disk but also the background itself can be non-stationary, because when deriving the system of equations 
(\ref{tw_phi_new}, \ref{tw_r_new}, \ref{tw_eq}), the only one assumption about the background, the smallness of $\delta\ll 1$, was used. 
Therefore, the twist equations enable us also to study the evolution of tilted/twisted gaseous tori/rings near the rotating black holes as they are spreading in the radial direction, in other words, as the non-stationary accretion goes on due to turbulent viscosity. 

In the particular case of a stationary, vertically isothermal background with $\alpha$-parametrization of the viscosity, 
the twist equations have been reduced to simpler equations (\ref{compl_A}), (\ref{compl_B}) and (\ref{compl_W})
for the complex amplitudes ${\bf A}$ and ${\bf B}$ describing the velocity perturbations, and ${\bf W}$ describing the disk geometry, which depend only on 
$r$ and $\tau$.  Here, the solution for a flat relativistic disk, which was presented in detail in Chapter 1, was utilized. 
The corresponding stationary problem can be described by a second-order linear differential equation for
${\bf W}$ (see equation (\ref{tw_stat}) ).
The analytic integration of this equation for a formally inviscid disk with $\tilde\delta\ll 1$ enabled us to find the singular resonance solutions for a discrete set $\tilde \delta_k$, which in fact correspond to an instability of a flat non-tilted disk, 
when the latter can acquire a twisted shape near the black hole, even with its outer part lying in the black hole equatorial plane. 
This instability, however, rapidly disappears already for $\alpha\sim 10^{-3}$ and for $\alpha >\tilde \delta$ provided that $\tilde\delta <0.1$, 
the numerical calculations show the Bardeen-Petterson effect. At the same time, already for $\alpha \sim \tilde\delta$, 
the alignment of the inner parts of the disk into the equatorial plane of the black hole is absent, and for $\tilde \delta \geq 0.1$ 
smooth but non-monotonic profiles $\beta(r)$ appear (see Fig. (\ref{fig_beta}) ), which suggests their stability under perturbations and 
the possibility of their realization in nature. The last effect is confirmed by the first numerical simulations of tilted thin relativistic accretion disks 
with $\delta \sim \alpha \sim a \sim 0.1$ carried out in recent papers \cite{Texeira_1} and \cite{Texeira_2}. 
In these papers, a comparison with the semi-analytic model based on the solution of the system of equations 
(\ref{tw_phi_new}, \ref{tw_r_new}, \ref{tw_eq}) was also done for a slightly tilted vertically barotropic torus.

Observational confirmations of the existence of twisted accretion disks around rotating black holes have just started emerging. 
Apparently, one of the most direct pieces of evidence of their existence is the observation 
of maser sources at subparsec scales in the disk around a supermassive black hole in the nucleus of NGC 4258 \cite{NM_1995}, \cite{Herr_1996}. 
The subsequent modeling in \cite{Martin_1} and \cite{Caproni_2007} 
showed that the disk twist in this case can be due to the Bardeen-Petterson effect. In the recent paper \cite{Wu_2013},
observations of jets in the nucleus of NGC 4248 were used to independently estimate the black hole Kerr parameter $a\sim 0.7$
and, in a similar model, to calculate the radius of the disk alignment into the equatorial plane of the black hole in agreement with observations. 
Additional but more indirect arguments favoring the presence of twisted disks in galactic nuclei were obtained, for example, in
\cite{Cadez_2003} and \cite{Cadez_2005}, where the observed profiles of X-ray iron line $K_\alpha$ 
were calculated for different accretion disk models. It was concluded that in many cases, the observed line profile can be more easily explained in the model of 
the twisted disk than in the model of a flat disk but, for example, with specific radial intensity distribution. In
\cite{Wu_2010}, a similar modeling of hydrogen Balmer lines was performed, which should arise due to the heating of the outer parts of 
a twisted disk by hard emission from its inner parts, which have much larger tilt than in the case of a flat disk. 
The presence of twisted disks is also suspected in binary stellar systems with black holes. 
For example, this can be the case in two microquasars, GROJ1655-40 and V4641 Sgr, in which the tilt of jets relative to the orbital plane was discovered
(see \cite{Martin_2} and \cite{Martin_3}).

As mentioned above, equations (\ref{tw_phi_new}), (\ref{tw_r_new}) and (\ref{tw_eq}) 
also describe the non-stationary dynamics of a torus tilted to the equatorial plane of a black hole. If
$\delta > \alpha$, the action of the gravitomagnetic force must lead to a solid-body precession of the torus, because in this case the twisted waves propagating 
at almost the speed of sound smear out the dependence of $\gamma$ on $r$ due to the Lense-Thirring effect. 
Similar non-stationary models are invoked to explain the variability of Balmer line profiles, as well as the precession of jets in active nuclei 
(see, e.g., \cite{Caproni_2004}).
In many papers, the precessing tori are used to explain low-frequency quasi-periodic oscillations in X-ray binary systems (see, e.g., \cite{Ingram_2013}). 
Of special interest is the modeling of observational appearances of a tilted accretion disk around the black hole in the center of our Galaxy
\cite{Dexter_2013}.

The theory of relativistic twisted disks presented here can also be successfully applied both to constructing self-consistent models of individual objects 
and to making further theoretical predictions on the dynamics of accretion flows around rotating black holes.

\section*{Acknowledgements}

The work is supported by the RSF grant 14-12-00146.

\end{document}